%% file: rpc.tex
\pdfoutput=1
\documentclass[a4paper,11pt]{article}
\usepackage{jinstpub}
\usepackage{subcaption}
\usepackage{multicol}
\usepackage{url}
\usepackage{hyperref}
\usepackage{hepnames}
\usepackage[sicmds,freestanding]{hepunits} 
\usepackage{booktabs}
\usepackage{chemformula}

\title{Description and stability of a RPC-based calorimeter in 
  electromagnetic and hadronic shower environments}

\author{\centering{\LARGE\bf The CALICE collaboration}\\
\vspace{0.3cm}
 }

\author{\centering
 J.\,Apostolakis, G.\,Folger, A.\,Ribon, E.\,Sicking
\\
{\it CERN, 1211 Gen\`{e}ve 23, Switzerland }\\
}

\author{\centering
 D.\,Boumediene$^a$, V.\,Francais
\\ 
{\it Universit\'e Clermont Auvergne, Universit\'e Blaise Pascal, CNRS/IN2P3, LPC, 4
Av. Blaise Pascal, TSA/CS 60026, F-63178 Aubi\`ere, France}\\
 }

\author{\centering 
K.\,Goto, K.\,Kawagoe, M.\,Kuhara, T.\,Suehara,
  T.\,Yoshioka, 
\\
{\it Department of Physics and Research Center for Advanced Particle Physics,
 Kyushu University, 744 Motooka, Nishi-ku, Fukuoka 819-0395, Japan}\\
 }

\author{\centering
A.\,Pingault, M.\,Tytgat 
\\ {\it Ghent University, Department of Physics and Astronomy, Proeftuinstraat
  86 (N3), B-9000 Gent, Belgium}\\
}

\author{\centering 
G.\,Garillot, 
G.\,Grenier, 
T. \,Kurca,  
I.\,Laktineh , 
B.\,Liu ,
B.\,Li,
L.\,Mirabito, 
A.\,Steen\footnote{Now at National Taiwan University, 10617, Taipei, Taiwan}
\\ {\it
Univ Lyon, Univ Claude Bernard Lyon 1, 
CNRS/IN2P3, IP2I Lyon, F-69622 
Villeurbanne, France}\\
 }

\author{\centering
R.\, \'Et\'e, K.\,Kr\"{u}ger, F.\,Sefkow
\\ {\it DESY, Notkestrasse 85, D-22603 Hamburg, Germany }\\
}

\author{\centering 
E.\,Calvo Alamillo,
C.\, Carrillo,
M.C.\,Fouz,
H.\, Garcia Cabrera,
J.\,Marin,
J.\,Navarrete,
J.\,Puerta Pelayo,
A.\,Verdugo
\\ {\it
CIEMAT, Centro de Investigaciones Energeticas, Medioambientales y Tecnologicas, Madrid, Spain}
}

\author{\centering
 F.\,Corriveau
 \\ {\it
 Department of Physics, McGill University, Ernest Rutherford Physics Bldg.,
 3600 University Ave., Montr\'{e}al, Qu\'{e}bec, Canada H3A 2T8}\\
 }

\author{\centering
 L.\,Emberger, C.\,Graf, F.\,Simon
 \\ {\it Max-Planck-Institut f\"ur Physik, F\"ohringer Ring 6, D-80805 Munich, Germany }\\
}

\author{\centering R.\,P\"oschl
 \\ {\it IJCLab, B\^at. 100, 15 rue Georges Clémenceau, CNRS / Université
   Paris-Saclay / Université Paris Cité, 91405 Orsay cedex, France }\\
}

\author{\centering D.W.\,Kim, S.W.\,Park
 \\ {\it Seoul National University Hospital, Bundang 13605, Republic of Korea
 }
}

\author{
\begin{flushleft}{\it
$^{a}$ Corresponding author\\
E-mail: djamel.boumediene@cern.ch}
\end{flushleft}}

\abstract{ 
The CALICE Semi-Digital Hadron Calorimeter technological prototype completed in 2011 is a sampling calorimeter using Glass Resistive Plate Chamber (GRPC) detectors as the active medium. This technology is one of the two options proposed for the hadron calorimeter of the International Large Detector for the International Linear Collider. The prototype was exposed in 2015 to beams of muons, electrons, and pions of different energies at the CERN Super Proton Synchrotron.  The use of this technology for future experiments requires a reliable simulation of its response that can predict its performance.  GEANT4 combined with a digitization algorithm was used to simulate the prototype. It describes the full path of the signal: showering, gas avalanches, charge induction, and hit triggering. The simulation was tuned using muon tracks and electromagnetic showers for accounting for detector inhomogeneity and tested on hadronic showers collected in the test beam. 
This publication describes developments of the digitization algorithm. It is used to predict the stability of the detector performance against various changes in the data-taking conditions, including temperature, pressure, magnetic field, GRPC width variations, and gas mixture variations. These predictions are confronted with test beam data and provide an attempt to explain the detector properties. The data-taking conditions such as temperature and potential detector inhomogeneities affect energy density measurements but have small impact on detector efficiency.
}

\def \degreecelsius {^\circ\mathrm{C}}

\begin{document}

\maketitle

\input{latex/introduction.tex} 

\input{latex/sdhcal.tex} 

\input{latex/simulation.tex} 

\input{latex/stability.tex} 

\input{latex/data.tex} 

\input{latex/conclusion.tex} 
\section{Acknowledgements}
This study was supported by the French ANR agency (DHCAL Grant) and the CNRS-IN2P3; by  the MCIN/AEI and the Programa Estatal de Fomento de la Investigaci\'on  Cientifica y T\'ecnica de Excelencia  Maria de Maeztu, grant MMDM-2015-0509; by the National Research Foundation of Korea, Grant Agreement 2019R1I1A3A01056616.   The authors would also like to thank the CERN accelerator staff for their precious help in preparing both the PS and the SPS beams.

\bibliographystyle{unsrt}
\bibliography{rpc}

\end{document}

%% file: latex/introduction.tex
\section{Introduction}

\begin{figure}[t]
 \begin{center}
  \begin{subfigure}[b]{0.49\textwidth}
    \includegraphics[width=\textwidth]{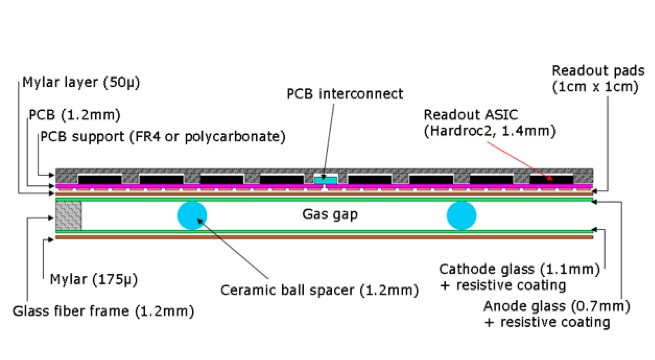}
    \caption{}
    \label{fig:rpc}
  \end{subfigure}
  \begin{subfigure}[b]{0.49\textwidth}
    \includegraphics[width=\textwidth]{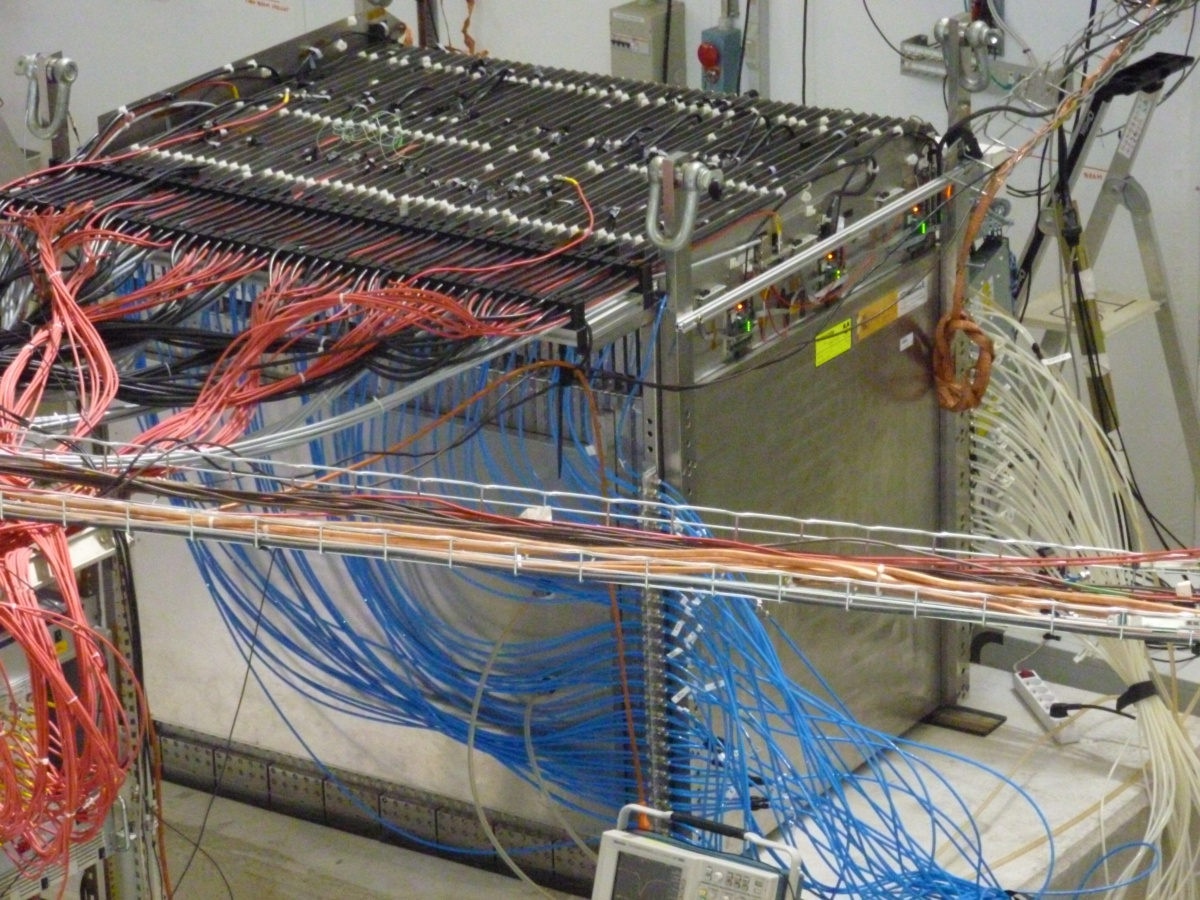}
    \caption{}
    \label{fig:proto}
  \end{subfigure}
  \caption{(\subref{fig:rpc}) Cross section of a GRPC
    chamber. (\subref{fig:proto}) The SDHCAL
    prototype in the CERN SPS area in 2015. The 48 layers are visible on the
    top side. The right side of the prototype was exposed to the beam.}  
  \end{center}
\end{figure}

The Semi-Digital Hadronic Calorimeter (SDHCAL)~\cite{sdhcal} is one of the
high-granularity calorimeter prototypes developed by the CALICE 
collaboration. This technology was optimised for the application of the
Particle Flow Algorithm~\cite{pfa} in an $\Ppositron\Pelectron$ collider environment. 
The SDHCAL is a sampling calorimeter where Glass Resistive Plate Chambers
(GRPC) are used as active medium while absorber layers are made of 2\,cm thick
stainless steel plates. The glass plates have a bulk resistivity of
  $10^{12}\,\ohm\centi{\meter}$ while the surface resistivity is
  $0.6-1\mega\ohm/\square$. The anode and cathode thicknesses are
  $0.7\,\milli{\meter}$ and $1.1\,\milli{\meter}$, respectively. The gas gap
  width is $1.2\,\milli\meter$. The geometry
  of the GRPC is described in Figure~\ref{fig:rpc}. The GRPC is
  placed inside a stainless steel cassette that plays the role of a Faraday
  cage. The longitudinal segmentation is given by 48 GRPC 
layers interleaved with absorbers, reaching a $1.3$\,m length and
6 interaction lengths ($\lambda_{\rm{I}}$). For each layer, the transverse segmentation is governed by the
$96\times96$ charge collection pads of $1\,{\rm cm^2}$ each, fine enough to
allow track reconstructions~\cite{hough} and particle identification using
machine-learning techniques~\cite{idmva}. The readout pads
  are isolated from the anode glass by a $50\,\micro{\meter}$ Mylar foil.
The charge collected by each pad is measured with a dedicated
ASIC~\cite{hardroc} that provides a three-threshold readout corresponding to
three amplitudes: about $0.1$, $5$ and $15$\,pC. The ASIC
  features a power-pulsing mode~\cite{powerpulsing}. It 
allows to place the readout electronics in an idle mode between two beam bunch
crossings. The active time corresponding to the beam bunch crossing in an 
$\Ppositron\Pelectron$ 
ILC-like experiment~\cite{ilc} is expected to
last $1\,\milli{\second}$ every $200\,\milli{\second}$. The power-pulsing
significantly reduces the power dissipation by a factor 100 to 200 such as no
active cooling is needed. The occupency rate is expected to be less than
$30\,\hertz/\centi{\meter}^{2}$ while the GRPC response is considered as
stable up to $100\,\hertz/\centi\meter^2$~\cite{rates}.

GRPC are usually used as tracking devices, while in the SDHCAL, we are confronted
with a large variety of particles and energies present in a hadronic shower.
Furthermore, the high energy density leads to multiple
  particles inducing a signal in the same pad.

The aim of the multi-threshold approach of the SDHCAL technology is to use
the energy density information. The three-amplitude information
provided by the detector is sensitive to the event type through the
distinction of multiple from single charged particle signal. It was shown that the SDHCAL provides precise 
energy measurements, especially when exploiting the semi-digital information, allowing to reach a 30\% 
improvement on the energy resolution with respect to a purely digital approach~\cite{1stresults}.  
However, the typical spread of the charge induced by a minimum ionizing particle (MIP) is significant compared to
its typical average value. 
Phenomena that bias the signal induced by the charge avalanches can
lead to a sizable change in the balance between the various threshold multiplicities.
This paper aims to study the dependency of the SDHCAL response 
using detailed detector simulations of the prototype
(Figure~\ref{fig:proto}.) and beam test data. Several 
phenomena that can affect the detector stability are reviewed. 

Section~\ref{sec:sdhcal} describes the prototype and beam test conditions.
The first set of results introduced in this publication is obtained by
modeling additional effects in the digitization procedure, based on
dedicated avalanche simulations (Section~\ref{sec:simu}).
Quantified estimates of the detector stability with respect to different
sources of signal bias are provided in Section~\ref{sec:stab}.
Section~\ref{sec:data} presents a second set of results based on
beam test data studies. 

%% file: latex/sdhcal.tex
\section{The SDHCAL Prototype at the CERN SPS beam test}
\label{sec:sdhcal}

The gas mixture used in the GRPC contains 93\% of TetraFluoroEthane
(\ch{C2H2F4}), 5\% of \ch{CO2}, which is a UV quencher gas and 2\% of \ch{SF6} which is electronegative gas that absorbs a fraction of the electrons in order to control the avalanche. The electric field is produced with a high voltage of about 6.9 kV.

For $1~\rm{m}^2$ of active layer, 144 HARDROC ASICS~\cite{hardroc} collect the signal from 9216 pads that are
located on the opposite face of the ASIC electronics board. Each ASIC handles 64
pads. The data are collected
by the ASIC until the RAM is full, i.e.\ until 127 events are recorded as
detailed in ref.~\cite{1stresults}. In each layer, the acquisition commands are sent to the ASICs by three
Detector InterFace (DIF) cards in charge of the data
acquisition. A masterboard controls these DIFs and is in charge of the data collection.

The SDHCAL prototype was exposed to beams of muons, electrons and pions during
beam test campaigns at the CERN Super Proton 
Synchrotron (SPS) in 2015. 
The data acquisition was performed in triggerless mode. The ASIC uses a
$5\,\mega\hertz$ clock to define a time slot. A time clustering method
is used. A time slot containing at
least 7 hits is selected. Hits belonging to the adjacent time slots are
aggregated to the selected time slot. They define a physical event whose
time length is $600\,\nano\second$~\cite{1stresults}. This method is optimised
for the rejection of intrinsic noise.

A cooling was needed due to the data-taking cycles at SPS
  that are longer than those of a full experiment and which limit the impact
  of the power-pulsing. The lateral sides of the calorimeter were equipped with an ad-hoc
 $10$\textdegree{}C water cooling system. 

\begin{figure}[t]
 \begin{center}
  \begin{subfigure}[b]{0.49\textwidth}
    \includegraphics[width=\textwidth]{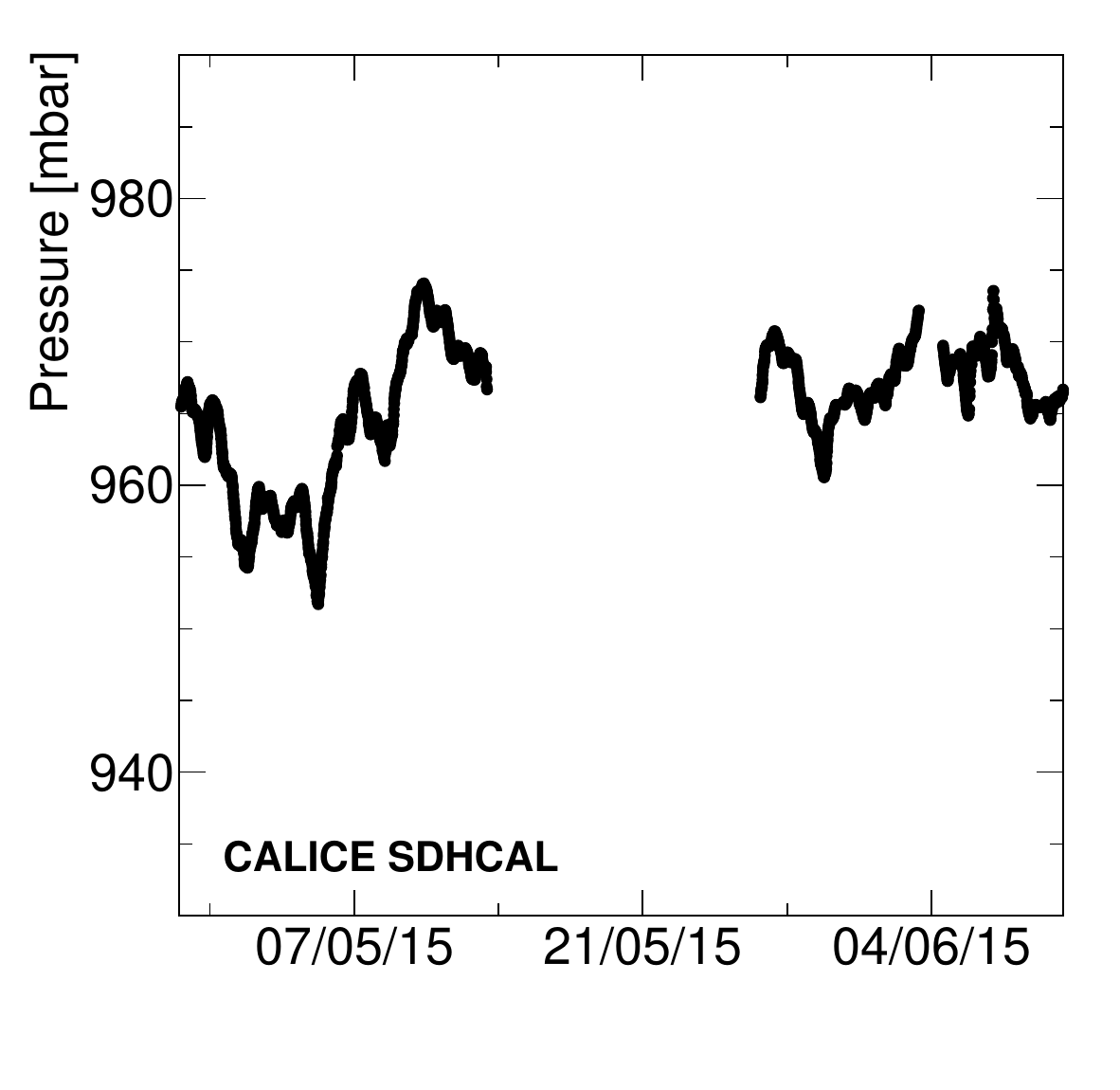}
    \caption{}
    \label{fig:p.data}
  \end{subfigure}
  \begin{subfigure}[b]{0.49\textwidth}
    \includegraphics[width=\textwidth]{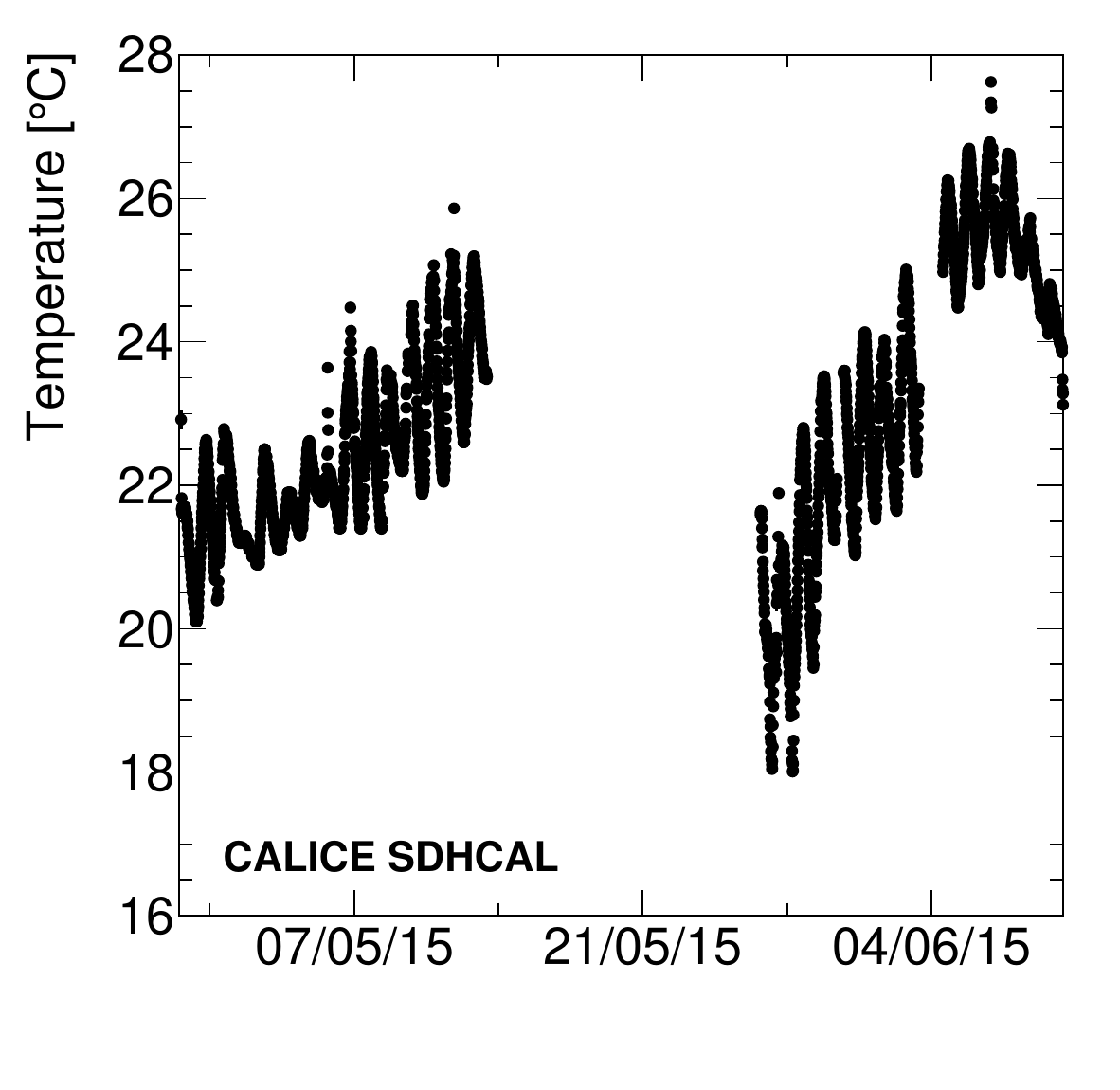}
    \caption{}
    \label{fig:t.data}
  \end{subfigure}
  \caption{ Evolution of the atmospheric pressure~(\subref{fig:p.data}) and of
    the temperature~(\subref{fig:t.data}) as measured during the 2015 beam test
    campaign at CERN SPS.}  
  \end{center}
\end{figure}

The atmospheric pressure, as well as the temperature, were monitored during
the 2015 data-taking campaigns. The temperature was measured on the outer side
of three GRPC chambers as an approximation of GRPC gas temperature.
The pressure and temperature variation are averaged over each hour and shown 
 in Figures~\ref{fig:p.data} and~\ref{fig:t.data}, respectively. The
temperature variations are noticeable, even during single data runs whose
duration is typically two to three hours.
 A set of data collected within 24 hours
is used to study the interplay between the stability of the 
response and temperature or pressure variations while the detector is operated
at a stable high voltage. 

\section{Data reconstruction}

The simulated, as well as the observed events, are reconstructed using the MARLIN
framework~\cite{marlin,marlinreco} within the ILC software framework~\cite{ilcsoft}.
The time of a hit is recorded and used to reject noise events and to select
the fired pads (hits) resulting from the particle interactions with the
detector material. 

Several beam energy points, from $7$ to $70\,\GeV$ were used to study the
observed or simulated detector response. 
Pion showers are selected from the data, and cuts are used to reject
electron and muon contaminations. These cuts include a threshold on the number
of hits: either no hit or more than three hits per layer. At least 30 layers have to contain a
hit. The layer at which the shower starts has to be at 
the $4^{th}$ position or more~\cite{1stresults}.
The high rate of the SPS beam induced efficiency losses that were observed
and corrected using a linear function of the time in the
spill~\cite{1stresults}.

The energy is reconstructed using a polynomial technique described
in ref.~\cite{1stresults}. It uses the amplitude threshold information that makes a
software compensation possible. It can also take advantage of the high granularity by
including reconstructed tracks~\cite{hough} and machine learning techniques~\cite{idmva}. 
The event reconstructed energy, $E_{\rm{event}}$, is calculated using the
polynomial combination method as follows:
\begin{equation}
E_{\rm{event}} = a(N_{\rm{tot}})\times N_{1} + b(N_{\rm{tot}})\times
N_{2} + c(N_{\rm{tot}})\times N_{3}
\label{eq:ereco}
\end{equation}
where $N_i$ is the number of hits exclusively associated to the $i^{\rm{th}}$ threshold and the
factors $a$, $b$ and $c$ are quadratic functions of the total number of
hits, $N_{\rm{tot}}$.
The factors are adjusted with a $\chi^2$ minimization method such as the
reconstructed energies fit to the expected ones:
\begin{equation}
\chi^2 = \sum\limits_{i=1}^{n} \left(E^i_{\rm{beam}}-E^i_{\rm{event}}\right)^2/E^i_{\rm{beam}}
\label{eq:chi2}
\end{equation}
where $n$ is the number of events used for the optimisation.
The mean reconstructed energy, $E_{\rm{reco}}$, and energy resolution,
$\sigma_{\rm{E}}$, are obtained from the mean and width of the Gaussian fit
to the energy distribution, respectively.
In the following, the calibration procedure refers to the measurement of $a$,
$b$ and $c$ from the data or simulated events. 

%% file: latex/simulation.tex
\section{Modeling of the detector response}
\label{sec:simu}
\subsection{Simulation of the SDHCAL prototype}

The detection process can be summarised by the following steps:
\begin{enumerate}
\item Interaction of the primary particle, mainly with the absorber,
  potentially leading to electromagnetic or hadronic showers;
\item Interaction of secondary charged particles within the particle shower with the GRPC gas medium leading to the ionization;
\item Charge avalanche development in the GRPCs;
\item Signal induction in the pads;
\item ASIC signal processing;
\end{enumerate}

The first two steps are simulated with the GEANT4 toolkit~\cite{geant4} as
described in Section~\ref{sec:geant4}. The three following ones are modeled 
with the so-called digitization algorithm described in
Section~\ref{sec:digitization}. Additional corrections of the charge avalanche
modeling are extracted from a dedicated simulation and added to the digitizer
as described in Section~\ref{sec:avalanche}. 

\subsection{Simulation of particle interactions with GEANT4}
\label{sec:geant4}
The particle interactions with the detector material are simulated using the version 9.6 of the GEANT4 toolkit, 
where the full SDHCAL geometry is
implemented. The \texttt{QGSP\_BERT\_HP} and \texttt{FTFP\_BERT\_HP} physics lists are used to simulate
hadronic and electromagnetic showers.
The GEANT4 simulated energy depositions in the detector active volumes are taken as the starting point of the
digitization procedure. They are associated with segments of a particle path, so-called steps.

\subsection{Digitization algorithm}
\label{sec:digitization}

The digitization algorithm determines the induced charge on each pad for each
particle crossing a gas gap. The method is described in detail in ref.~\cite{digit}.
The algorithm proceeds with the selection of the GEANT4 steps as follows:
\begin{itemize}
\item rejection of the energy depositions that are created more than $1\,\micro\second$ later than the primary
  particle generation time,
\item rejection of steps with a null length,
\item random step selection in order to reproduce a given efficiency, $\epsilon$, defined as the fraction of avalanches above the detection threshold of 0.1\,pC,
\item association of a total induced charge, $Q$, produced by an avalanche with each energy deposition,
\item application of a correction, $A(\theta)$, to the total induced charge, based on the
  angle $\theta$ between the step direction and the GRPC plane, whereby the dependence
  of $A$ on $\theta$ is extracted using muon data~\cite{digit},
\item application of corrections, $\rho$ and $\rho^\prime$, to the total
  induced charge and to the efficiency based on the avalanche
  modeling, whereby the dependence of $\rho$ and $\rho^\prime$ on the energy deposit and data-taking conditions is
  described in Section~\ref{sec:avalanche},
\item a cut-off distance is defined, such as an avalanche development is cancelled
if a second avalanche with more charges is present within this distance. This
scale is tuned using particle showers to reproduce the observed multiplicities~\cite{digit},
\item distribution of the charges over the pads.
\end{itemize}
The amount of induced charges, $Q$, is used to populate the three-amplitude categories following
three thresholds: $0.1$, $5$ and $15$\,pC. A charge $Q_0$ is first modeled by a Polya
distribution~\cite{polya} using the data:
\begin{equation}
P(Q_0) = \frac{1}{\Gamma\left(1+{\delta}\right)}{\left(\frac{1+\delta}{\bar{Q}}\right)}^{1+\delta}
Q_0^{\delta} e^{\left[ \frac{-Q_0}{\bar{Q}(1+\delta)} \right]}
\label{eq:polya}
\end{equation}
where $\Gamma$ is the Gamma function. $\bar{Q}$ and $\delta$ are the average charge and the width of the charge distribution. They are derived from dedicated high energetic muon
beams with a threshold scan method~\cite{digit}.  The remaining parameters of
the digitization are tuned in order to reproduce the number of hits observed
in electromagnetic showers.

The simulated charge, including all the corrections, is given by:
\begin{equation}
Q_{\rm MC}=Q_0\times\rho\times{A(\theta)}
\label{eq:qcor}
\end{equation} 
Similar correction, $\rho^\prime$, is applied to the efficiency:
\begin{equation}
\epsilon_{\rm MC}=\epsilon_0\times{\rho^\prime}
\label{eq:effcor}
\end{equation} 
where $\epsilon_0$ is the efficiency measured using high energetic muon beams.  

The digitization procedure was tuned for isolated tracks or
showers. The impact of two consecutive showers on the performance of the
detector is neglected due to the low expected particle rate. Moreover, it
was assumed that the variations modelled in the next Section, can be
factorised, i.e., $\rho$ is independent from $Q$. $Q_0$ is measured from
the data and can be updated separately.

\subsection{Simulation of the avalanches}
\label{sec:avalanche}

The mathematical distribution used to predict the GRPC signal in ref.~\cite{digit} oversimplifies
and neglects some physical processes that will be considered for the first time in this study.
The Monte Carlo simulation of the avalanche described in this section
provides corrections applied hit by hit to the charge $Q_0$ and to the efficiency in the
digitization procedure. The charge correction $\rho$ is defined by:
\begin{equation}
\rho=\frac{<Q(\Delta{C})>}{<Q(\rm{nominal})>}
\label{eq:rho}
\end{equation}
where $\Delta{C}$ represents a variation in data-taking condition (e.g.\
temperature or the energy deposit),
$<Q>$ is the average of the total induced charge estimated using the full avalanche
simulation described in this section.
 The properties of the GRPC that are taken into account for
the avalanche simulation are listed in Table~\ref{tab:rpc}. They are used to
define $<Q(\rm{nominal})>$.

 The implementation of the avalanche simulation is described
in refs.~\cite{vincentphd,vincentrpc} and adapted to the SDHCAL properties.
This Monte Carlo simulation models the
amplitude and the efficiency.

\begin{table}[h]
  \centering
  \caption{Avalanche simulation parameters with their nominal
  values. $\epsilon_0$ refers to the vacuum electric permitivity.\label{tab:rpc}}
\begin{tabular}{llr}
\toprule
Parameter &  & Value \\ \hline
Width     & Gap      & $0.12\,$cm      \\
          & Anode    & $0.07\,$cm      \\ 
          & Cathode        & $0.11\,$cm       \\ \hline
Permitivity  & Anode   & $7\epsilon_0$      \\ 
             & Cathode & $7\epsilon_0$ \\ \hline
Gas Mixture & \ch{C2H2F4}  & 93\%       \\  
            & \ch{CO2}     & 5\% \\ 
            & \ch{SF6}     & 2\% \\ \hline
Electric Field & & $57500\,{\rm V{cm^{-1}}}$ \\ \hline
Temperature & & $293.15\,\kelvin$ \\ \hline
Pressure & & $1$\,atm \\ \bottomrule
\end{tabular}
\end{table}

\subsubsection{Primary ionization}
\label{sec:priion}
When running the complete digitization of SDHCAL, the primary ionization is described by
GEANT4 in terms of energy deposits. In order to have an event-by-event simulation of the
electron distribution in the gas, the HEED simulation program is
used~\cite{heed}. Primary and secondary ionisations are simulated by HEED
including emissions of photo-electron and auto-ionization Auger electrons. 
About 8 charge clusters are 
typically produced per $\milli{\meter}$ in the GRPC when a charged particle
traverses it, containing 21 electrons in average.

 The correlation between the number
of ionisation electrons and energy deposit for the gas mixture used in the SDHCAL
prototype is determined with the HEED package. An average energy of
$29.5$\,eV is required for an electron-ion pair production.  In the digitization process, this value is used to associate a number of ionisations with each GEANT4 step.

\subsubsection{Electronic avalanche}

The electric field induces the electron drift towards the anode in the GRPCs.
An electron avalanche is the consequence of the charge multiplication in the GRPC while the
electrons interact with gas molecules. 
This phenomenon is described in the simulation using the
Riegler-Lippman-Veenhof model~\cite{rpcmodel}.

Two coefficients are used to characterize the avalanche development: the
Townsend coefficient, $\alpha$, and the attachment coefficient, $\eta$.
 
 Three additional parameters are considered: the electron drift
velocity, the longitudinal and the transverse diffusion coefficients. The
drift velocity and the diffusion amplitude account for the avalanche development.
The diffusion has a significant impact on the amplification process as it
increases the average charge path.

 Before each avalanche simulation, for each condition of
  temperature, pressure, and gas composition, the coefficients $\alpha$,
  $\eta$, the electron velocity and diffusion parameters are estimated using
  the Magboltz~9.01 package~\cite{magboltz}. They are mapped as 
  a function of the electric field value. Since the electric field is a function
of the position in the gas gap and a function of time, these parameters will
also vary depending on the position and time as detailed below.

The avalanche is simulated in a gas gap divided into longitudinal intervals of
$0.5\,\micro{\meter}$ each.
The average number of electrons in each interval is governed by the 
coefficients and the initial number of charges. The average number of
electrons, $\bar{n}(x)$, and ions, $\bar{p}(x)$, 
produced by one electron after a step of lengh $x$ are modeled as follows
\begin{equation}
\label{eq:nbar}
\bar{n}(x) = e^{(\alpha-\eta)x}
\end{equation} 
and
\begin{equation}
\label{eq:pbar}
\bar{p}(x) =
\frac{\alpha}{\alpha-\eta}(e^{(\alpha-\eta)x}-1)
\end{equation}
An iterative procedure is used to estimate the number of charges in each detector interval
until all the electrons have reached the anode, after 4500
iterations in average~\footnote{The avalanche simulations used in this paper required
  $10^{4}$ CPU hours.}. The evolution of the number of electrons is given by

\begin{equation}
\label{eq:nelectrons}
 n(x) = \left\{
   \begin{array}{ll}
      0, & s<k\frac{\bar{n}(x)-1}{\bar{n}(x)-k}\\
1+{\rm{ln}}\left(\frac{(\bar{n}(x)-k)(1-s)}{\bar{n}(x)(1-k)}\right)\frac{1}{{\rm{ln}}\left(1-\frac{1-k}{\bar{n}(x)-k}\right)},
& s>k\frac{\bar{n}(x)-1}{\bar{n}(x)-k}
\end{array}
\right.
 \end{equation}
where $s$ is a random number $\in [0,1)$ from the uniform distribution and $k=\eta/\alpha$.

The diffusion of each electron is simulated at each iteration.

The electrons that reach the resistive anode lead to an
  accumulation of charges at the surface of the anode. Their relaxation time is higher than the avalanche development time. The space charge effect is estimated while the influence of the charges
produced in the avalanche on the electric field is computed. The numbers and
positions of electrons or ions, number of electrons on the anode, combined to
the applied electric field are used 
to compute the electric field as a function $x$ at each time iteration. 
Any change in the electric field requires 
the various coefficients to be updated, especially $\alpha$ and $\eta$.

Finally, the induced current is computed using Ramo's theorem generalized to
resistive materials accounting for the properties of the resistive layers~\cite{rpcmodel}. 
The transverse profile of the avalanche is assumed to be negligible compared to
the pad size. While the avalanche simulation predicts 
a total signal induced by a given avalanche, the digitization
procedure models its distribution on multiple pads.
This modeling of the avalanches does not describe the signal in regimes with very 
  high number of charges like streamers.

\subsubsection{Simulation output}

The main outputs of the avalanche simulation are:
\begin{itemize}
\item the total induced charge, $Q$, at the level of the pads.
 For the stability studies, only the relative variation of the mean values of $Q$ are used;
\item the efficiency, $\epsilon$;
\item streamer probability: an empirical and raw way to estimate the
  streamer probability is to monitor the fraction of simulated
  events for which the amplification exceeds a factor of $e^{22} \simeq 4.85\times
  10^{8}$~\cite{polya,streamers}.
\end{itemize}

\begin{figure}[t]
 \begin{center}
    \includegraphics[width=0.5\textwidth]{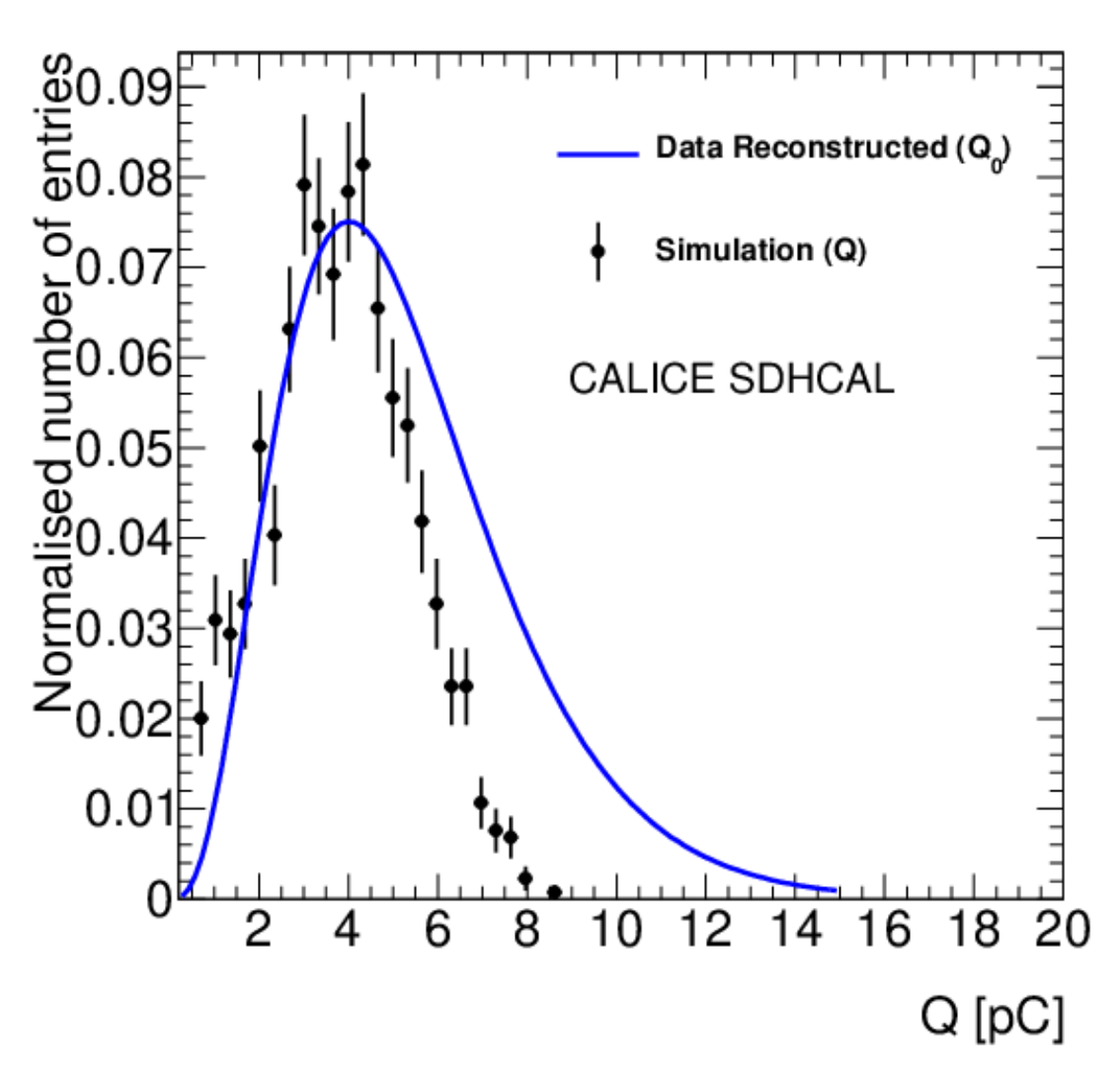}
  \caption{\label{fig:qsimvsdata}Distribution of total induced charge for
      a simulated RPC (dots). The total induced charge was reconstructed
      from the data using the threshold scan method
      assuming a Polya distribution and describes all the channels of the
      prototype. The simulation assumes a homogeneous detector at nominal conditions.} 
  \end{center}
\end{figure}

  The number of electron charges produced in the avalanches is
  about $15$\,pC on average, while the total induced charges are $3.7$\,pC on
  average. The induced current reaches a maximum after
  about $11$\,ns. The distribution of simulated $Q$ is compared to $Q_0$ 
  reconstructed from the data and shown in Figure~\ref{fig:qsimvsdata}. $Q_0$ is
  the superimposition of the different channels, thus combining different readout channels and different
  planes. The overall amplitude of $Q$ is coherent with the measured one and the spread
  observed in the data is due to the inhomogeneity of the detector
  channels.


%% file: latex/stability.tex
\section{Predicted Stability of the SDHCAL Prototype Response}
\label{sec:stab}
Instabilities in the prototype response to electromagnetic or
hadronic showers have been reported in ref.~\cite{1stresults}. These variations
are seen between different shower types, 
between different data taking periods, between different layers or even inside
the same chamber. In this section, several effects that can induce variations
in the detector response are reviewed using simulations. Some are related to the calibration
method (use of high energy muons to model the avalanche in the digitizer),
some are related to intrinsic properties of the prototype (GRPC gap
homogeneity) or data taking conditions (temperature, pressure, magnetic field, gas
mixture).
In order to estimate the impact of each effect on the number of hits, a
Monte Carlo simulation of electrons, pions and muons is produced. The effects
described in this section are modeled using the avalanche simulation
technique described in Section~\ref{sec:avalanche} and summarized in terms of
variations in the mean total induced charge and efficiency.
These two variations are used to scale the 
charge and efficiency in the digitization 
procedure described in Section~\ref{sec:digitization} when producing the full
Monte Carlo simulation of the SDHCAL response to electron or pion showers.

\subsection{Universality of the MIP-based calibration}

\begin{figure}[t]
 \begin{center}
  \begin{subfigure}[b]{0.495\textwidth}
    \includegraphics[width=\textwidth]{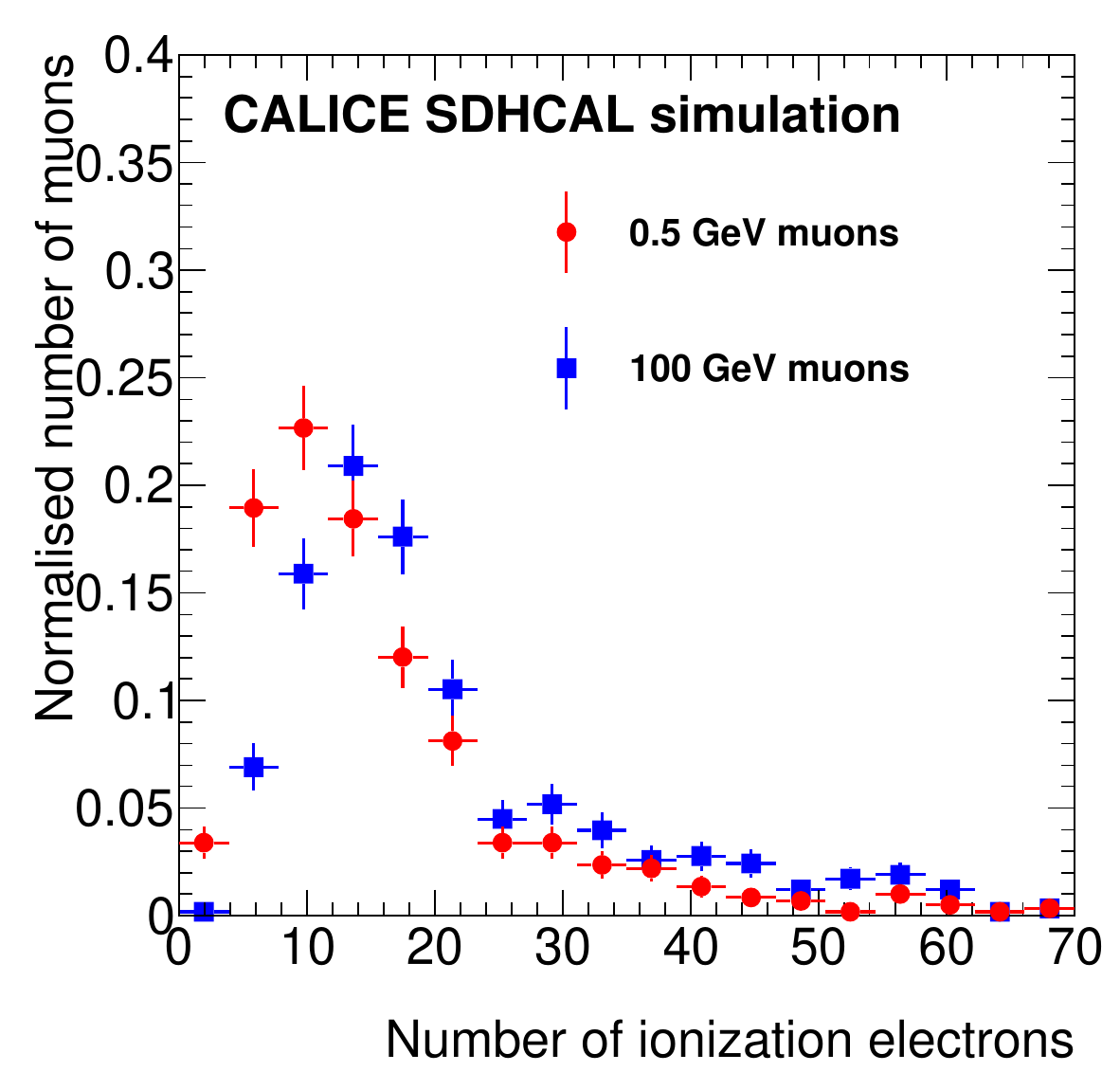}
    \caption{}
    \label{fig:qi.dist}
  \end{subfigure}
 \begin{subfigure}[b]{0.495\textwidth}
    \includegraphics[width=\textwidth]{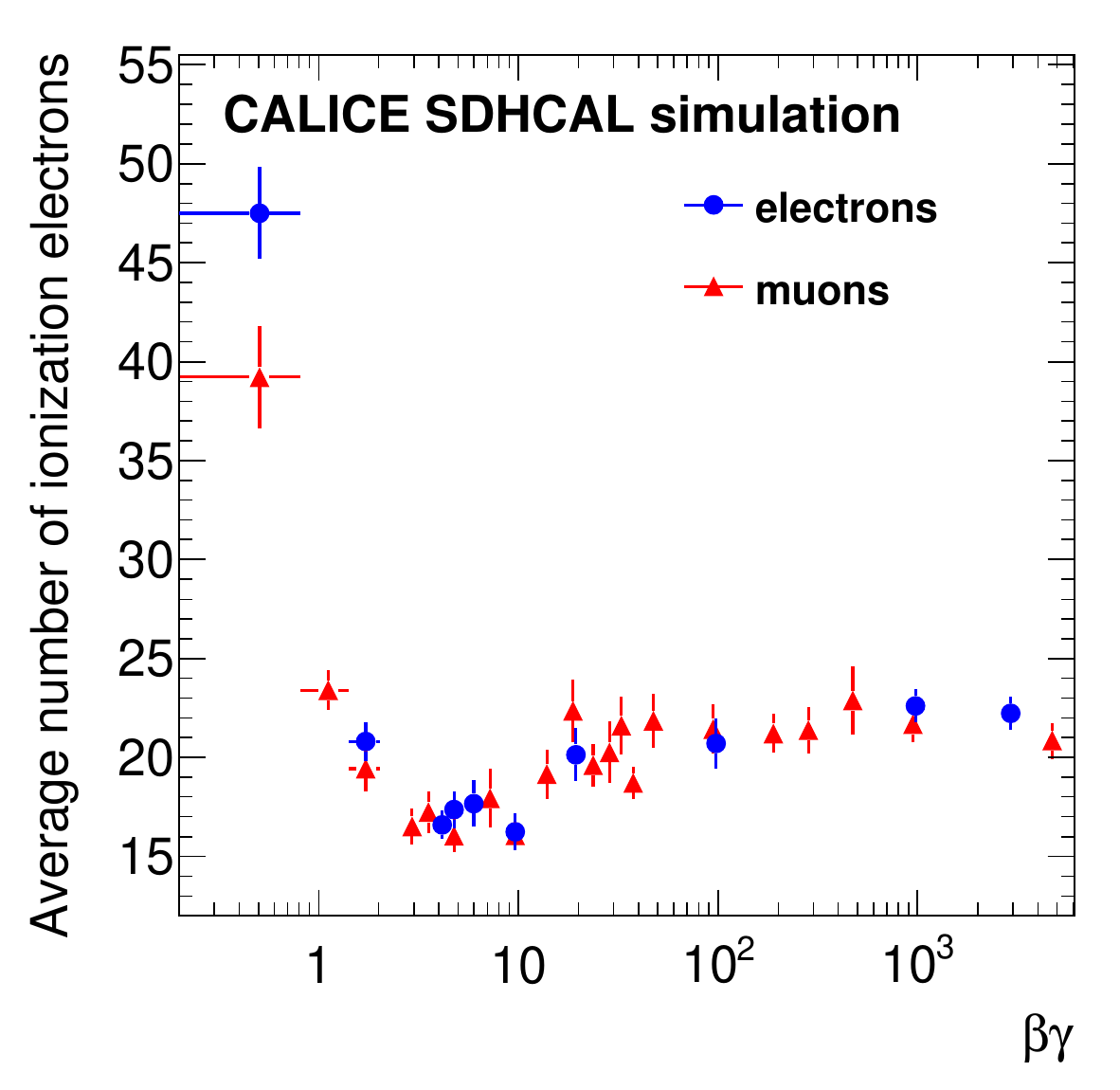}
    \caption{}
    \label{fig:qi.vs.bg}
  \end{subfigure}
  \caption{(\subref{fig:qi.dist}) Normalized distribution of number of
    ionisation electrons produced
    in the gas mixture by 0.5 and $100\,\GeV$ muons.
(\subref{fig:qi.vs.bg}) Average number of ionisation electrons produced in the gas mixture by
    electrons or muons as a function of the Lorentz factors $\beta\gamma$.}
  \end{center}
\end{figure}

For a given type of particles, the energy deposit in the GRPC gas mixture depends on
their momenta and fluctuates from one interaction to another.
Figure~\ref{fig:qi.dist} illustrates the distribution of the number of
ionization electrons created by 0.5 and $100\,\GeV$ muons. Typically, an
average of 15 electrons
are created in the SDHCAL chamber due to ionization. A systematic difference
between the two distributions is observed.

As shown in Figure~\ref{fig:qi.vs.bg}, the number of electrons (and ions)
created in the gas mixture follows the Bethe-Bloch
formula~\cite{bethe}. Ideally, the total induced charge measured after the
avalanche should not depend on the 
properties of the MIP. A possible dependence on the number of
initial charges, and therefore on the energy and type of the charged particle
that ionized the gas mixture, is investigated.  
The digitizer charge distribution is tuned using the average response 
of the GRPCs to $100\,\GeV$ muons. The relative response of a GRPC to these muons
with respect to the GRPC response to low energetic charged particles\footnote{The
  typical energy of interacting charged particles in an electromagnetic or hadronic
  shower covers a large scale from hundreds of keV to GeV.} thus has to be
estimated. This relative response can be used to 
extrapolate the muon calibration to any particle type or energy or
estimate the bias induced by this modeling.

\begin{figure}[h]
 \begin{center}
  \begin{subfigure}[b]{0.49\textwidth}
    \includegraphics[width=\textwidth]{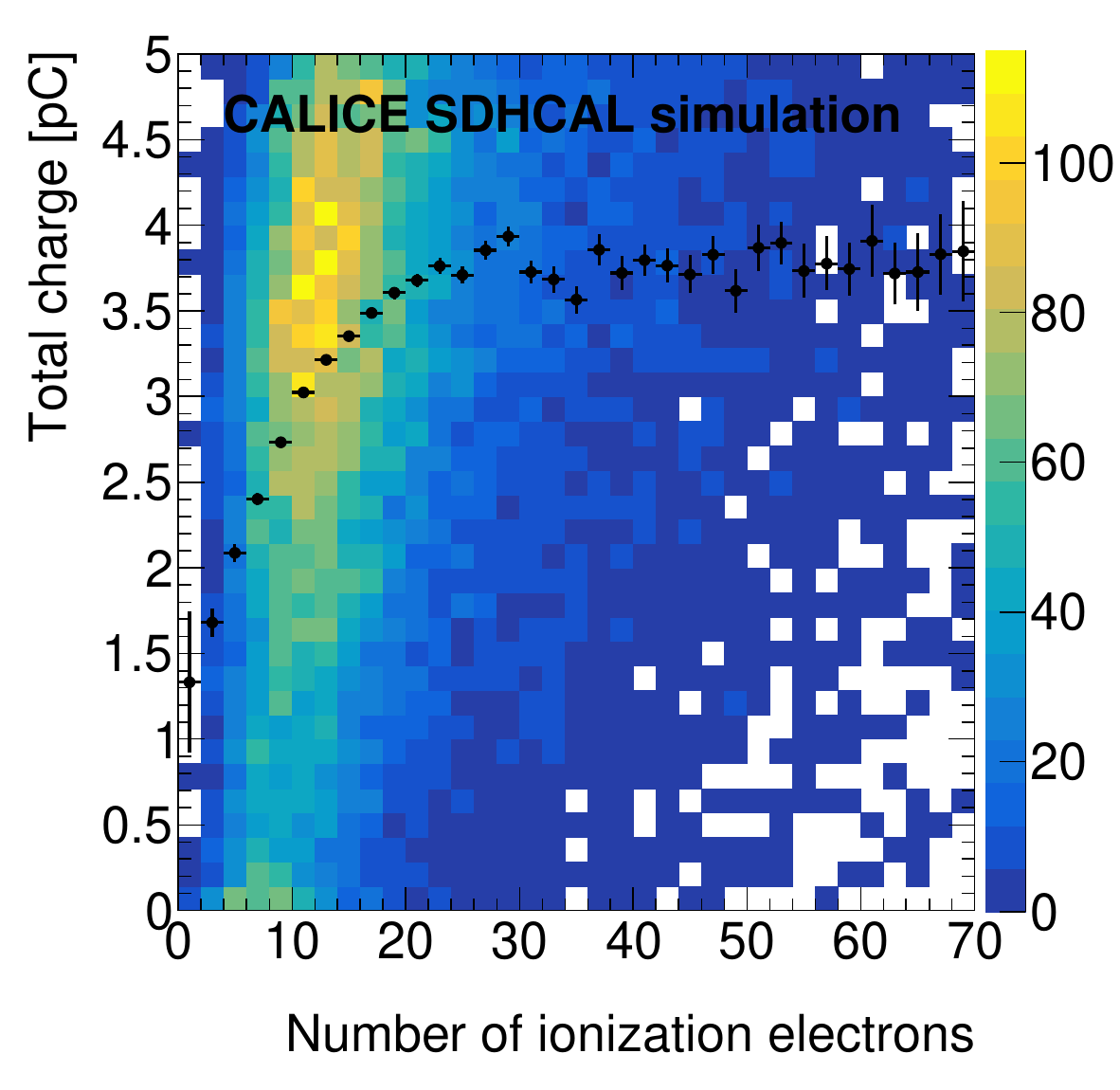}
    \caption{}
    \label{fig:q.vs.qi}
  \end{subfigure}
 \begin{subfigure}[b]{0.49\textwidth}
    \includegraphics[width=\textwidth]{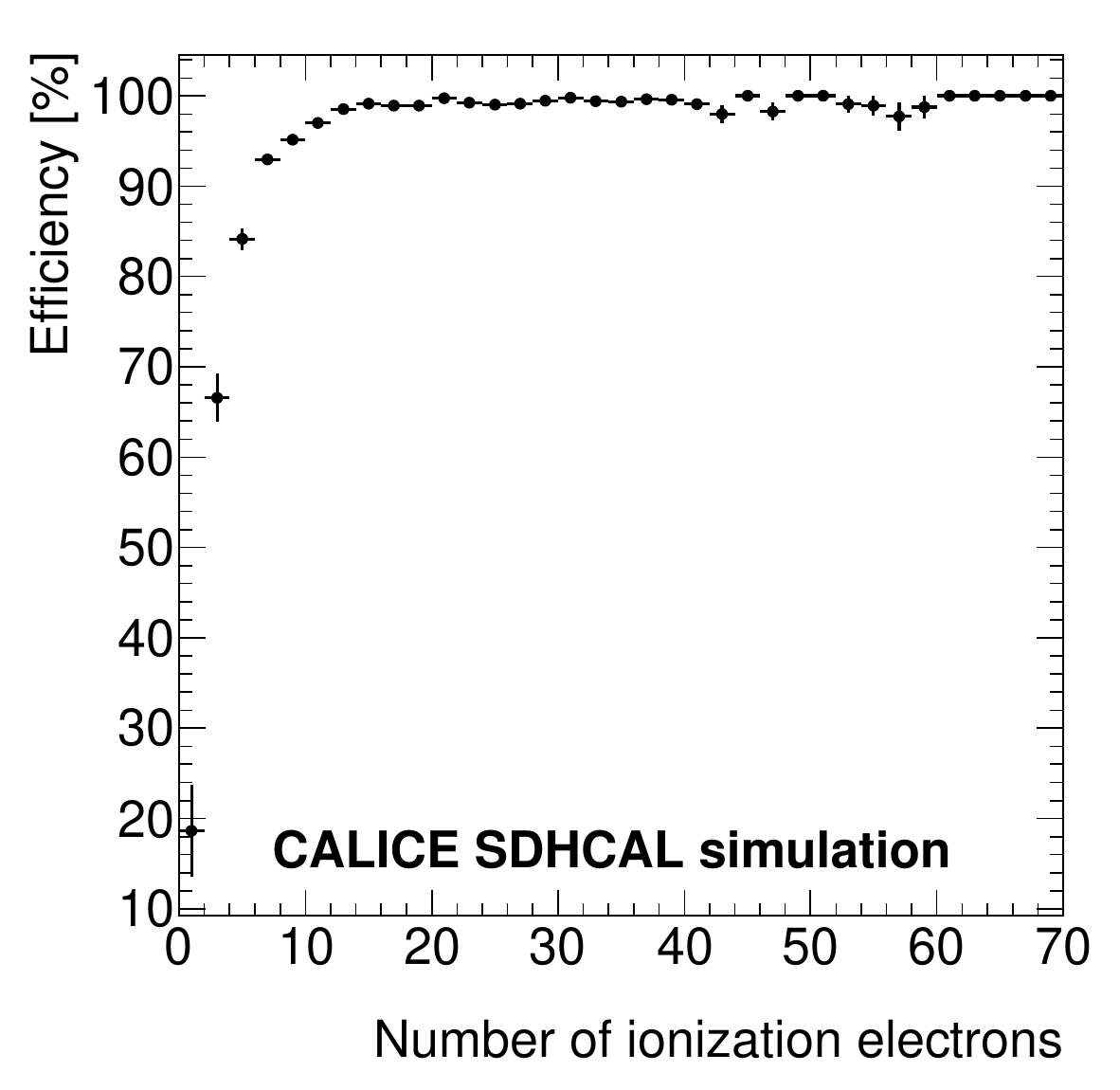}
    \caption{}
    \label{fig:eff.vs.qi}
  \end{subfigure}
  \caption{Total induced charge (\subref{fig:q.vs.qi}) and efficiency
    (\subref{fig:eff.vs.qi}) induced
    by an avalanche as a function of the number of
    ionisation electrons produced in the gas mixture by 
    the ionizing particle. Total induced charge is averaged on each bin and shown as
    black dots in (\subref{fig:q.vs.qi}).
} 
  \end{center}
\end{figure}

Electron and muon induced avalanches are
generated
 using different momenta values from 100\,keV to 500\,GeV. The overall
 dependence of the total induced charge on the number of ionisations is shown on 
Figure~\ref{fig:q.vs.qi}. 
For a number of initial electrons greater than 20, a plateau is observed. In
this saturated regime, the total induced charge does not depend on the number of ionisations.
Below 10 electrons, a significant drop is observed.
Furthermore, the probability that an avalanche reaches the detection threshold is expressed in terms
of efficiency and is shown on Figure~\ref{fig:eff.vs.qi}. An efficiency plateau
is reached above 10 electrons. For example, the efficiency is 66\%
on average if the incoming charged particle creates 3 electrons by ionising
the gas mixture.

The dependence of the total induced charge
on the initial number of ionisations is used to model a correction, $\rho(E/E_0)$
defined as
\begin{equation}
\rho(E/E_0) = \frac{<Q(E/E_0)>}{<Q({\rm{nominal}})>}
\label{eq:rhoioncor}
\end{equation}
where $E$ is the energy deposit associated to a given GEANT4 step, $E_0$ is
equal to $29.5\,\eV$ (section~\ref{sec:priion}), ${<Q(E/E_0)>}$ is the average
total induced charge derived from Figure~\ref{fig:q.vs.qi}
 and ${<Q({\rm{nominal}})}>$ is the average total induced charge expected for $100\,\GeV$
muons under nominal conditions. This correction 
extrapolates the $100\,\GeV$ muon Polya calibration to the variety of particles that are
detected by the GRPC, especially in electromagnetic and hadronic showers.
A correction is also defined for the avalanche efficiency:
\begin{equation}
\rho^\prime(E/E_0) = \frac{\epsilon(E/E_0)}{\epsilon({\rm{nominal}})}
\label{eq:rhoprimeioncor}
\end{equation}
where $\epsilon(E/E_0)$ is the efficiency derived from Figure~\ref{fig:eff.vs.qi} and $\epsilon({\rm{nominal}})$ is the efficiency expected for $100\,\GeV$ muons under nominal conditions. 

\begin{figure}[]
 \begin{center}
 \begin{subfigure}[b]{0.31\textwidth}
    \includegraphics[width=\textwidth]{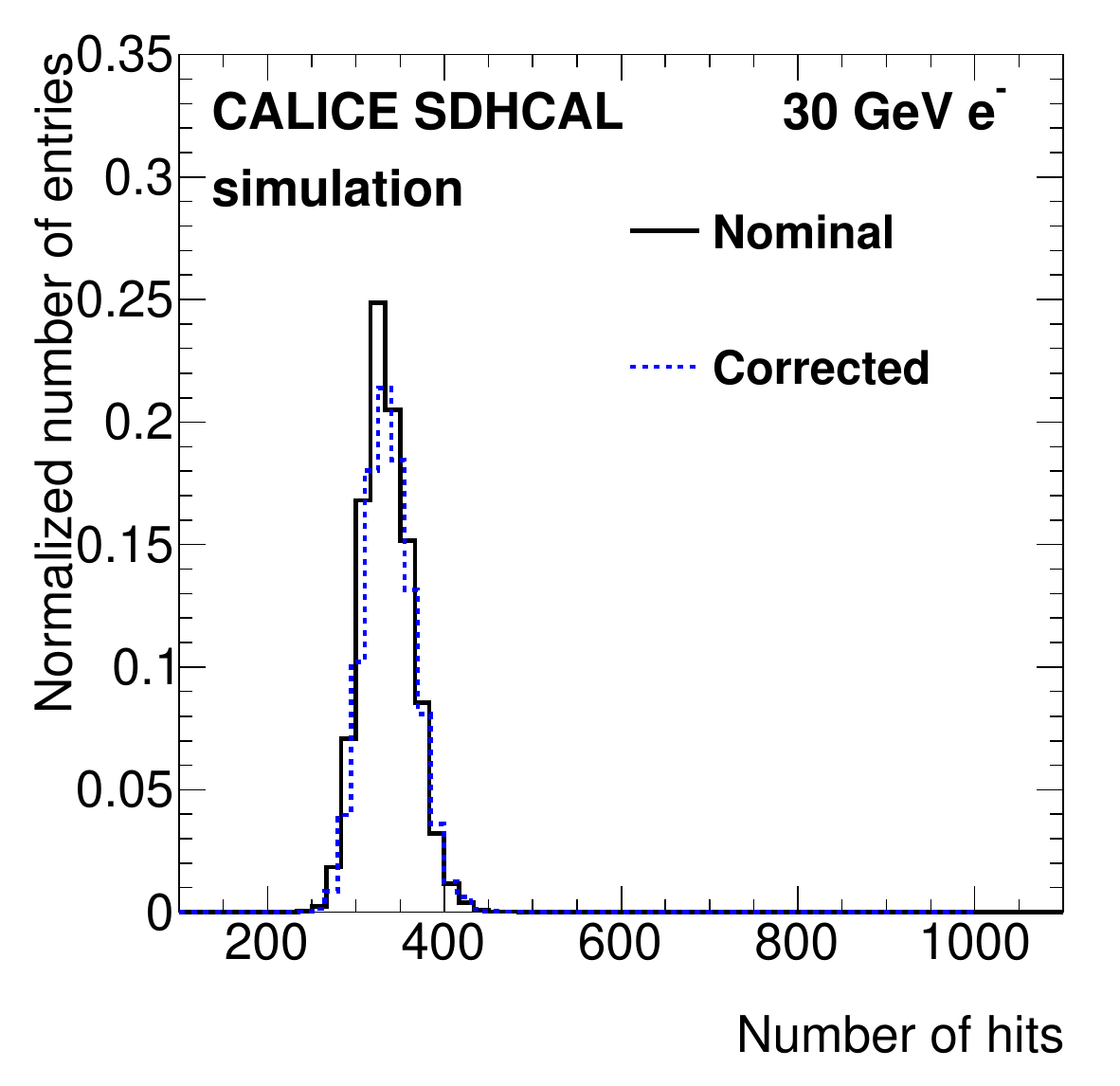}
    \caption{}
    \label{fig:nhcor.vs.nocor.1.e}
  \end{subfigure}
  \begin{subfigure}[b]{0.31\textwidth}
    \includegraphics[width=\textwidth]{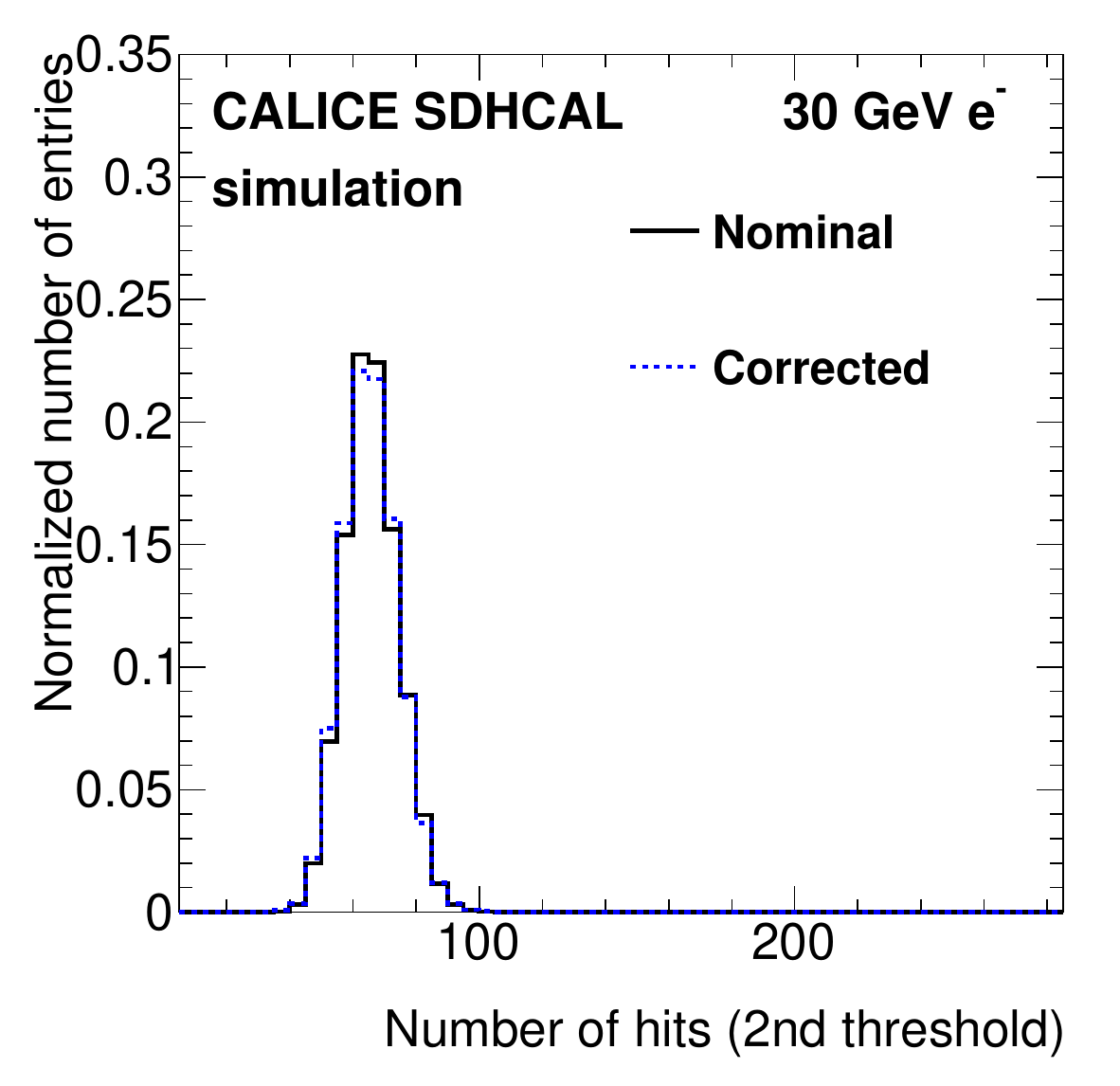}
    \caption{}
    \label{fig:nhcor.vs.nocor.2.e}
  \end{subfigure}
  \begin{subfigure}[b]{0.31\textwidth}
    \includegraphics[width=\textwidth]{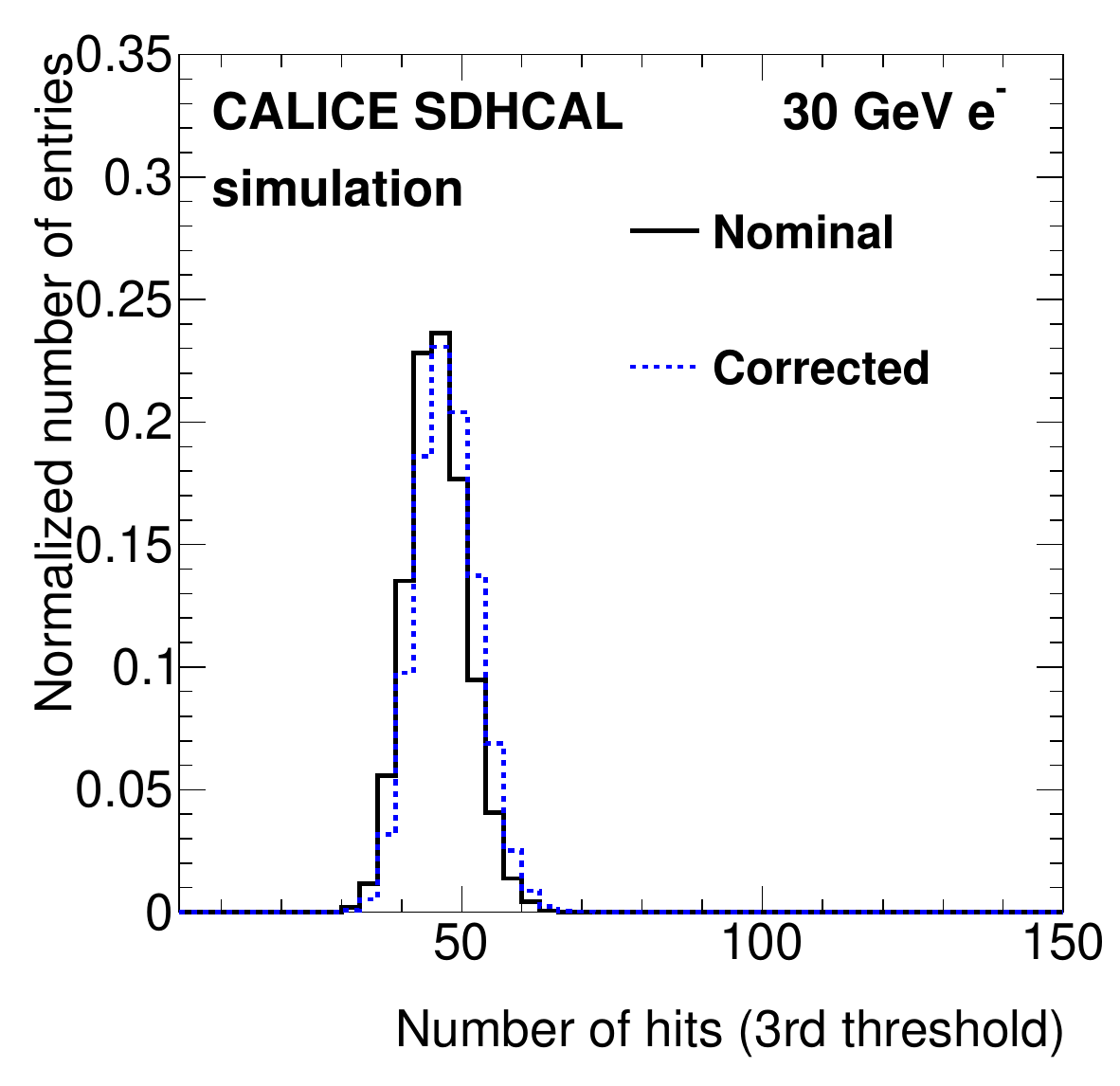}
    \caption{}
    \label{fig:nhcor.vs.nocor.3.e}
  \end{subfigure}
 \begin{subfigure}[b]{0.31\textwidth}
    \includegraphics[width=\textwidth]{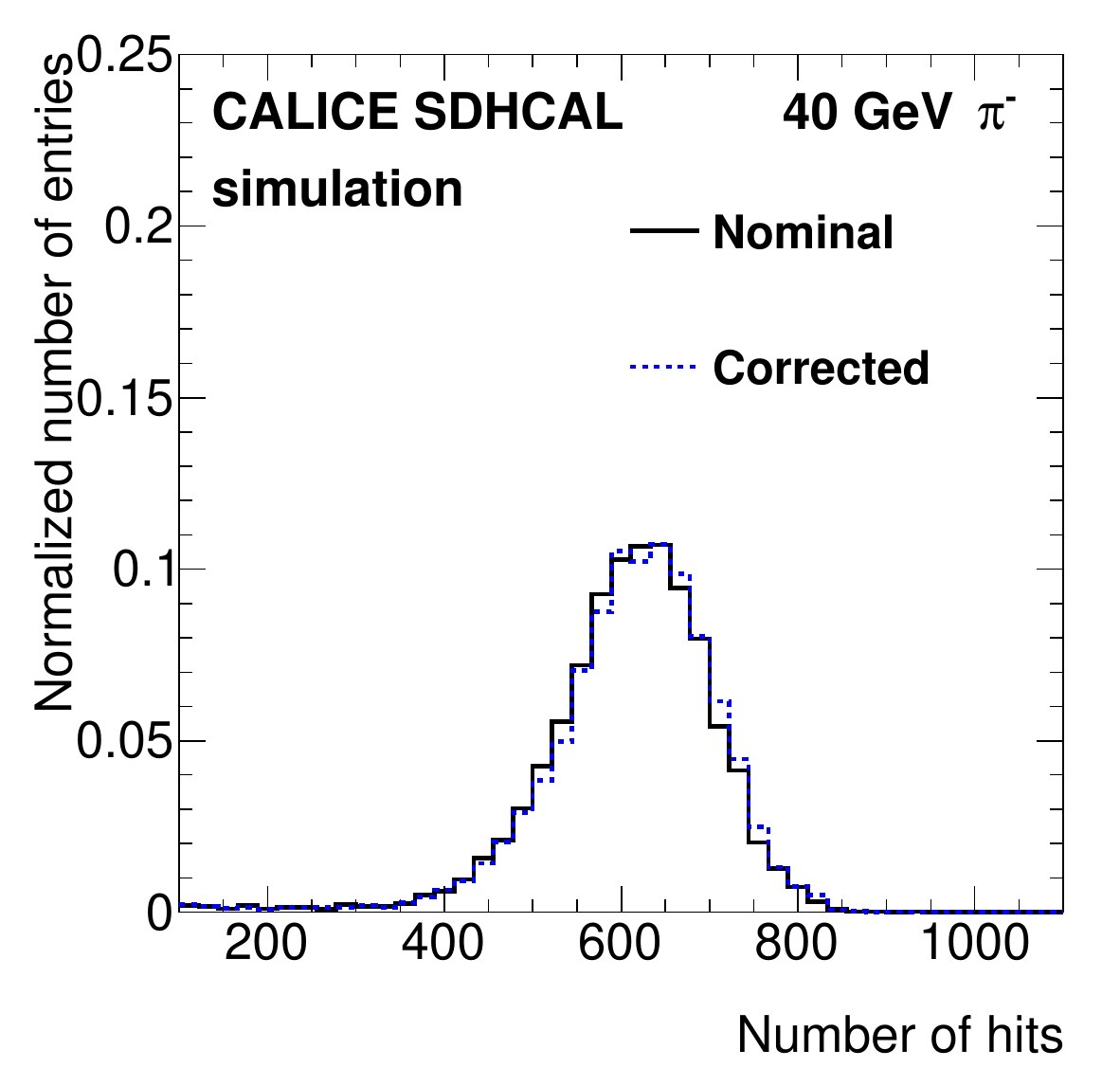}
    \caption{}
    \label{fig:nhcor.vs.nocor.1}
  \end{subfigure}
  \begin{subfigure}[b]{0.31\textwidth}
    \includegraphics[width=\textwidth]{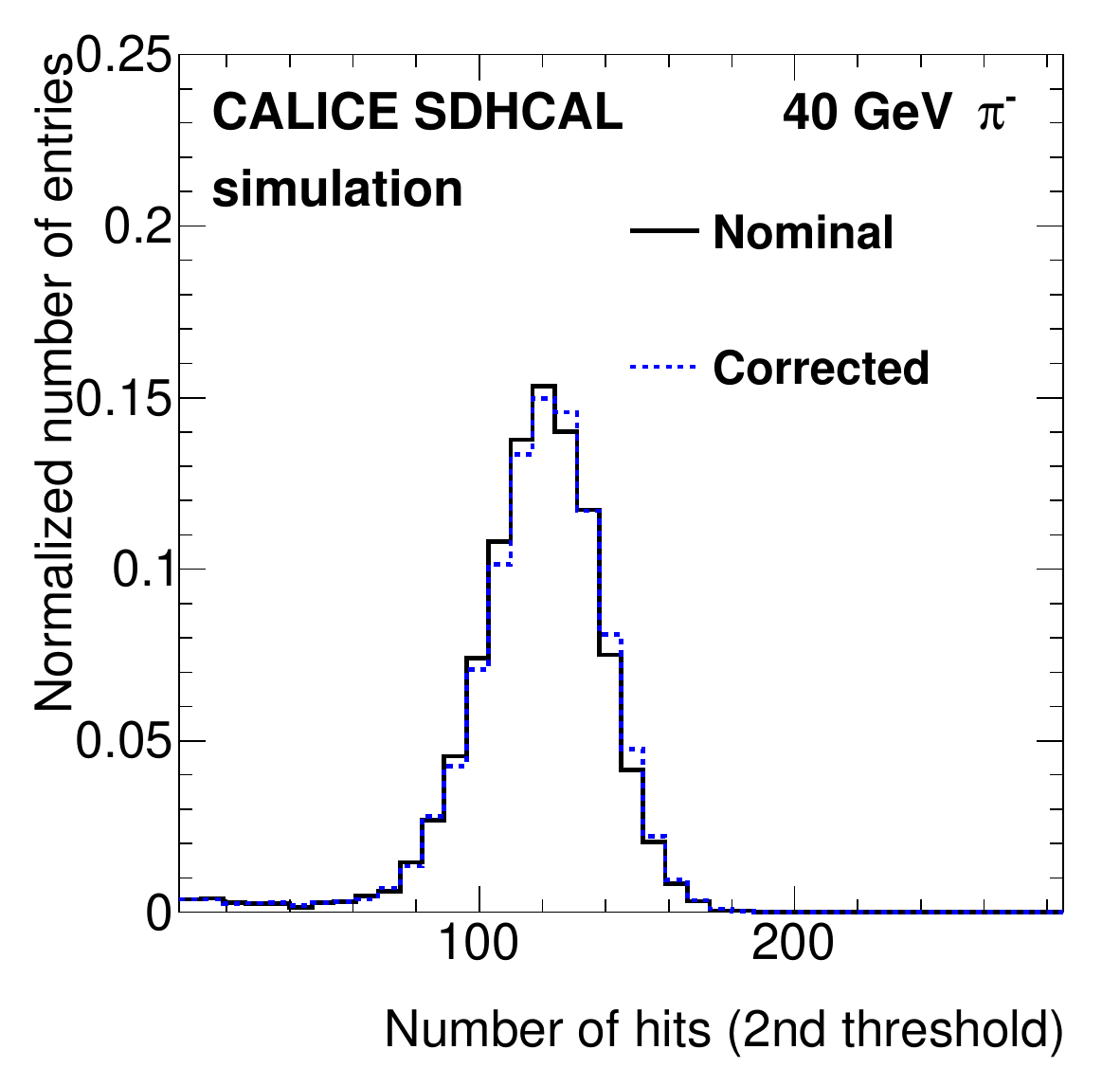}
    \caption{}
    \label{fig:nhcor.vs.nocor.2}
  \end{subfigure}
  \begin{subfigure}[b]{0.31\textwidth}
    \includegraphics[width=\textwidth]{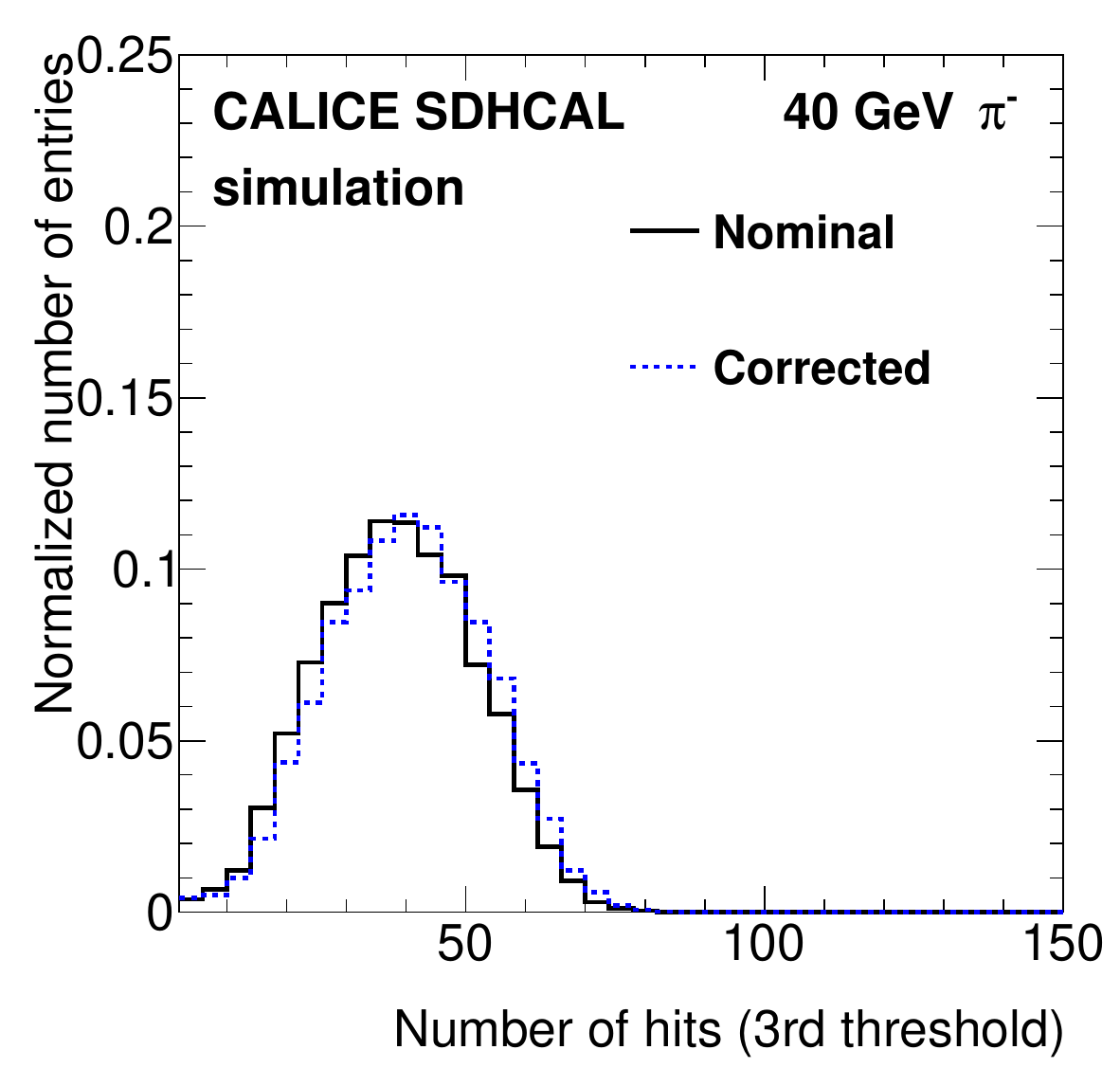}
    \caption{}
    \label{fig:nhcor.vs.nocor.3}
  \end{subfigure}
  \caption{ Distribution of the number of hits passing the first
    threshold at
    $0.1$\,pC~(\subref{fig:nhcor.vs.nocor.1.e},\subref{fig:nhcor.vs.nocor.1}), the
    second threshold at
    $5$\,pC~(\subref{fig:nhcor.vs.nocor.2.e},\subref{fig:nhcor.vs.nocor.2}) and the
    third threshold at
    $15$\,pC~(\subref{fig:nhcor.vs.nocor.3.e},\subref{fig:nhcor.vs.nocor.3}) for
    simulated  30\,GeV electrons (top) or 40\,GeV pions (bottom). The full
    GEANT4 simulation was performed with digitization modeling that does not
    account for energy deposit in the GRPC gas (solid black histogram) or a modeling
    that accounts for it (dashed blue histogram).\label{fig:nh.vs.ioncor}} 
  \end{center}
\end{figure}

This correction addresses potential mismodelings of the avalanche, leading to a
bias on the reconstructed energy in the simulation depending on the type
of shower. In order to estimate the simulation bias, the full
digitization procedure described in Section~\ref{sec:digitization} is applied. It
is completed with the efficiency and amplitude correction described above 
which takes as input GEANT4 step energies. The impact of this correction on
the average number of hits is estimated and is shown in
Figure~\ref{fig:nh.vs.ioncor}. 
The relative variation of the average total number of hits is typically
1\% for both electrons and pions as well as the average number of hits
passing the second threshold. For the third threshold this variation is
$2.9\pm0.5\,\%$ ($5.3\pm0.5$\,\%) for electrons (pions).

The energy reconstruction is tuned, using the polynomial combination
method. The factors $a$, $b$ and $c$ from eq.~\eqref{eq:ereco} are adjusted
with a $\chi^2$ minimization method as given in
eq.~\eqref{eq:chi2}. These factors are extracted using simulations of
electron (or pion) showers with energy deposit correction.  
They are used to reconstruct the electron (or pion) energies in the simulation
with and without energy deposit correction in an energy range of $7$ to $70\,\GeV$.
 The largest variation in the reconstructed energy is $4\%$ for
 electromagnetic showers and $2\%$ for hadronic showers.
In the following, the dependence on the energy deposit is included in the
digitizer algorithm.

\input{latex/section-width.tex}

\clearpage
\input{latex/section-temp.tex}
\input{latex/section-bfield.tex}

\clearpage
\input{latex/section-gas.tex}
\clearpage

\subsection{Summary of detector effects simulation}
The dependence of the detector response on the initial ionizing particle was described and included in the digitization procedure.
It was shown that a magnetic field has no significant impact on the SDHCAL performance.
However, other scenarios tested in the simulation can lead to non negligible
effects as summarized in Table~\ref{tab:stab.pi}. The uncertainties on the variation in the number of 
hits correspond to the finite Monte Carlo size and to the
uncertainty on the total induced charge variation. 
It is also checked that in the situation where these effects are definite 
and known, e.g.\ a static detector deformation, updating the energy calibration
factors for each homogeneous data sample allows to restore the linearity of 
detector response with some resolution loss.
A simulation that is tuned with beam test data (e.g.\ muon events to tune the
charge distribution) in a given condition of temperature, pressure or
mechanical homogeneity will wrongly model the data collected under different
conditions, both in terms of amplitude and resolution.

\begin{table}
  \centering{\small
\caption{Most significant effects and their corresponding impact on the
  detector response to $40\,\GeV$ pions.\label{tab:stab.pi}
  $\Delta{Q}/Q$ is the relative bias of the single MIP induced charge with respect
  to its nominal value. $\Delta{N_{\rm{tot}}}/N_{\rm{tot}}$ is the variation in the total number of
  hits. $\Delta{N_2}/N_2$ is the variation in the number of hits passing the
  second threshold, and  $\Delta{N_3}/N_3$ for the number of hits
  passing the third threshold. 
  $\Delta\rm{E}/\rm{E}$ is the relative bias of the
  reconstructed energy when applying an energy reconstruction defined on the nominal
  detector simulation to simulated data at different conditions.}
\begin{tabular}{ll c ccc ccc cc}
\toprule
& & ${\Delta}Q/Q$ [\%] & $\Delta N_{\rm{tot}}/N_{\rm{tot}}$ [\%] & $\Delta{N_2}/N_2$ [\%]& $\Delta{N_3}/N_3 [\%]$ & $\Delta\rm{E}/\rm{E}$ [\%] \\\hline

Gap & $+10\,{\micro\meter}$ & $-7.2\pm0.3$ & $-3.5\pm0.2$ & $-8.0\pm0.3$ & $-12.3\pm0.5$ & $-8.6\pm0.5$ \\ 

 & $\pm100\,\micro\meter$ &  & $-7.9\pm0.2$ & $-13.7\pm1.8$ & $-19.2\pm0.2$& $-6.9\pm0.2$ \\ \hline
T & $+1\degreecelsius$ & $4.1\pm0.4$ &  $1.9\pm0.2$ & $4.3\pm0.4$ & $7.5\pm0.7$ & $4.2\pm1.1$\\ \hline
P & $+10\,{\rm mbar}$ & $-11.1\pm0.9$ & $-4.5\pm0.3$ &  $-11.3\pm0.4$ &
$-18.7\pm0.9$ & $-11.9\pm0.8$\\ \hline
\ch{SF6} & $+5\%$ & $-6.4\pm0.4$ & $-2.8\pm0.2$ &  $-6.5\pm0.2$ & $-11.6\pm0.2$ & $-7.2\pm0.7$\\
\bottomrule
\end{tabular}}
\end{table}

%% file: latex/section-width.tex
\subsection{Impact of mechanical homogeneity}

The SDHCAL design is challenging due to the large number of GRPC
plates and their large size. The large-scale prototype is produced with a tolerance of
$40\,\micro\meter$ on the thickness of the individual GRPC plates. 
Variations in the GRPC gap width have a direct impact on the 
avalanche development and can lead to inhomogeneous efficiencies or
signal amplitudes. A gap width deformation would lead to two opposite
effects: a variation in the charge path and therefore of the charge
multiplication and a variation in the electric field.
In order to quantify the impact of a potential gap width
inhomogeneities on the avalanche production process, different geometries
are simulated using $100$\,GeV muons. The high-voltage is assumed to be stable and
the gap width variations are then propagated to the electric field.

\begin{figure}[t]
 \begin{center}
 \begin{subfigure}[b]{0.495\textwidth}
    \includegraphics[width=\textwidth]{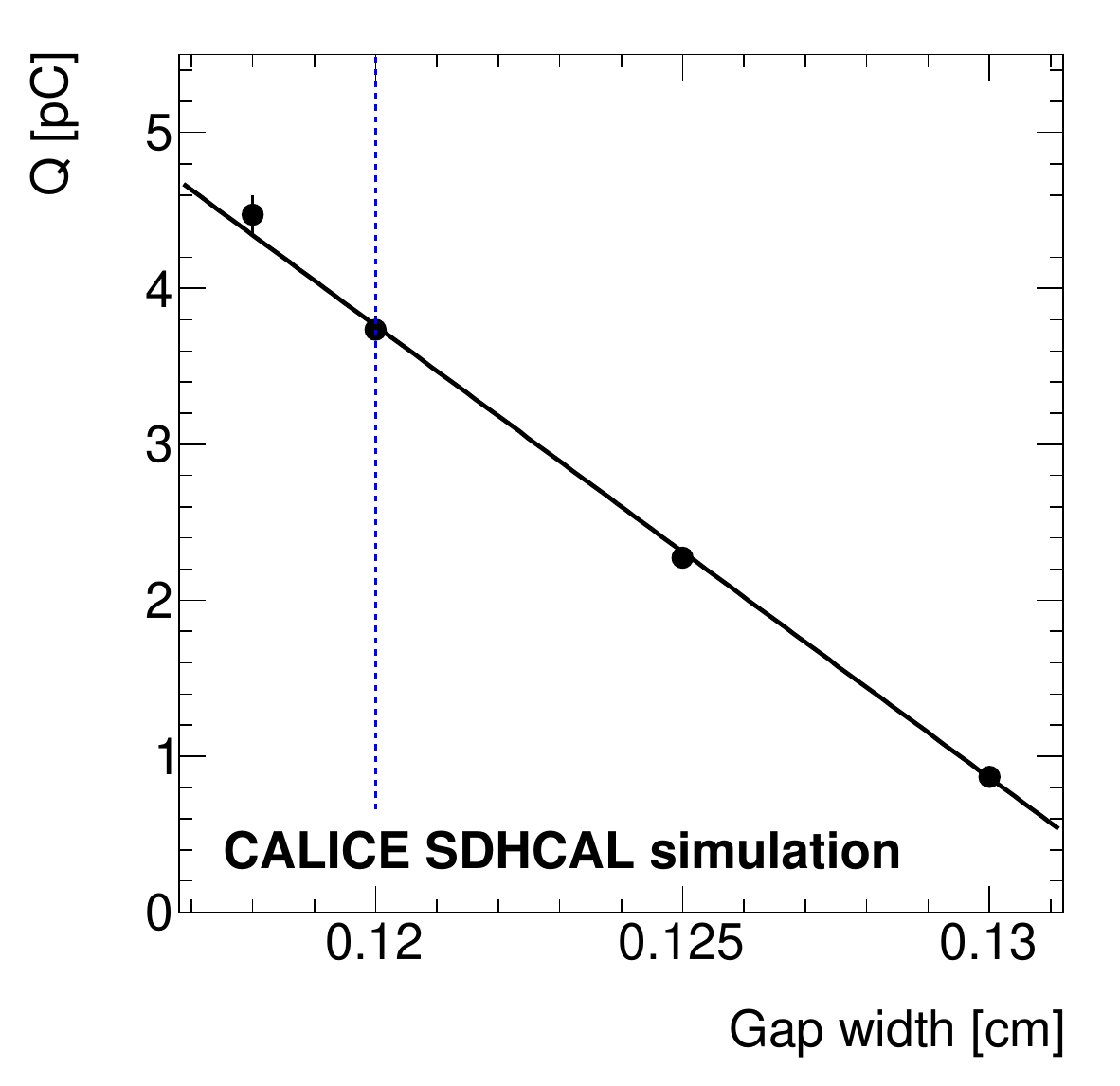}
    \caption{}
    \label{fig:q.vs.d.mu}
  \end{subfigure}
  \begin{subfigure}[b]{0.495\textwidth}
    \includegraphics[width=\textwidth]{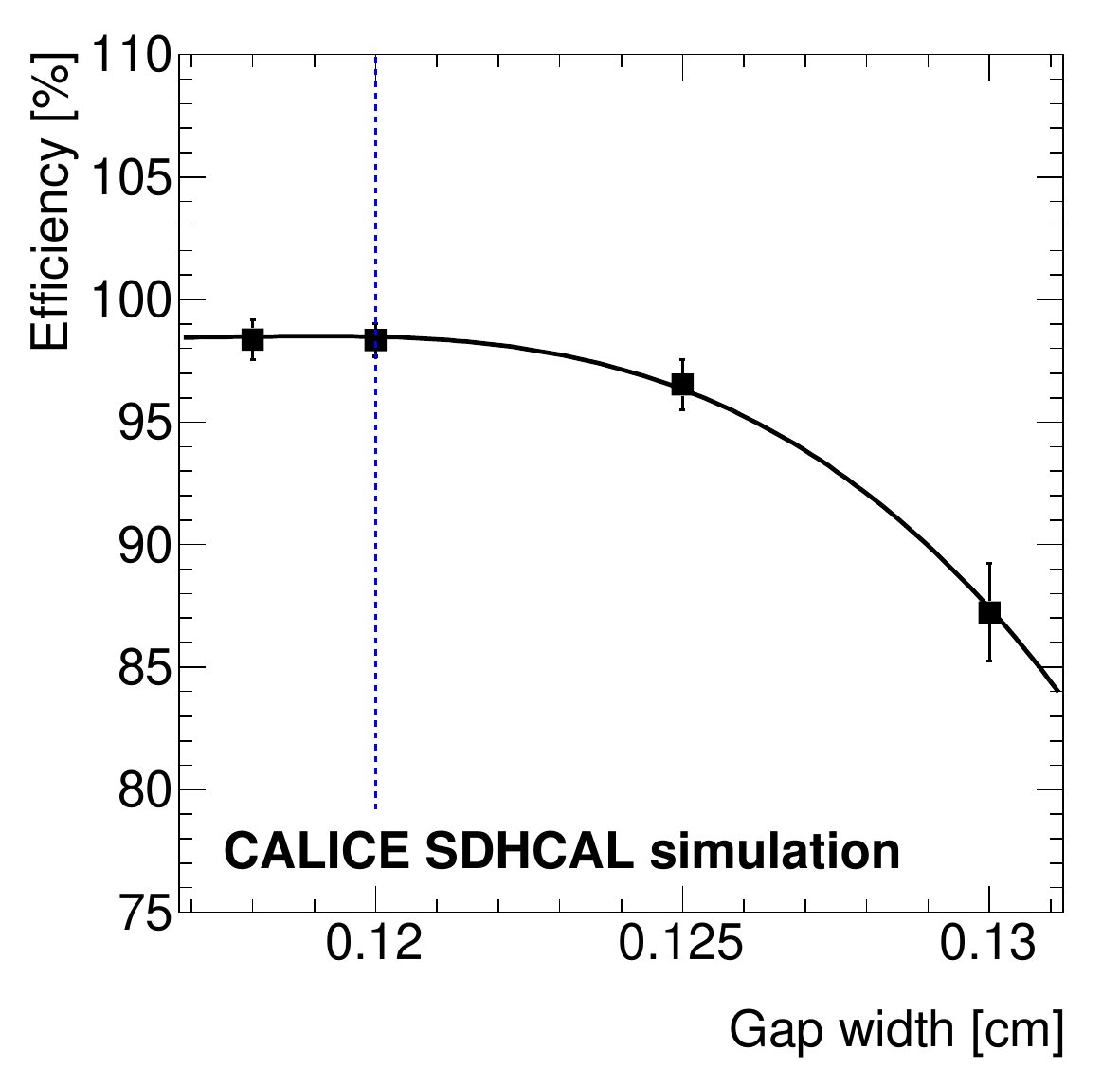}
    \caption{}
    \label{fig:eff.vs.d.mu}
  \end{subfigure}
  \caption{Mean total induced charge (\subref{fig:q.vs.d.mu}) and efficiency
    (\subref{fig:eff.vs.d.mu}) as a function of the gap width for hits
    initiated by simulated 100\,GeV muons. The nominal gap width is
  indicated by a vertical dashed blue line. The function that fits
  the variation is represented by a black solid line.}
  \end{center}
\end{figure}

The signal induced in the GRPC increases significantly for smaller gap sizes.
Figure~\ref{fig:q.vs.d.mu} shows the  
evolution of the total induced charges as a function of the GRPC gap width.
Above $0.12\,\centi\meter$, the nominal thickness of the SDHCAL GRPC gaps, a
sizable loss of signal is expected. 
A gap deformed by $+50\,\micro\meter$ will lead to a reduction of the signal amplitude of
about $38\%$. Such a variation will induce a significant effect on the
second and third threshold hit multiplicities.
The hit efficiency, defined as the fraction of
ionizations that induce a signal above the first
threshold, is estimated as a function of the gap width
(cf. Figure~\ref{fig:eff.vs.d.mu}). It decreases in the case of gap inflation.
An efficiency loss of about $1\%$ ($2\%$) is expected if the GRPC gap is deformed
by $+10\,\micro\meter$ ($+50\,\micro\meter$).

In order to estimate the dependence of the detector response regarding gap
width variations, a full simulation of electron and pion showers is
produced. Particles with energies from 7 to 70\,GeV interacting in the SDHCAL are simulated.
In order to account for the effect of a gap width variation, the dependence of
the efficiency and the signal amplitude on the gap width is interpolated
using Figures~\ref{fig:q.vs.d.mu} and~\ref{fig:eff.vs.d.mu}. This dependence
is implemented in the digitization step of the simulation.

Different detector geometries are compared: the nominal geometry and pessimistic
scenarios where the GRPC gaps are inflated following different
values, from 
$10$ to $50\,\micro\meter$. It is assumed that all the detector layers are deformed
coherently (referred to as "$+10\,\micro\meter$").
An alternative scenario is considered where the gap chambers are randomly
varied following a flat distribution from $-100\,\micro\meter$ to
$+100\,\micro\meter$ (referred to as "$\pm100\,\micro\meter$"). In this
case, the deformation of each layer gap is kept constant in all Monte Carlo
events to model an intrinsic design inhomogeneity.
The chosen value of $100\,\micro\meter$ is pessimistic regarding the tolerance of $40\,\micro\meter$ imposed on the design of each GRPC plate.

\begin{figure}[t]
 \begin{center}
 \begin{subfigure}[b]{0.31\textwidth}
    \includegraphics[width=\textwidth]{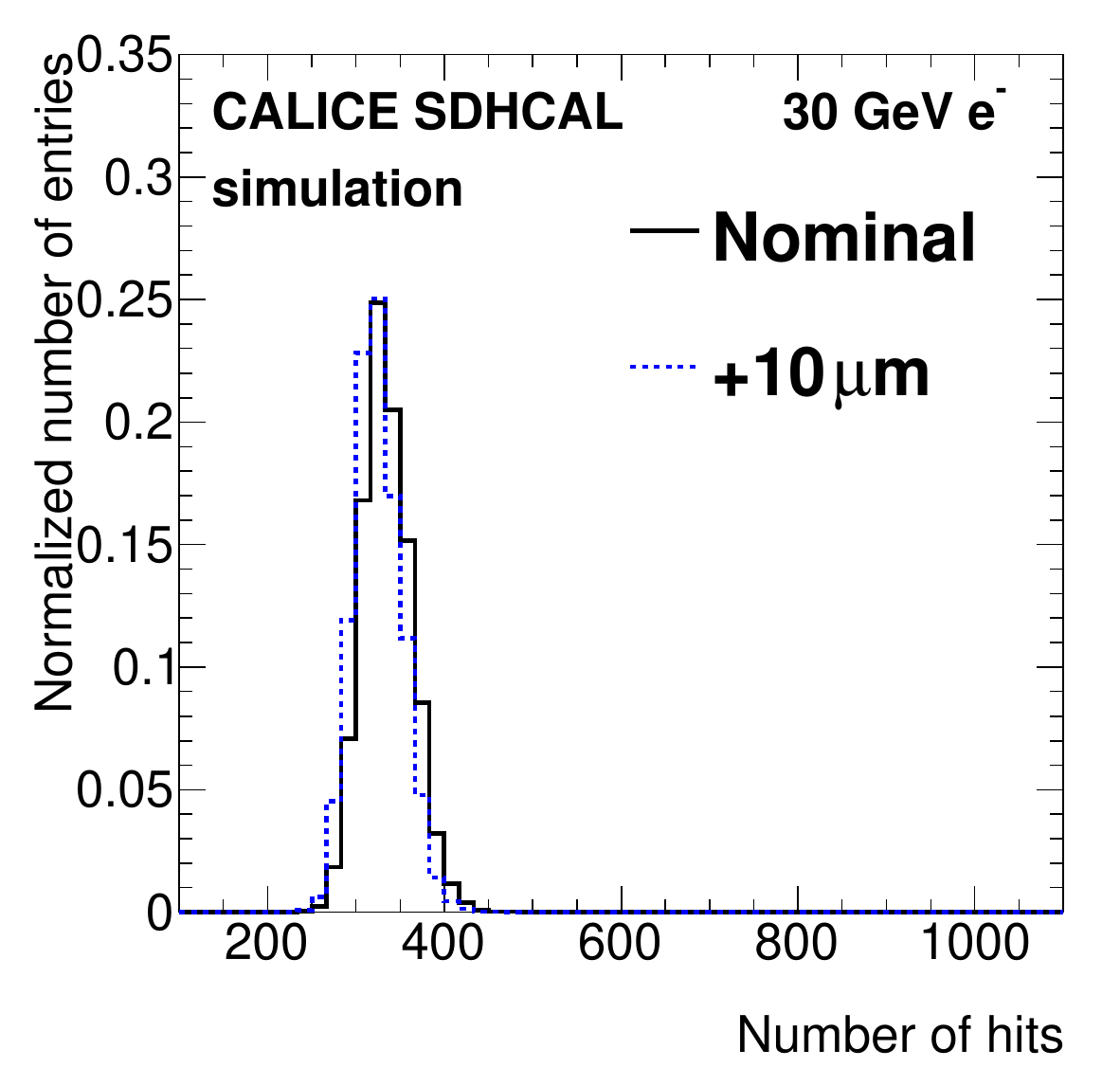}
    \caption{}
    \label{fig:nh.vs.width.1}
  \end{subfigure}
  \begin{subfigure}[b]{0.31\textwidth}
    \includegraphics[width=\textwidth]{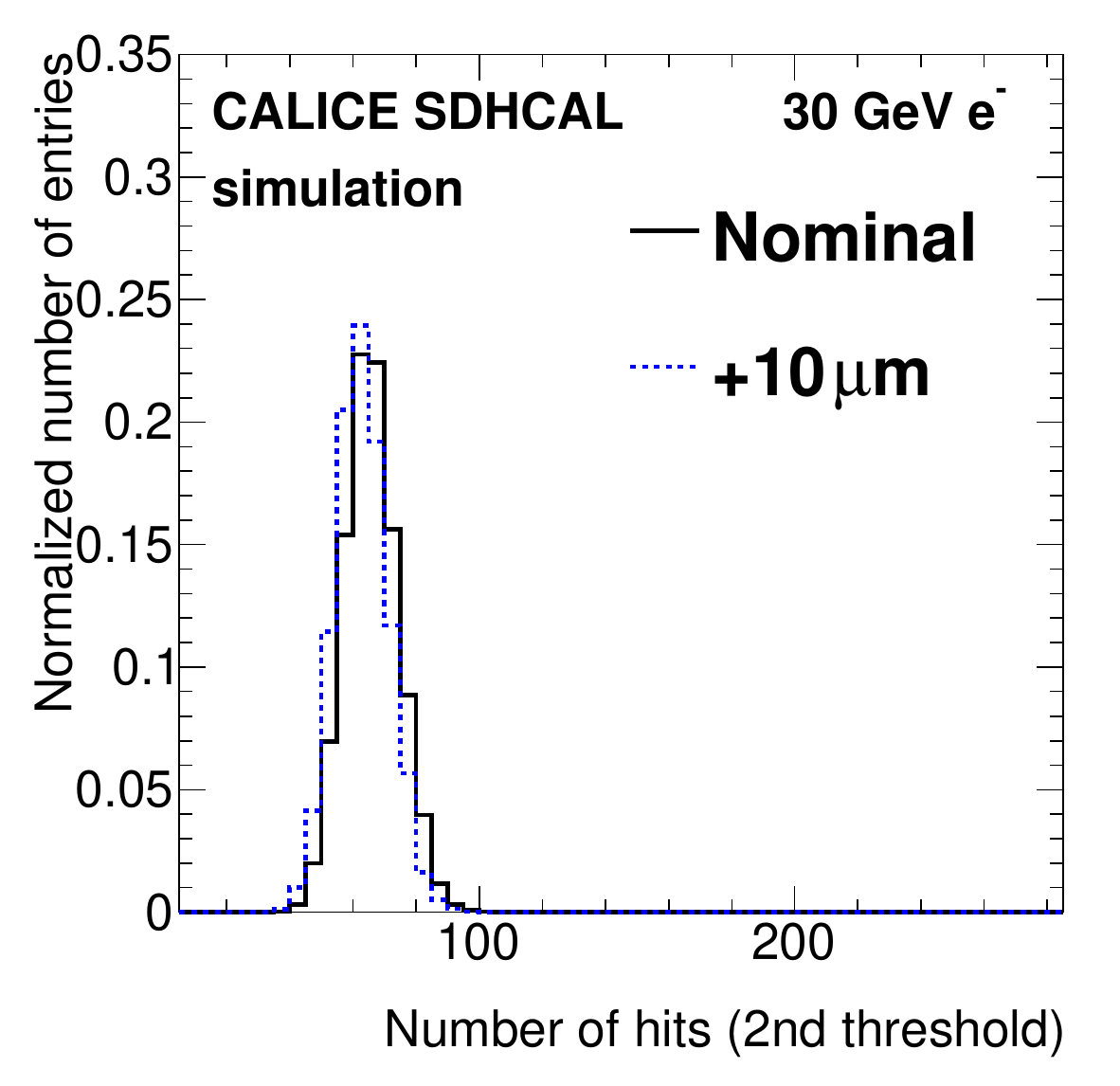}
    \caption{}
    \label{fig:nh.vs.width.2}
  \end{subfigure}
  \begin{subfigure}[b]{0.31\textwidth}
    \includegraphics[width=\textwidth]{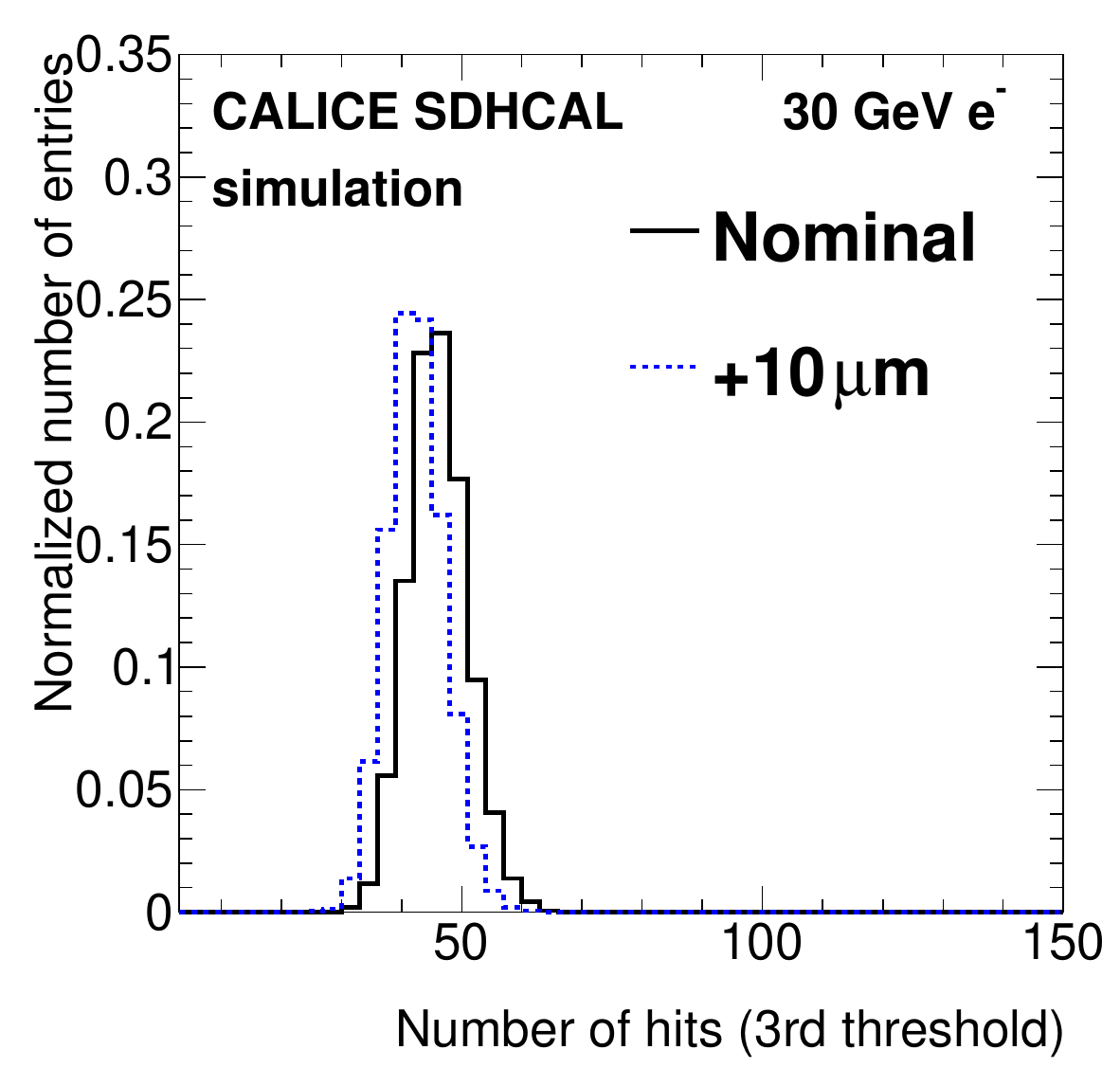}
    \caption{}
    \label{fig:nh.vs.width.3}
  \end{subfigure}
  \begin{subfigure}[b]{0.31\textwidth}
    \includegraphics[width=\textwidth]{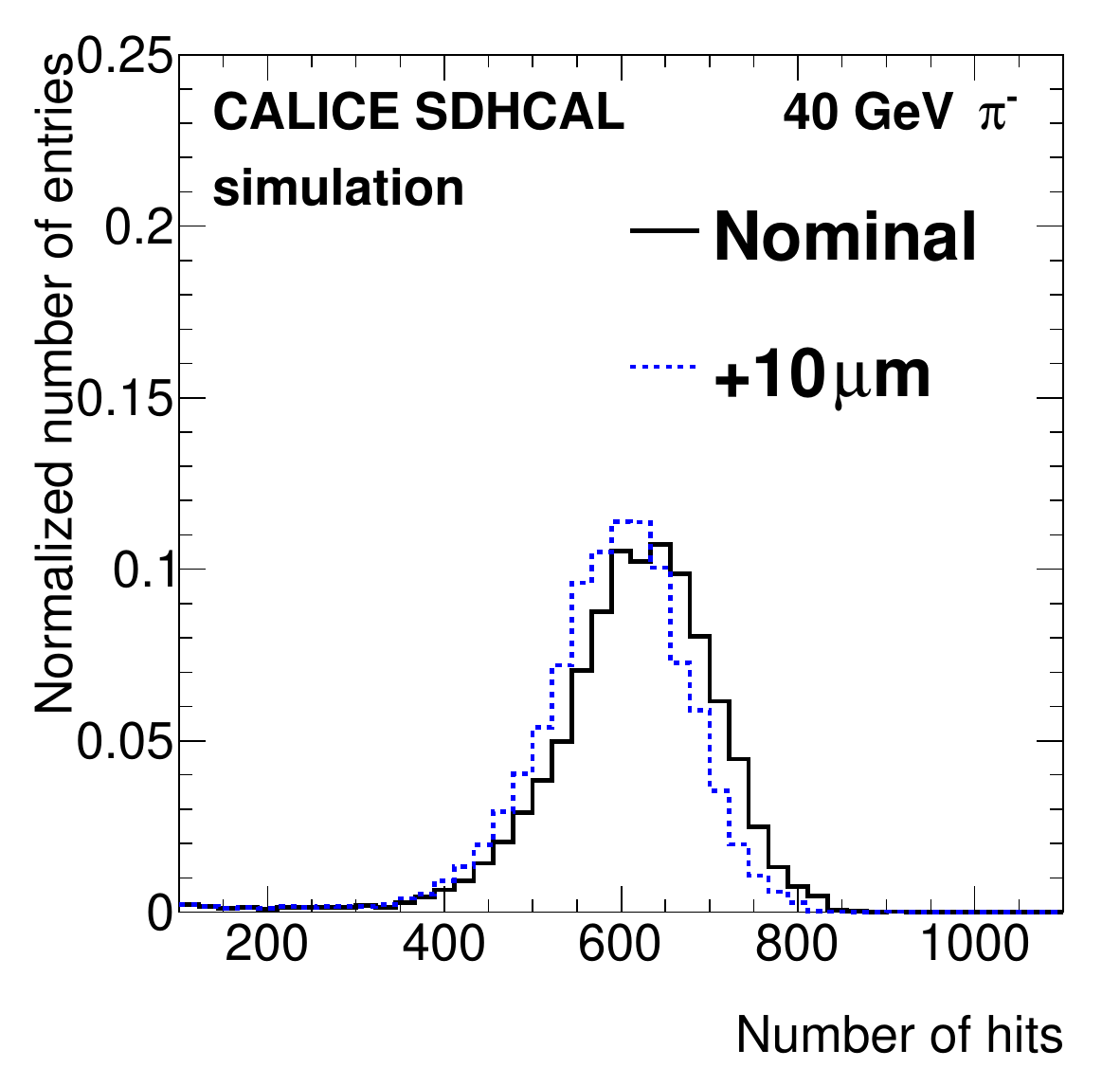}
    \caption{}
    \label{fig:nh.vs.width.pi1}
  \end{subfigure}
  \begin{subfigure}[b]{0.31\textwidth}
    \includegraphics[width=\textwidth]{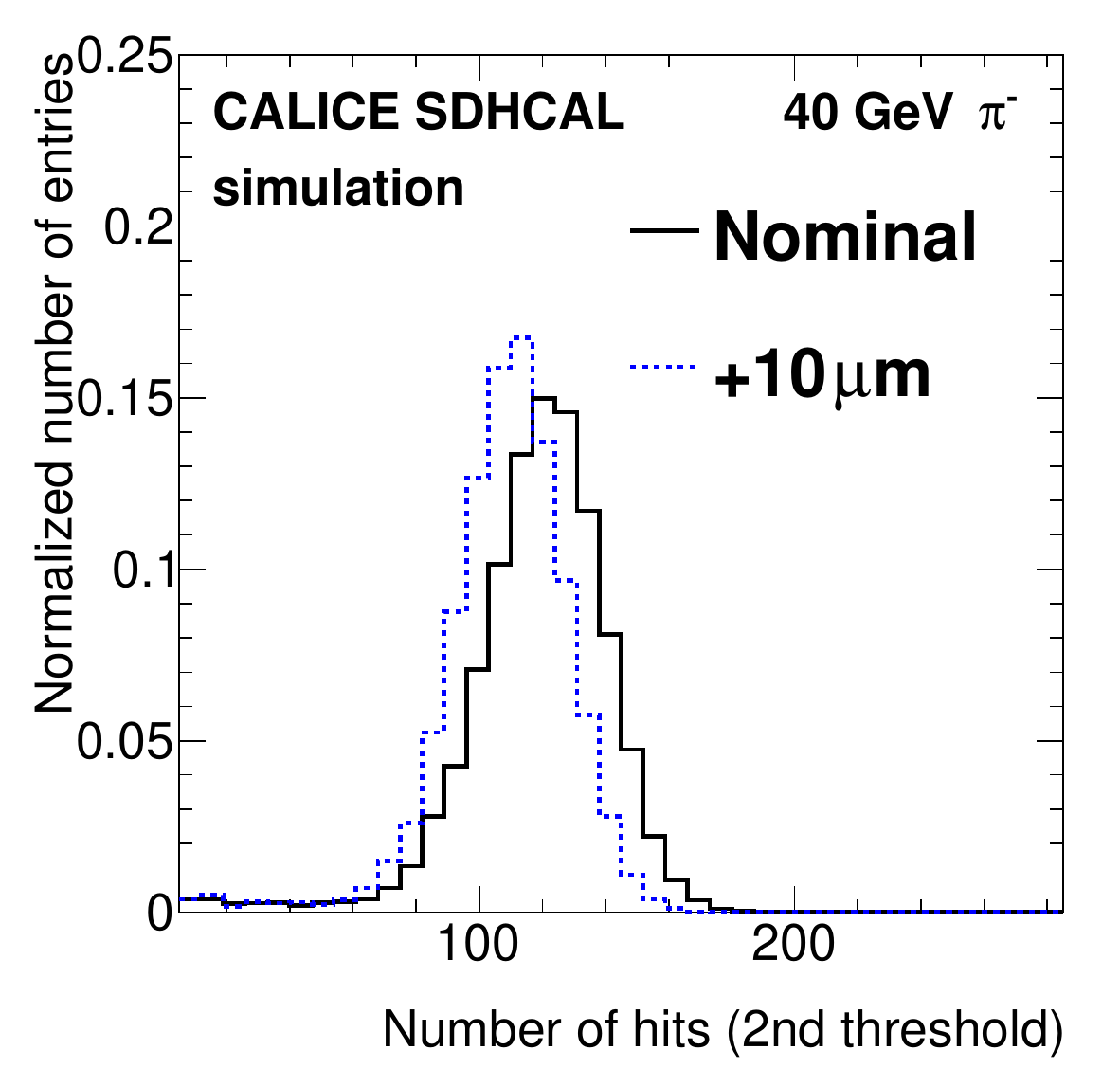}
    \caption{}
    \label{fig:nh.vs.width.pi2}
  \end{subfigure}
  \begin{subfigure}[b]{0.31\textwidth}
    \includegraphics[width=\textwidth]{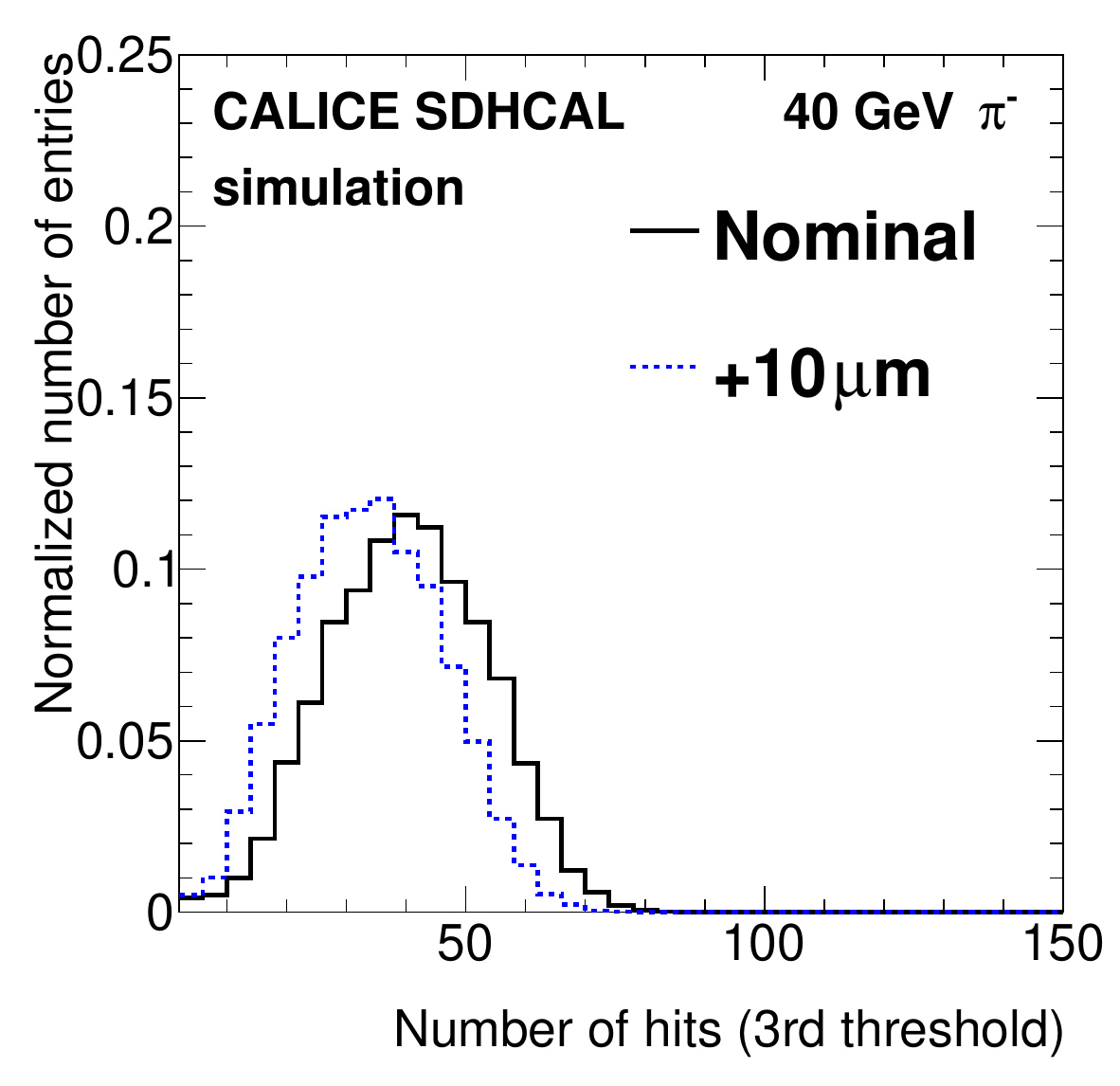}
    \caption{}
    \label{fig:nh.vs.width.pi3}
  \end{subfigure}
  \caption{ Distribution of the number of hits passing the first
    threshold at $0.1$\,pC~(\subref{fig:nh.vs.width.1},\subref{fig:nh.vs.width.pi1}), the second
    threshold at $5$\,pC~(\subref{fig:nh.vs.width.2},\subref{fig:nh.vs.width.pi2}) and the third
    threshold at $15$\,pC~(\subref{fig:nh.vs.width.3},\subref{fig:nh.vs.width.pi3}) for simulated  
    30\,GeV electrons (top) or 40\,GeV
    pions (bottom). The full GEANT4 simulation was performed with digitization modeling using the nominal geometry (solid black histograms) or
    a geometry where the gaps are coherently inflated by $+10\,\micro\meter$ (dashed blue histogram).\label{fig:nh.vs.width}} 
  \end{center}
\end{figure}

The number of hit distributions are reported on
Figure~\ref{fig:nh.vs.width} for the nominal
geometry and for a $+10\,\micro\meter$
inflated gap using electrons and 
pions. 
A gap width variation of $10\,\micro\meter$ leads to
a drop of $3\%$ in the average hit number. When considering the second and third threshold hit multiplicities, the drops are about $6\%$ and $8\%$, respectively.

The $\pm100\,\micro\meter$ tolerance scenario leads to a loss in
the expected number of hits. Even though the width variations sum to zero on average, the effect 
on the number of hits is significant. This is explained by threshold effects and
the non-linear dependence of the efficiency on the gap width. 
Moreover, such a random width variation induces a layer-to-layer smearing of
the detector response.

\begin{figure}[t]
 \begin{center}
    \includegraphics[width=0.6\textwidth]{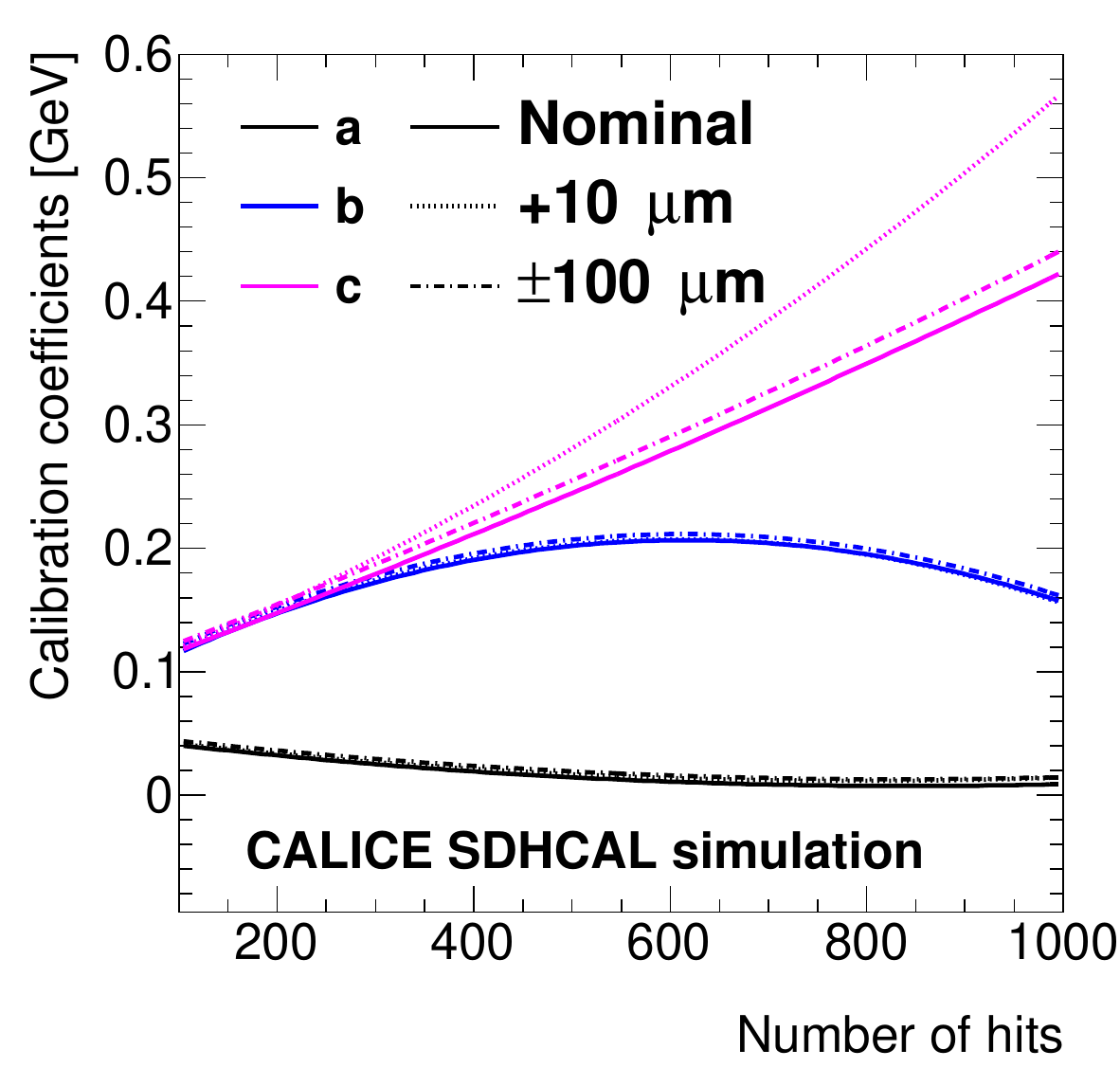}
  \caption{
    Dependences of the $a$, $b$ and $c$ energy reconstruction factors on the
    number of hits. The factors are determined from a simulated sample of pion
    showers using the nominal geometry (solid lines), geometry with coherent
    gap inflation by $+10\,\micro\meter$ (dotted lines) and geometry with
    random gap variations by $\pm100\,\micro\meter$ (dash-dotted
    lines).\label{fig:alpha.beta.gamma.pion.d}} 
  \end{center}
\end{figure}

Although the overall number of hits is relatively stable, the variation of high
amplitude hits will affect the performance of the SDHCAL, where the energy
reconstruction benefits from the shower density information provided by the
second and third threshold hits.
In order to estimate the stability of the reconstructed energy as a function
of the detector gap deformation, the entire reconstruction procedure is
applied to simulated pion showers at various beam energies, from 7 to 70\,GeV.
The $a$, $b$ and $c$ parameters from eq.~\eqref{eq:chi2} are computed for both types of
showers and different configurations. Figure~\ref{fig:alpha.beta.gamma.pion.d}
illustrates the dependence of these factors on the gap width. Only the
variations in the $c$ factor value is significant. 

Two scenarios are considered:
\begin{itemize}
\item The energy calibration is performed on a given detector geometry, then
  data are collected after a gap width variation: the energy bias and the
  loss in resolution induced by this case is 
obtained by applying the nominal reconstruction factors, i.e., calibrated on the
nominal geometry, on events simulated with a deformed geometry. 
\item The detector geometry is different from the nominal one but stable and
  known: in this case, the energy reconstruction factors are derived from and
  applied to each geometry. 
\end{itemize}

\begin{figure}[t]
 \begin{center}
 \begin{subfigure}[b]{0.45\textwidth}
    \includegraphics[width=\textwidth]{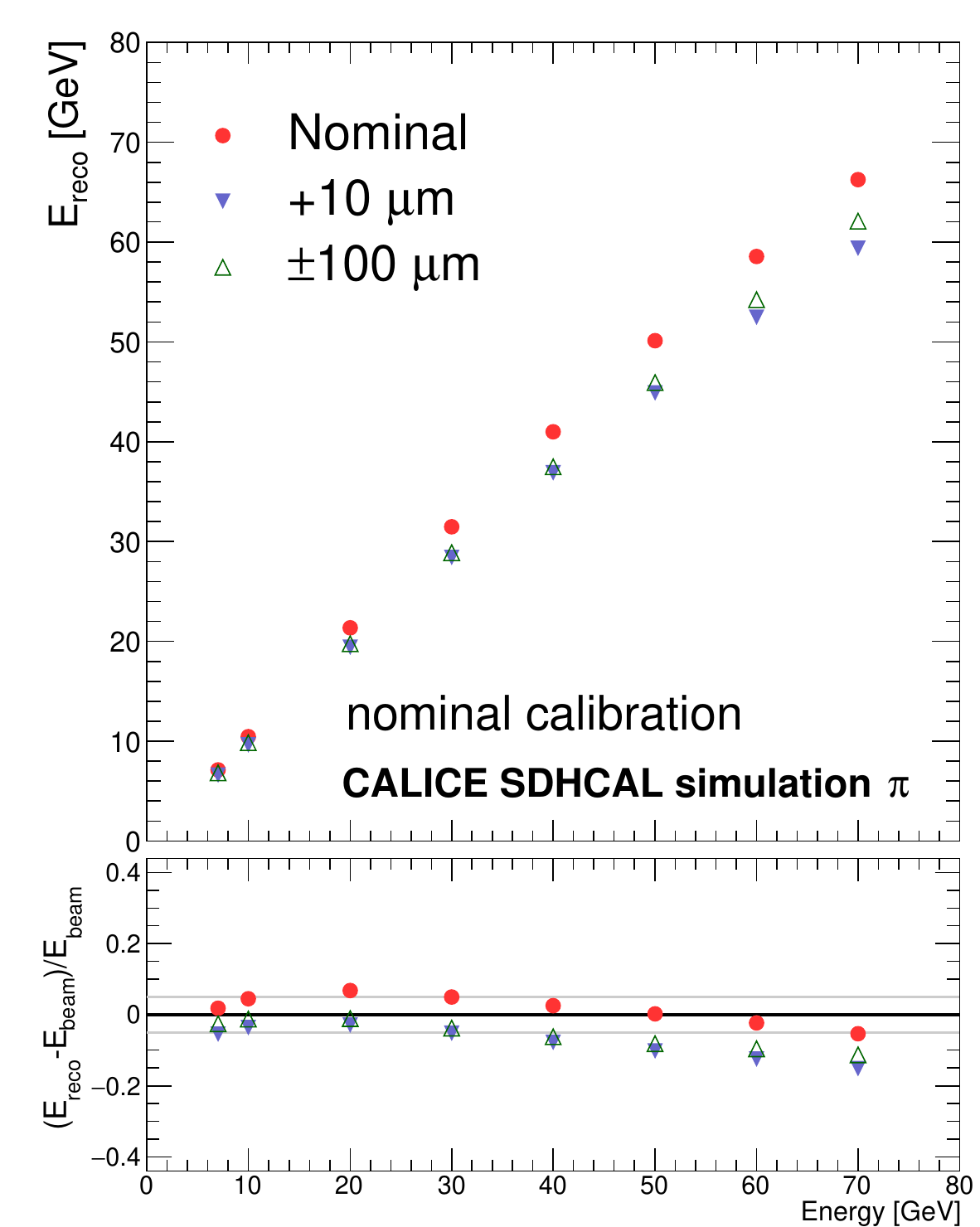}
    \caption{}
    \label{fig:ereco.pi.d.bias}
  \end{subfigure}
 \begin{subfigure}[b]{0.45\textwidth}
    \includegraphics[width=\textwidth]{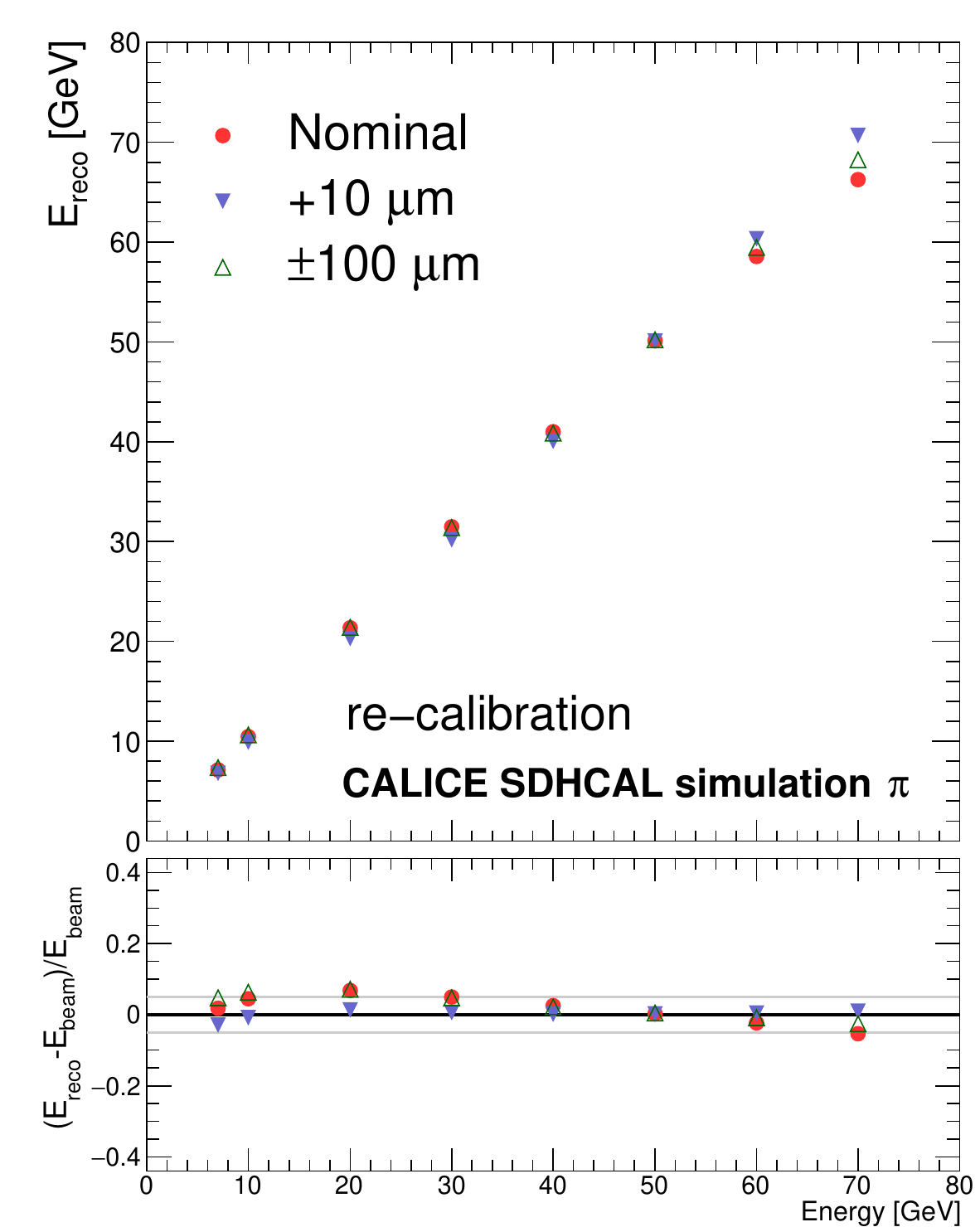}
    \caption{}
    \label{fig:ereco.pi.d.optimal}
  \end{subfigure}
  \caption{Pion reconstructed energy as a function of the generated energy for
    simulations using nominal gap (red circles), coherent gap variation by
    $+10\,\micro\meter$ (blue filled triangles) and random gap variation by
    $\pm100\,\micro\meter$ (green open triangles). 
    For each of the three simulation options, the energy reconstruction factors
    are optimised using the nominal simulation for all cases or~(\subref{fig:ereco.pi.d.bias}) 
    re-calibrated for each geometry~(\subref{fig:ereco.pi.d.optimal}).     
    \label{fig:ereco.lin.d}}
  \end{center}
\end{figure}

In the first scenario, 
if the detector chambers are coherently inflated by
$+10\,\micro\meter$,
the reconstructed energy 
is shifted by  5\% to 17\% when considering 7 to 70\,GeV 
pion showers (cf. Figure~\ref{fig:ereco.pi.d.bias}).

\begin{figure}[t]
 \begin{center}
 \begin{subfigure}[b]{0.495\textwidth}
    \includegraphics[width=\textwidth]{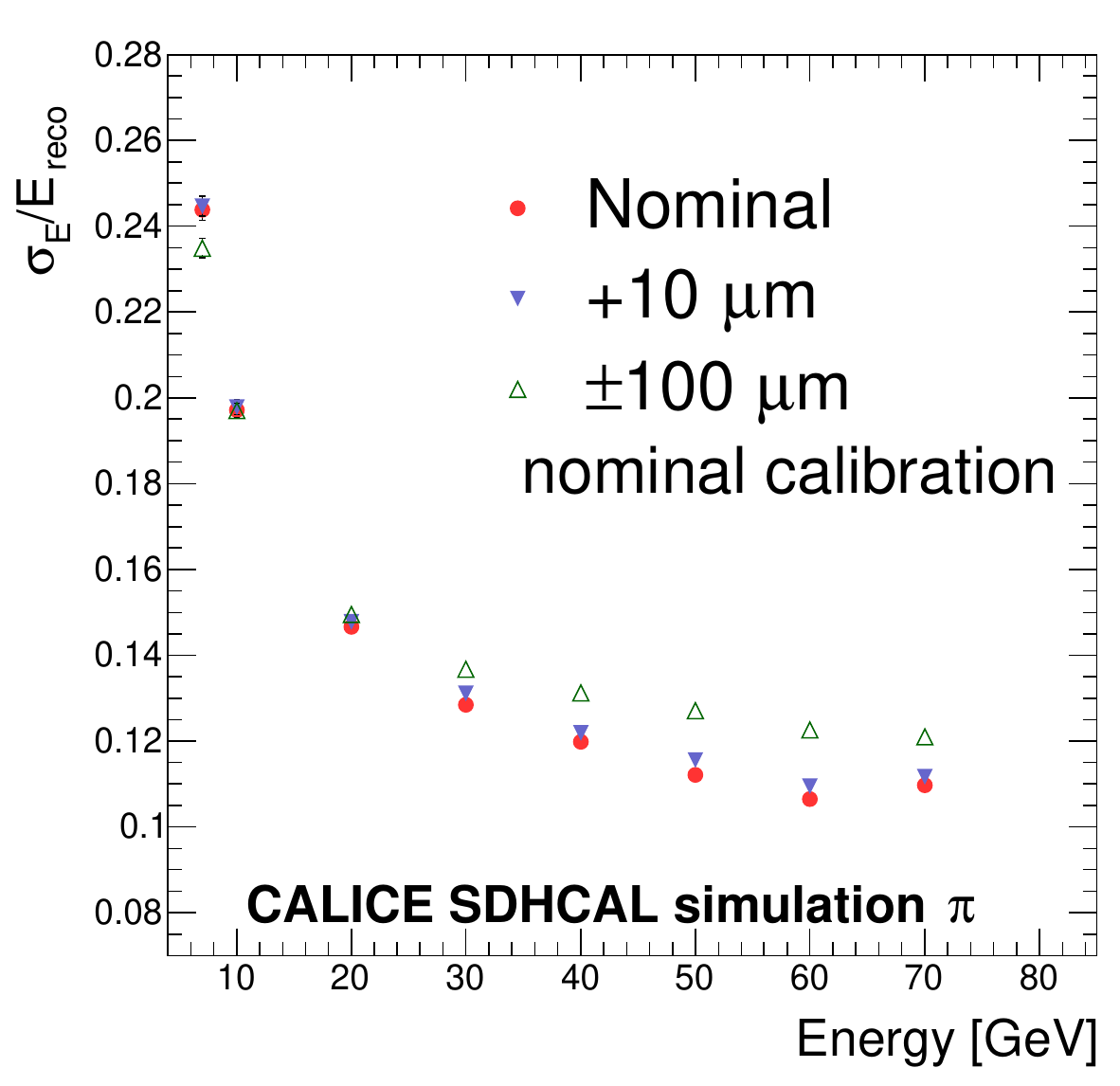}
    \caption{}
    \label{fig:resol.d.bias}
  \end{subfigure}
  \begin{subfigure}[b]{0.495\textwidth}
    \includegraphics[width=\textwidth]{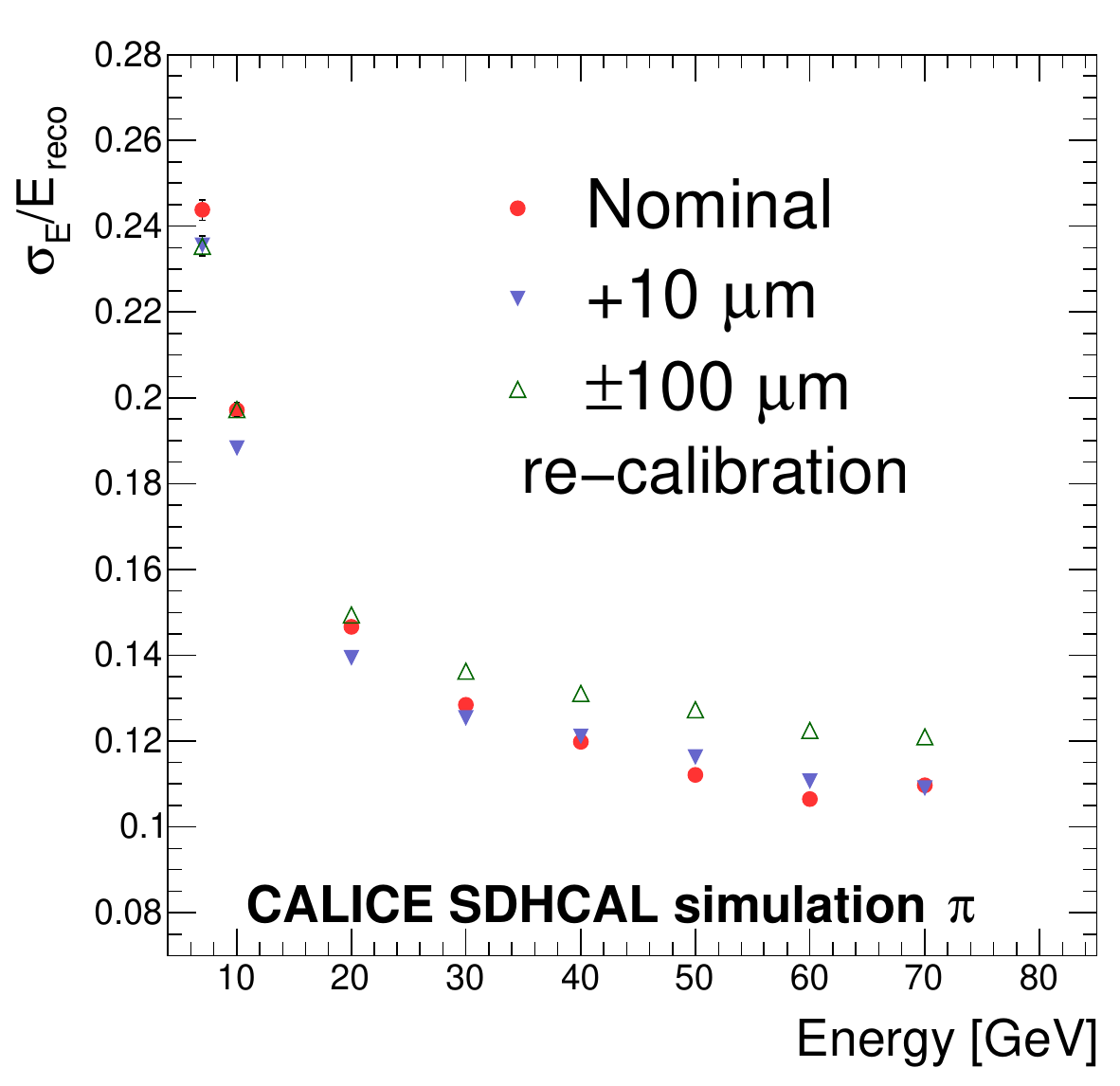}
    \caption{}
    \label{fig:resol.d.opt}
  \end{subfigure}
  \caption{ Relative energy resolution for pions versus generated energy for
    simulations using nominal gap (red circles), coherent gap variation by
    $+10\,\micro\meter$ (blue filled triangles) and random gap variation by
    $\pm100\,\micro\meter$ (green open triangles). 
    For each of the three simulation options, the energy reconstruction factors
    are~(\subref{fig:resol.d.bias}) optimised using the nominal simulation for
    all cases or~(\subref{fig:resol.d.opt}) re-calibrated for each geometry. 
    \label{fig:ersol.d}}  
  \end{center}
\end{figure}

If the energy determination is adapted to the deformed geometry (second
scenario), the new $a$, $b$ and $c$ factors absorb the hit
multiplicity changes and restore the linearity within $\pm5\%$ 
(cf. Figure~\ref{fig:ereco.pi.d.optimal}). 
When considering a tolerance of $\pm100\,\micro\meter$, the layer-to-layer gap
width variation translates into a spread in the reconstructed energy the $a$,
$b$ and $c$ factors cannot absorb.  The worsening of energy resolution by about $2\%$ is observed as shown in Figure~\ref{fig:ersol.d}.

%% file: latex/section-temp.tex
\subsection{Impact of temperature and pressure}

\begin{figure}[t]
 \begin{center}
 \begin{subfigure}[b]{0.495\textwidth}
    \includegraphics[width=\textwidth]{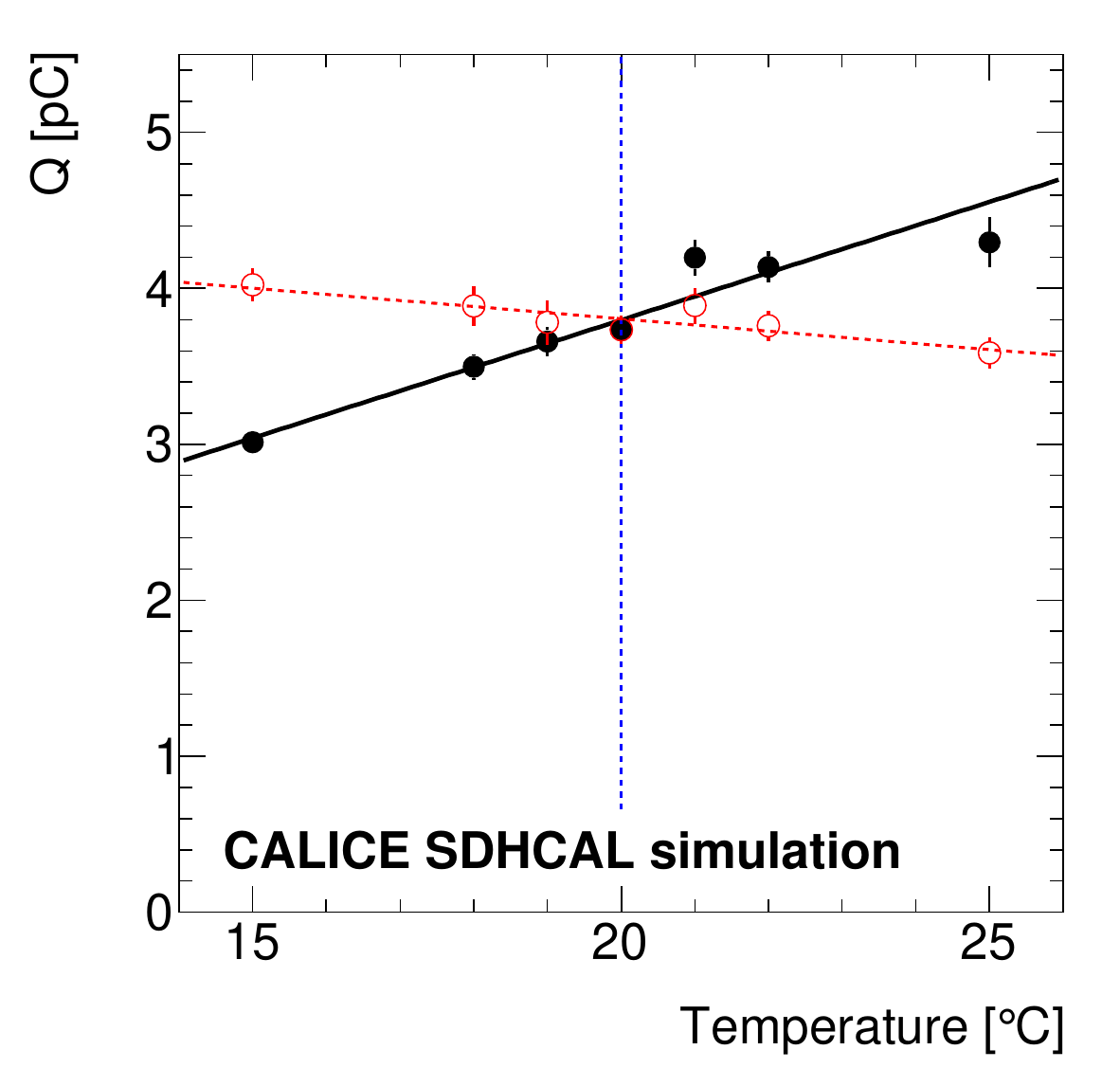}
    \caption{}
    \label{fig:q.vs.t}
  \end{subfigure}
  \begin{subfigure}[b]{0.495\textwidth}
    \includegraphics[width=\textwidth]{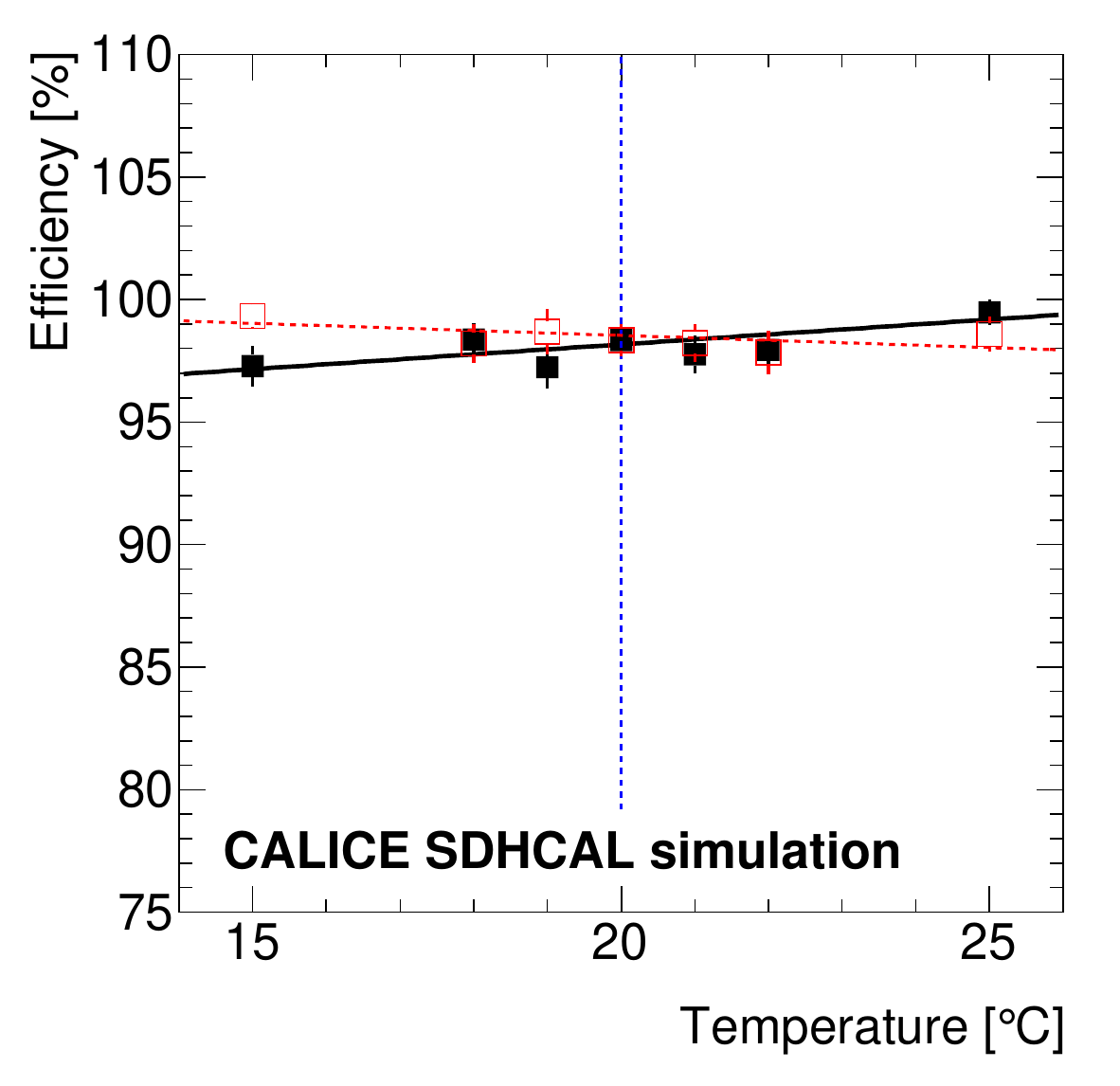}
    \caption{}
    \label{fig:eff.vs.t}
  \end{subfigure}
  \caption{
    Mean total induced charge (\subref{fig:q.vs.t}) and efficiency
    (\subref{fig:eff.vs.t}) as a function of the gas temperature for hits
    initiated by simulated 100\,GeV muons.
  The nominal temperature is indicated by a vertical dashed blue line. High voltage is assumed to be stable (solid markers) or rescaled with the temperature (empty markers). The linear function that fits the variations is represented by a black solid line and a red dotted line.\label{fig:qeff.vs.t}}
  \end{center}
\end{figure}

\begin{figure}[t]
 \begin{center}
 \begin{subfigure}[b]{0.31\textwidth}
    \includegraphics[width=\textwidth]{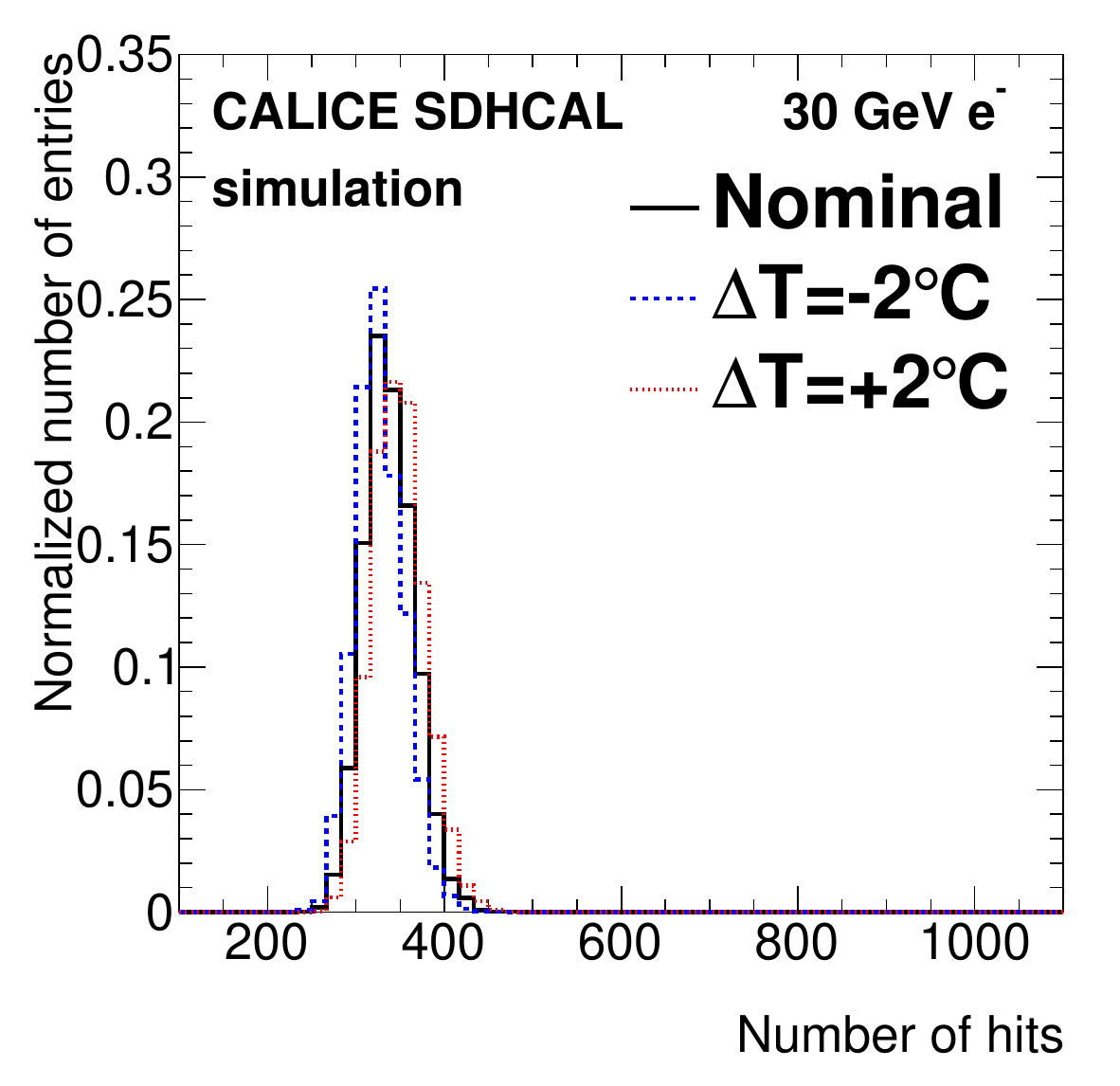}
    \caption{}
    \label{fig:nh.vs.t.ele.1}
  \end{subfigure}
  \begin{subfigure}[b]{0.31\textwidth}
    \includegraphics[width=\textwidth]{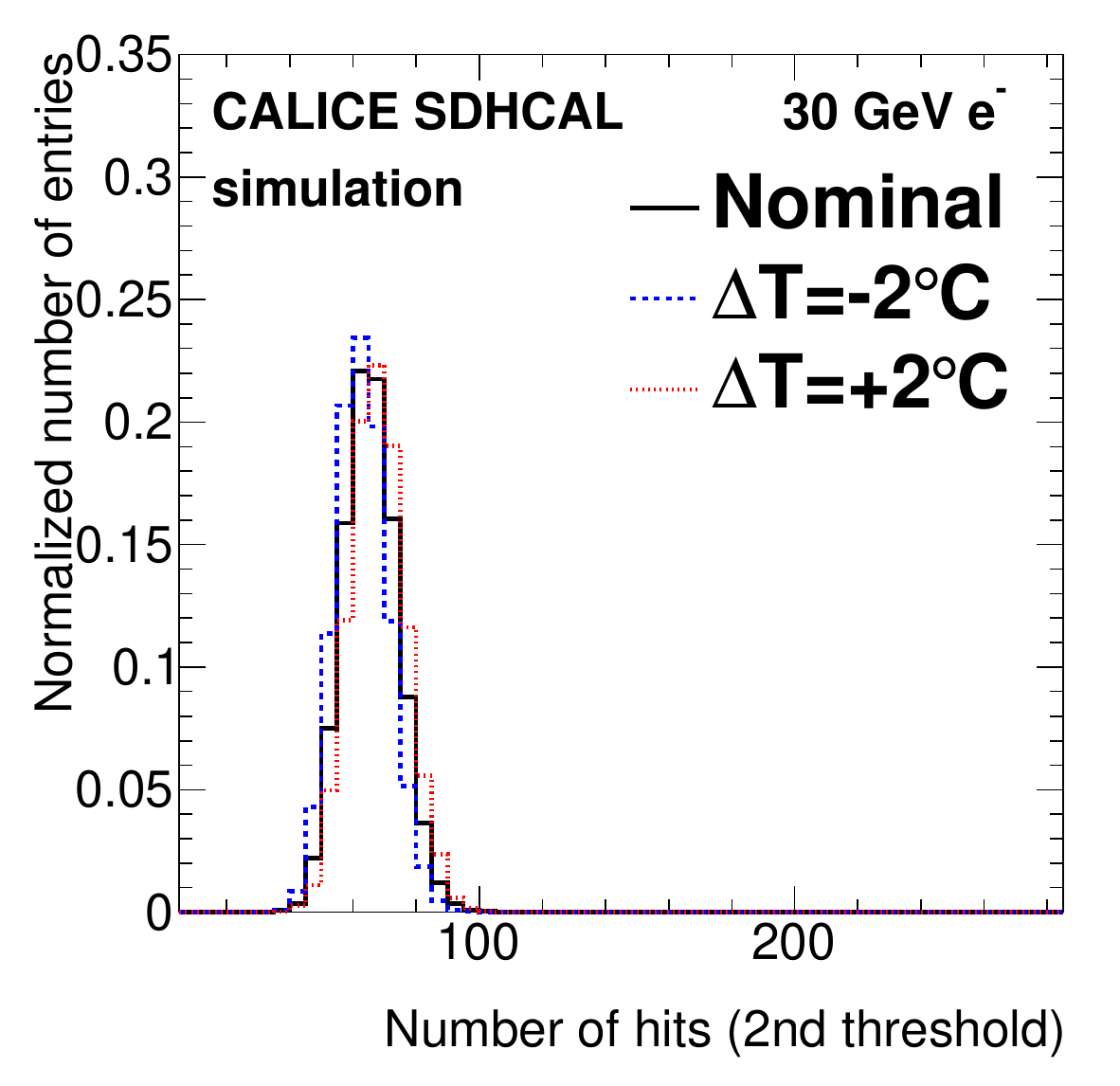}
    \caption{}
    \label{fig:nh.vs.t.ele.2}
  \end{subfigure}
  \begin{subfigure}[b]{0.31\textwidth}
    \includegraphics[width=\textwidth]{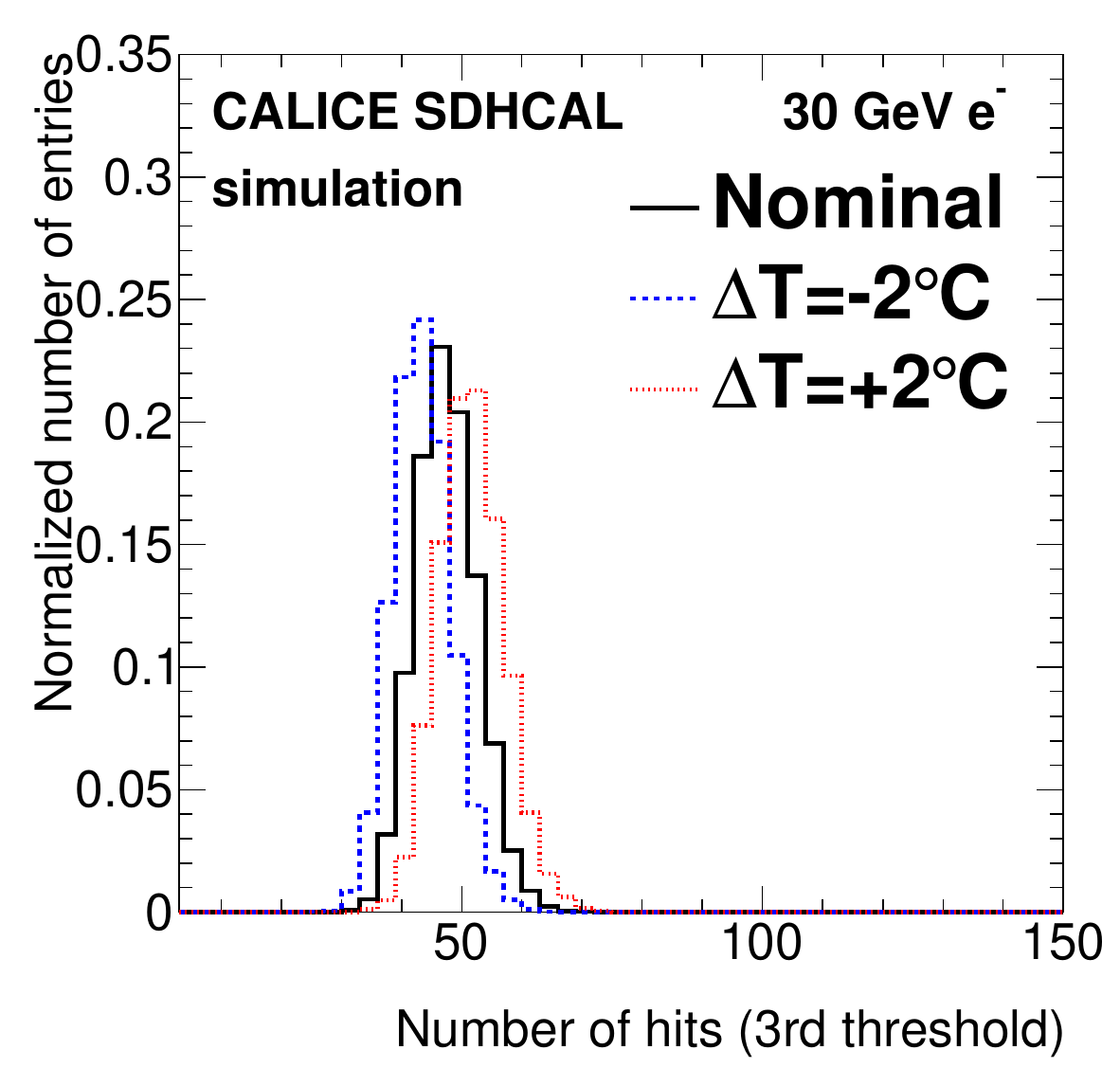}
    \caption{}
    \label{fig:nh.vs.t.ele.3}
  \end{subfigure}
 \begin{subfigure}[b]{0.31\textwidth}
    \includegraphics[width=\textwidth]{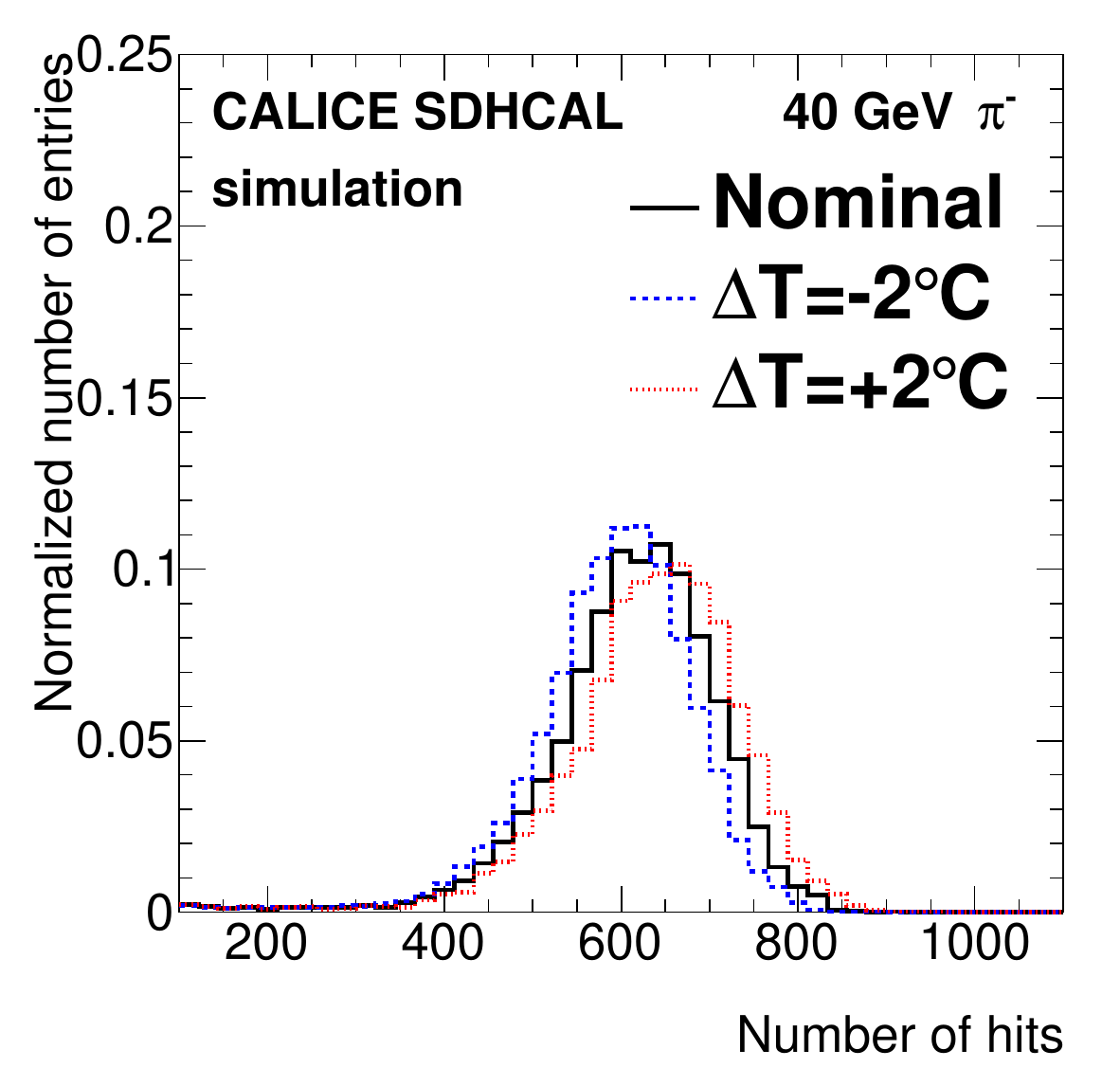}
    \caption{}
    \label{fig:nh.vs.t.1}
  \end{subfigure}
  \begin{subfigure}[b]{0.31\textwidth}
    \includegraphics[width=\textwidth]{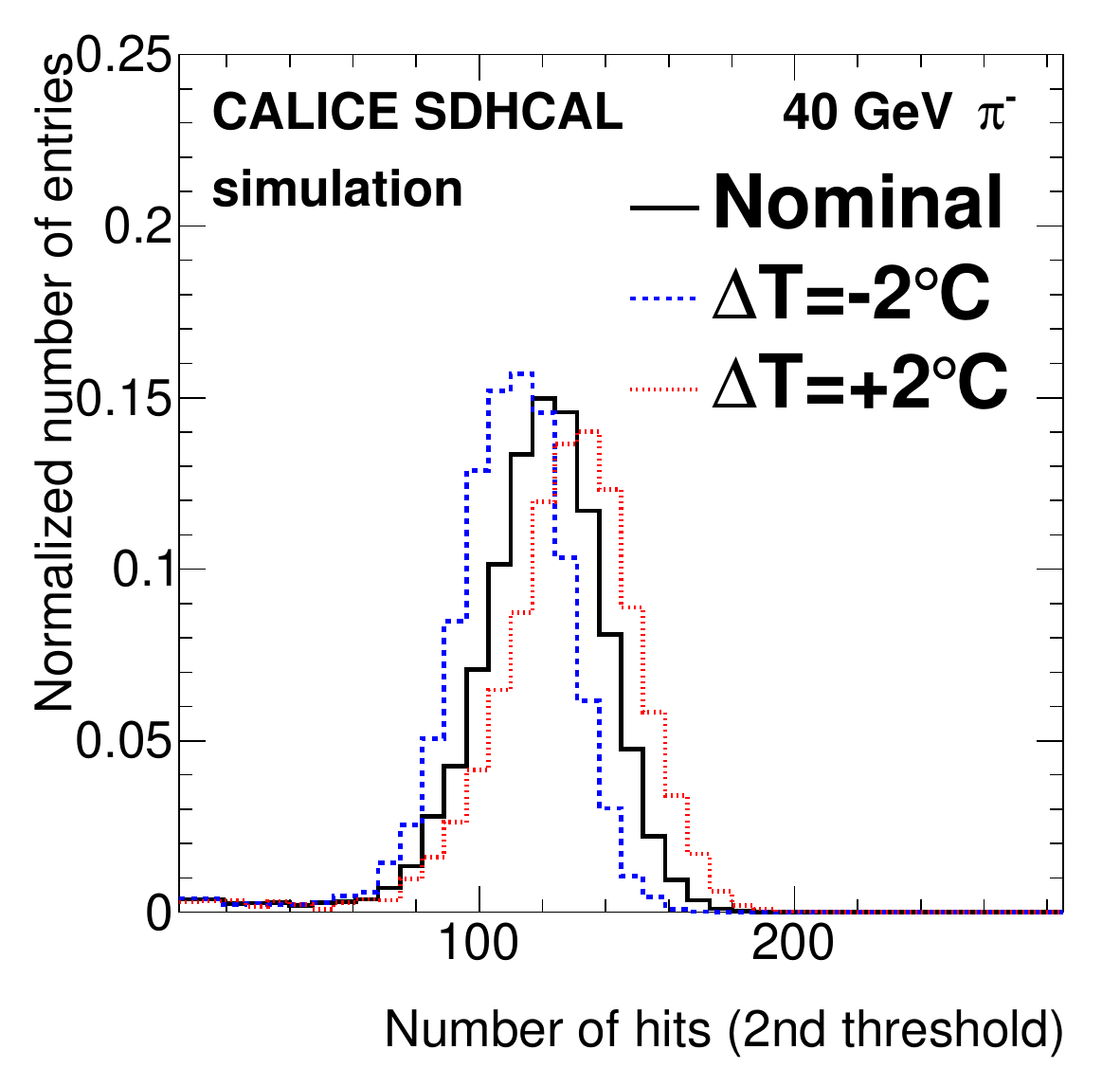}
    \caption{}
    \label{fig:nh.vs.t.2}
  \end{subfigure}
  \begin{subfigure}[b]{0.31\textwidth}
    \includegraphics[width=\textwidth]{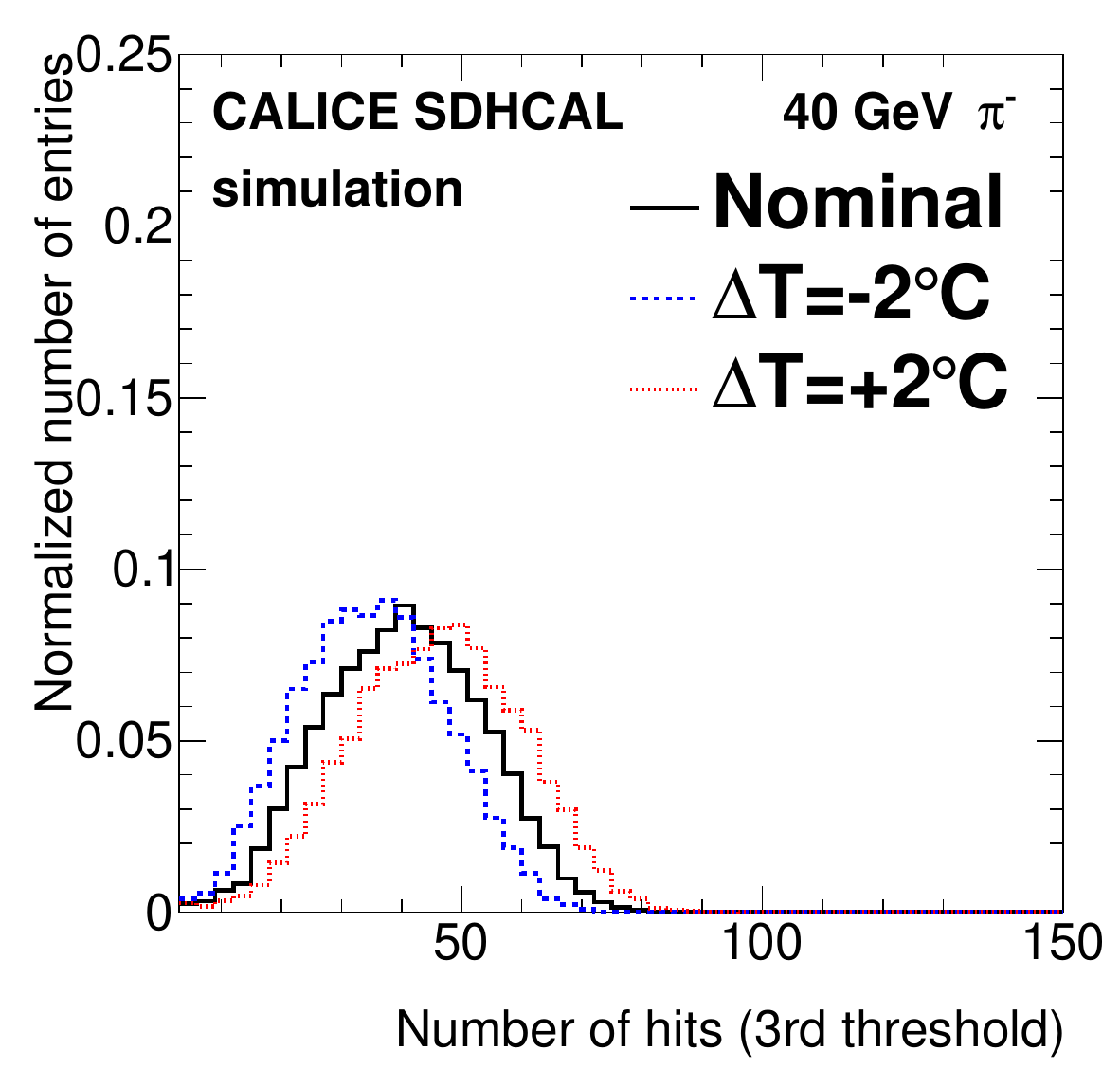}
    \caption{}
    \label{fig:nh.vs.t.3}
  \end{subfigure}
  \caption{Distribution of the number of hits passing the first threshold at
    $0.1$\,pC~(\subref{fig:nh.vs.t.ele.1},\subref{fig:nh.vs.t.1}), the second
    threshold at $5$\,pC~(\subref{fig:nh.vs.t.ele.2},\subref{fig:nh.vs.t.2})
    and the third threshold at
    $15$\,pC~(\subref{fig:nh.vs.t.ele.3},\subref{fig:nh.vs.t.3}) for
    simulated  30\,GeV electrons (top) or 40\,GeV pions (bottom). The full
    GEANT4 simulation was performed with digitization modeling using the
    nominal temperature (solid black histograms), using a temperature varied
    by $-2\degreecelsius$ (dashed
    blue histogram) or $+2\degreecelsius$ (dotted red histogram).\label{fig:nh.vs.t}}
  \end{center}
\end{figure}

At a given pressure, the temperature affects the gas density and therefore
the charge multiplication and the absorption probabilities,
charge velocity as well as the diffusion amplitude.
The multiplication and absorption probabilities scale linearly with the
temperature. 
In order to quantify the stability of the detector response in an evolving
environment, the temperature and pressure are varied in the simulation as
follows, starting with the variation of the temperature only first and assuming a stable high voltage.  
The amplitude and the efficiency of the signal are estimated and reported in
Figures~\ref{fig:q.vs.t} and~\ref{fig:eff.vs.t}.
It is found that a drop of the temperature by $5\degreecelsius$ is
accompanied by an efficiency drop of $1\%$ while the amplitude of the signal
is reduced by $20\%$. These effects are propagated to the number of hits as
shown in Figure~\ref{fig:nh.vs.t}.

\begin{figure}[h]
 \begin{center}
    \includegraphics[width=0.55\textwidth]{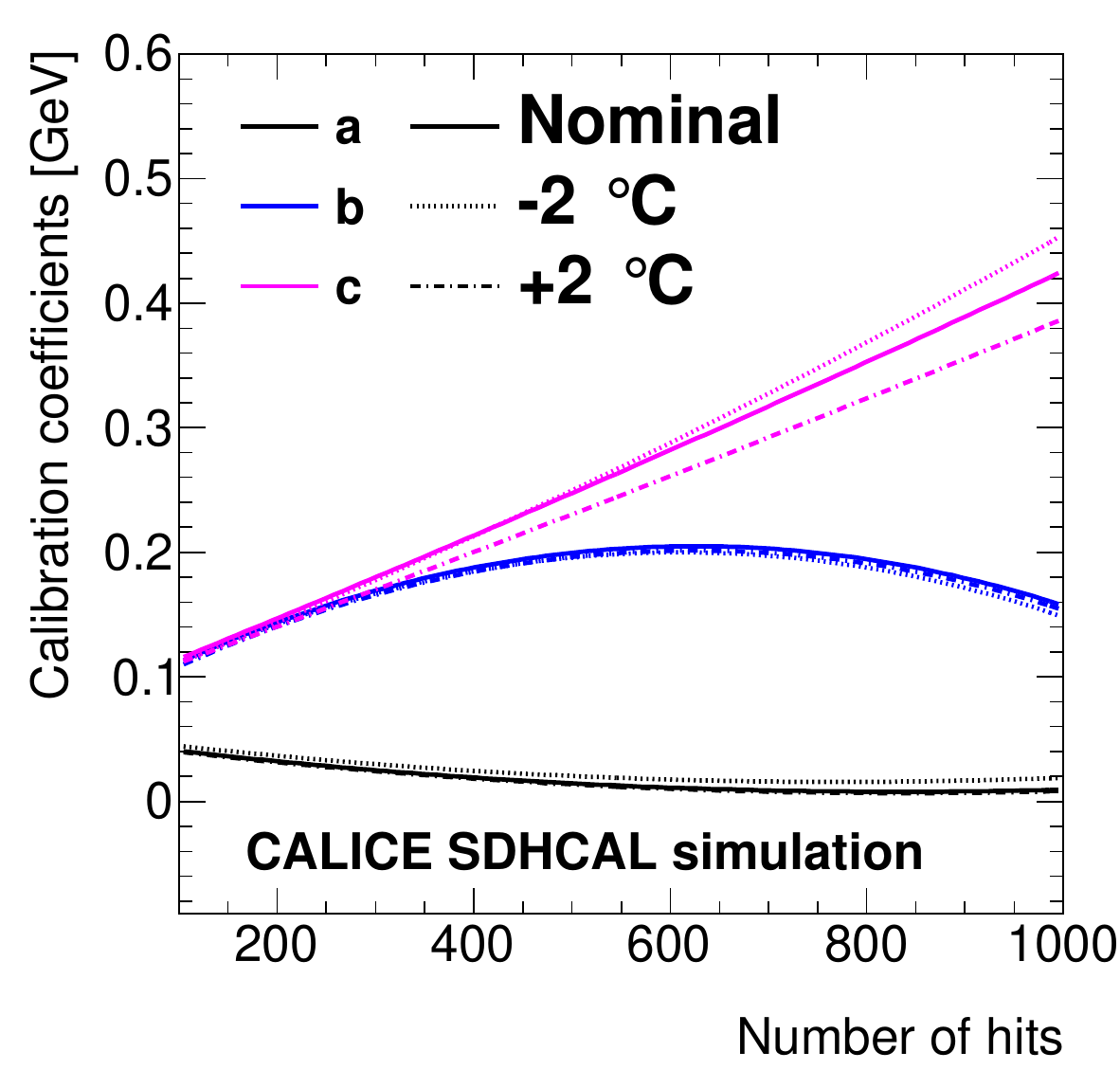}
  \caption{
      Dependences of the $a$, $b$ and $c$ energy calibration reconstruction factors on the
      number of hits. The factors are determined from a simulated sample of
      pion showers with the GRPCs at the nominal temperature (solid lines), a
      variation of $-2\degreecelsius$ (dotted lines) and
      $+2\degreecelsius$ (dash-dotted
      lines).\label{fig:alpha.beta.gamma.pi.t}}
  \end{center}
\end{figure}

\begin{figure}[h]
 \begin{center}
 \begin{subfigure}[b]{0.49\textwidth}
    \includegraphics[width=\textwidth]{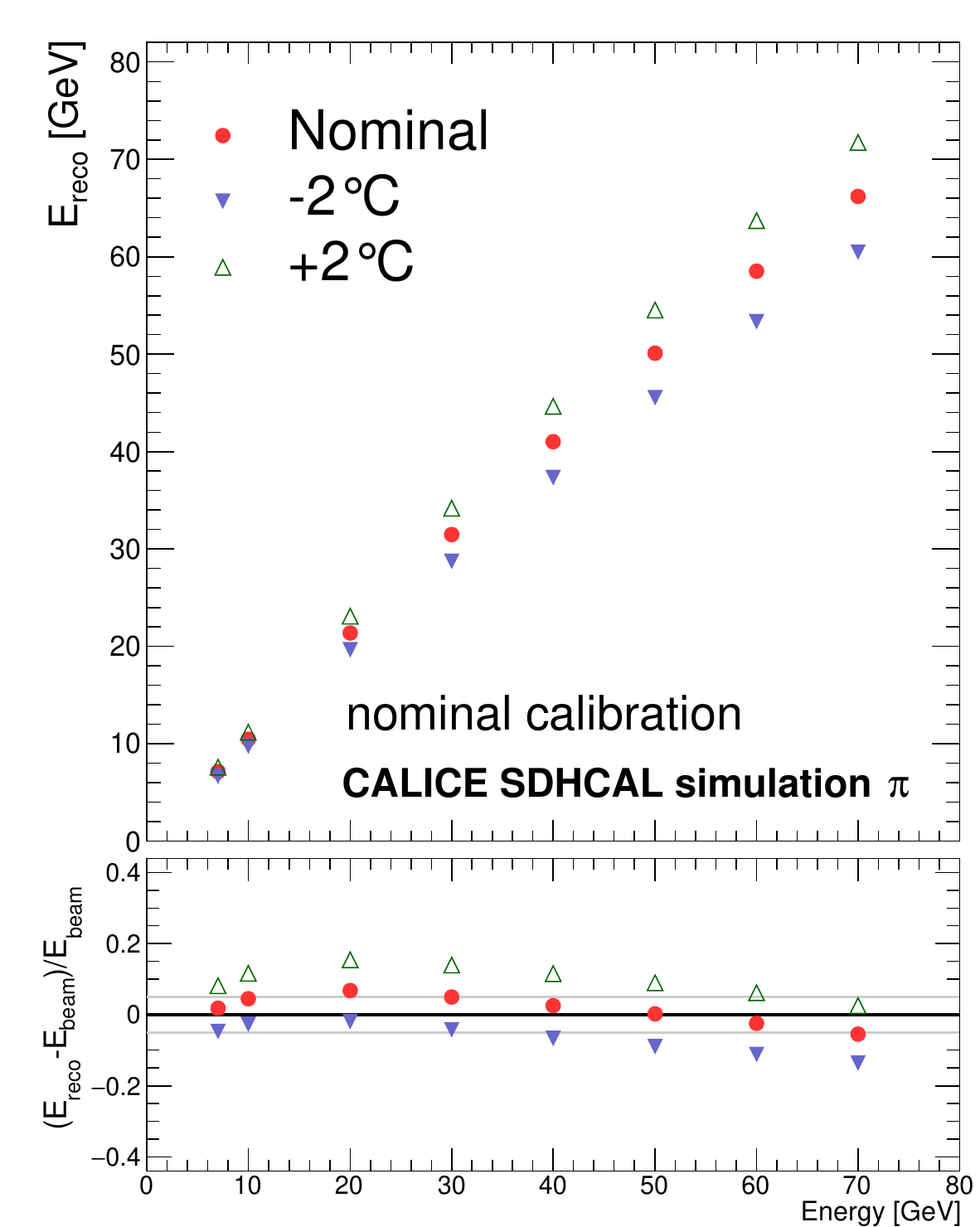}
    \caption{}
    \label{fig:ereco.pion.t.bias}
  \end{subfigure}
 \begin{subfigure}[b]{0.49\textwidth}
    \includegraphics[width=\textwidth]{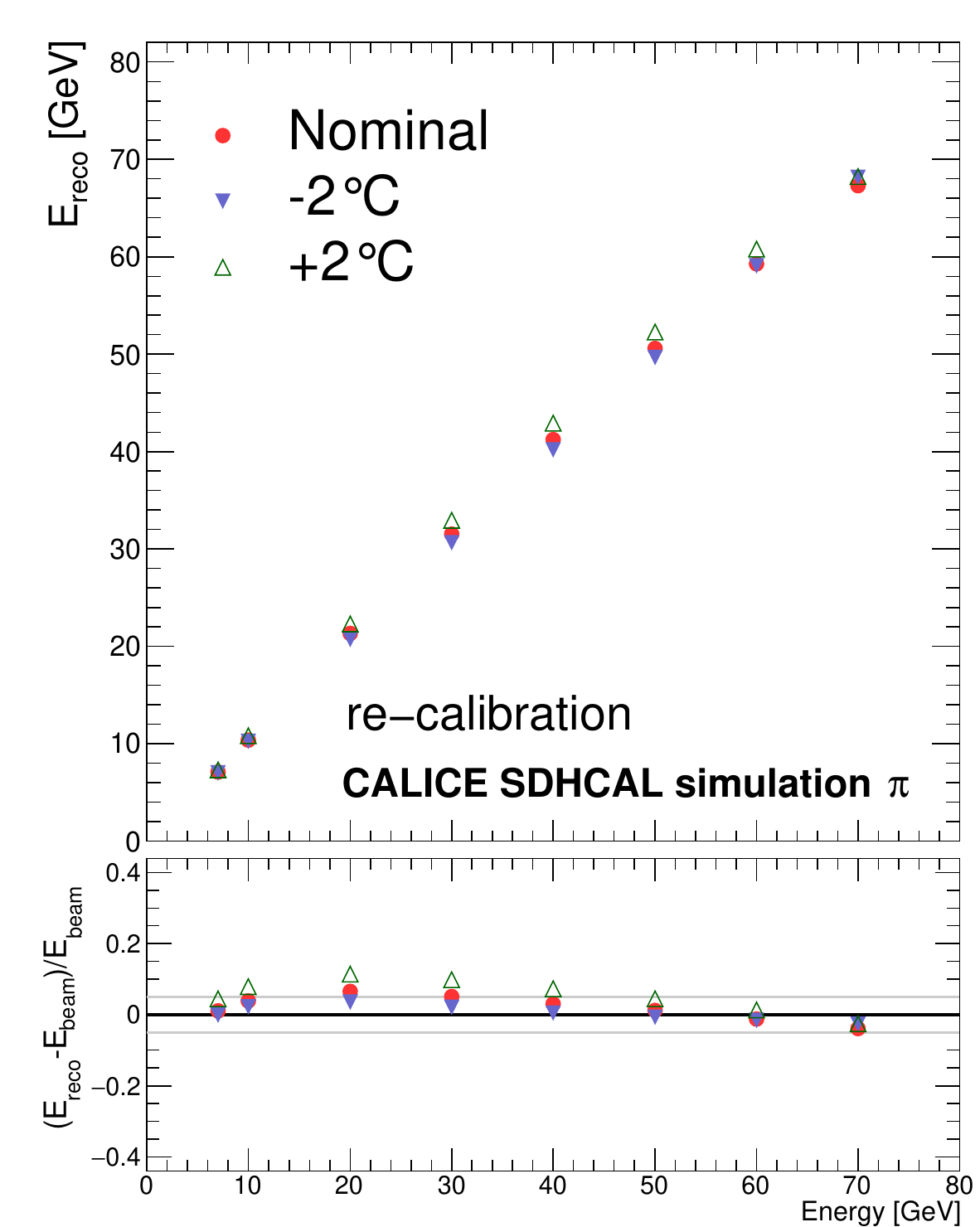}
    \caption{}
    \label{fig:ereco.pion.t.optimal}
  \end{subfigure}
  \caption{Pion reconstructed energy as a function of the generated energy for
    simulations with the GRPCs at the nominal temperature (red circles), a
    temperature variation by $-2\degreecelsius$ (blue filled triangles) and
    $+2\degreecelsius$ (green open triangles). The energy calibration reconstruction factors are calibrated using the nominal simulation for all cases~(\subref{fig:ereco.pion.t.bias}) or re-calibrated for each of the three simulations~(\subref{fig:ereco.pion.t.optimal}).\label{fig:ereco.t}}
  \end{center}
\end{figure}

\begin{figure}[h]
 \begin{center}
 \begin{subfigure}[b]{0.495\textwidth}
    \includegraphics[width=\textwidth]{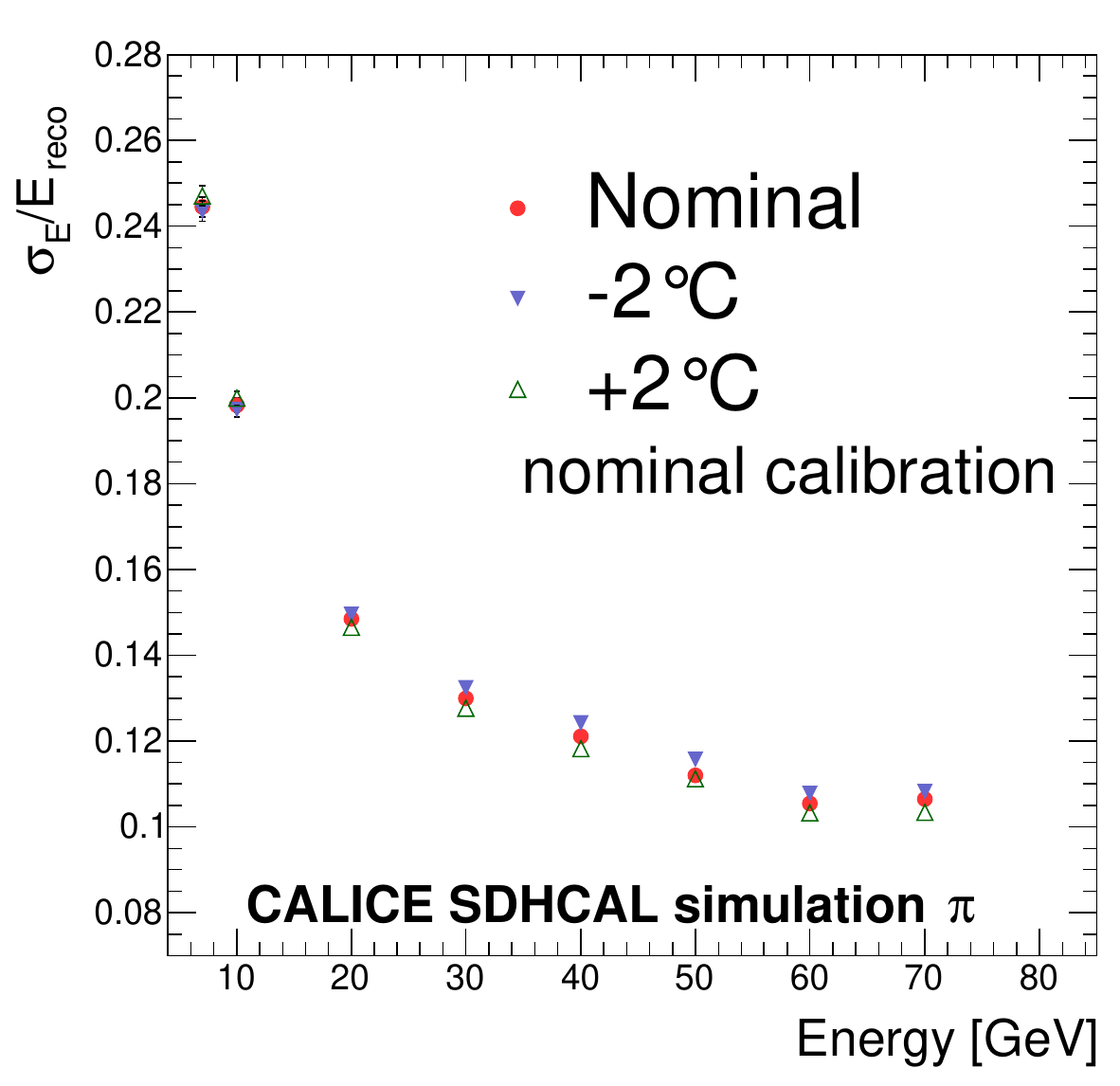}
    \caption{}
    \label{fig:resol.t.bias}
  \end{subfigure}
  \begin{subfigure}[b]{0.495\textwidth}
    \includegraphics[width=\textwidth]{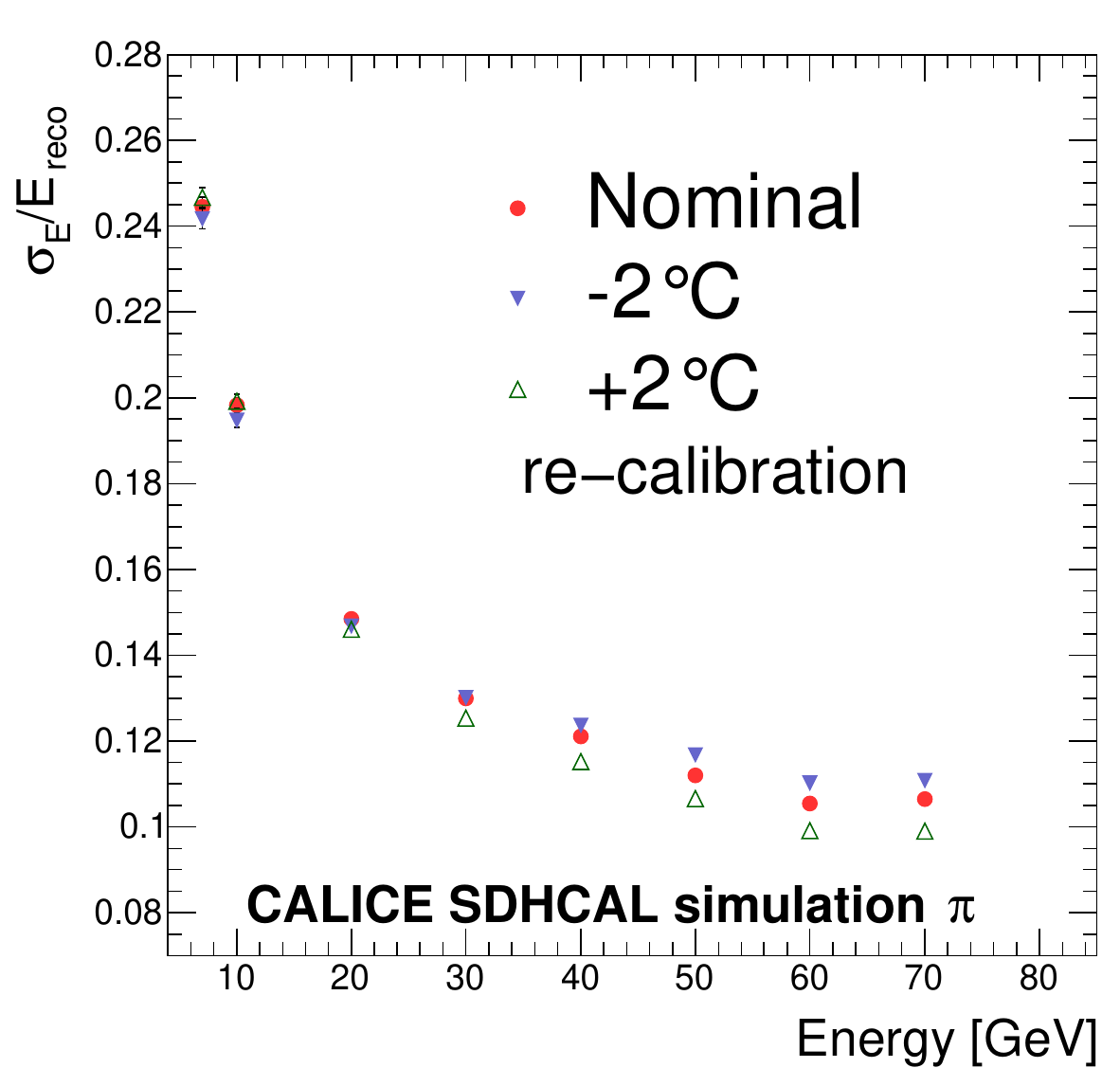}
    \caption{}
    \label{fig:resol.t.opt}
  \end{subfigure}
  \caption{Relative energy resolution for pions versus generated energy for
    simulations with GRPCs at the nominal temperature (red circles), a
    temperature variation by $-2\degreecelsius$ (blue filled triangles) and
    $+2\degreecelsius$ (green open triangles). The energy reconstruction
    factors are optimised using the nominal simulation for all
    cases~(\subref{fig:resol.t.bias}) or re-calibrated for each of the three 
    simulation options~(\subref{fig:resol.t.opt}).\label{fig:resol.t}} 
  \end{center}
\end{figure}

The energy calibration reconstruction factors, $a$, $b$ and $c$, are
compared for pions 
in Figure~\ref{fig:alpha.beta.gamma.pi.t} at three detector temperatures.
The energy is reconstructed, using the nominal factors, from simulated events
at different temperatures (see Figure~\ref{fig:ereco.t}). 
In the absence of re-calibration (Figure~\ref{fig:ereco.pion.t.bias}), the reconstructed energy
of a 40\,GeV pion varies by 9\% if the temperature is varied by 
$2\degreecelsius$. When the energy calibration reconstruction factors are adapted to each
simulation (Figure~\ref{fig:ereco.pion.t.optimal}), the linearity of the energy is restored.
The impact on the energy resolution was also checked. The relative energy
resolution varies by less than one $1\%$ if the temperature is varied by
$\pm2\degreecelsius$ as shown in Figure~\ref{fig:resol.t}

\begin{figure}[t]
 \begin{center}
 \begin{subfigure}[b]{0.495\textwidth}
    \includegraphics[width=\textwidth]{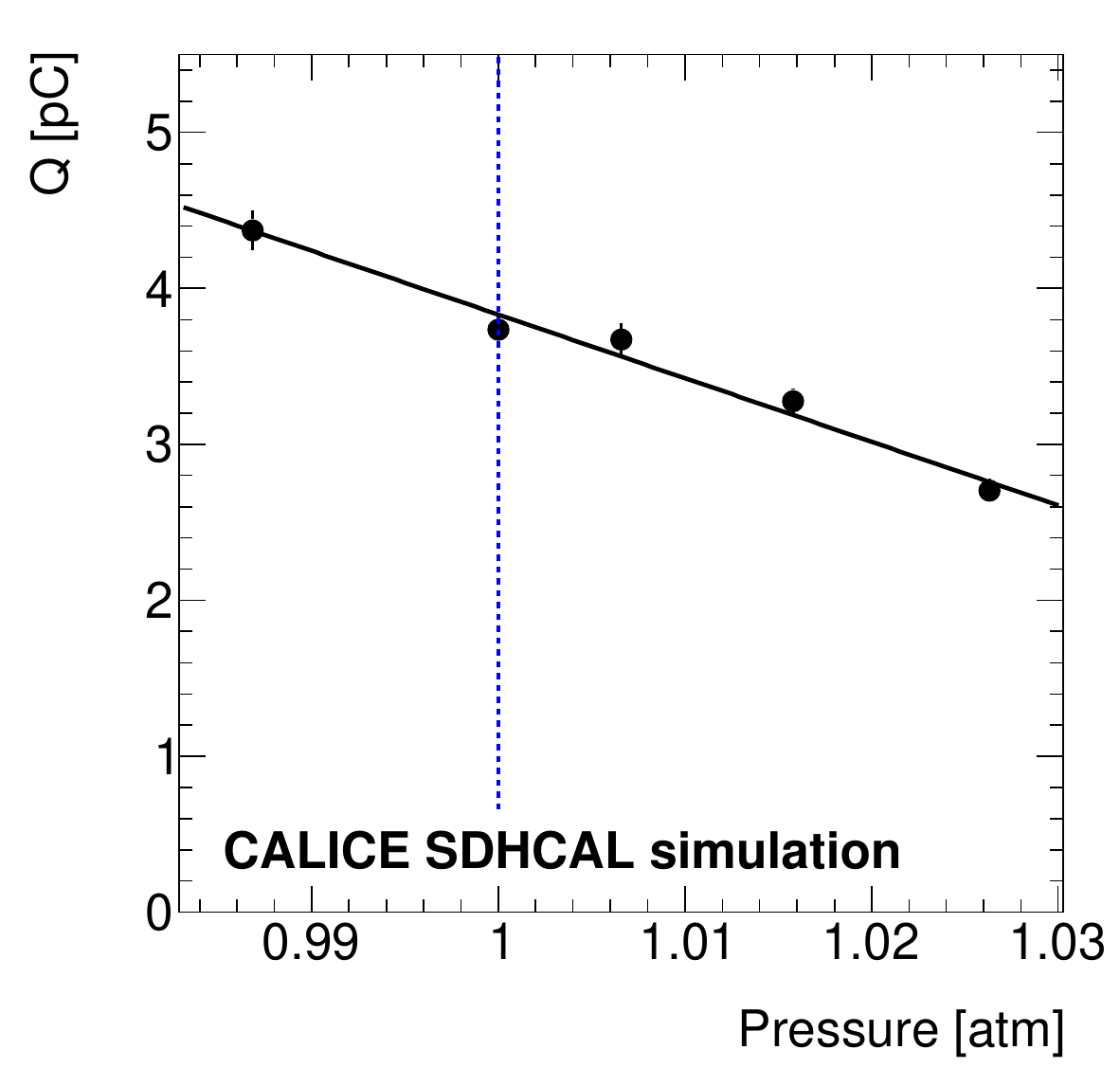}
    \caption{}
    \label{fig:q.vs.p}
  \end{subfigure}
  \begin{subfigure}[b]{0.495\textwidth}
    \includegraphics[width=\textwidth]{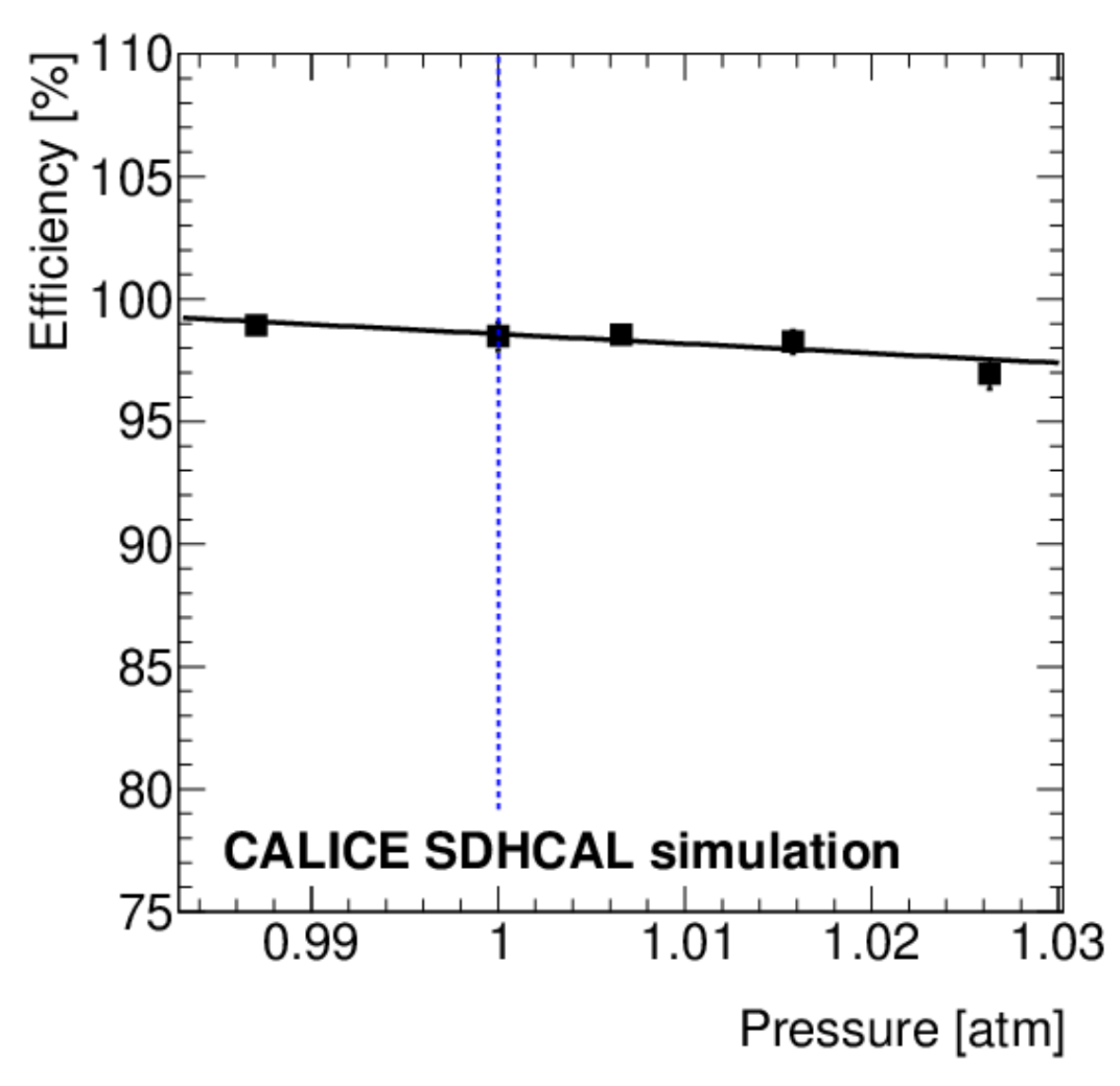}
    \caption{}
    \label{fig:eff.vs.p}
  \end{subfigure}
  \caption{Mean total induced charge (\subref{fig:q.vs.p}) and efficiency
    (\subref{fig:eff.vs.p}) as a function of the gas pressure for hits
    initiated by simulated 100\,GeV muons. The nominal pressure is
  indicated by a vertical dashed blue line.  The linear function that fits to
  the variations is represented by a black solid line.}
  \end{center}
\end{figure}

A similar study is performed by varying the atmospheric pressure in the
simulation. The amplitude of the signal and the efficiency are estimated for
different GRPC gas pressures and reported in 
Figures~\ref{fig:q.vs.p} and~\ref{fig:eff.vs.p}.  A typical pressure increase of $10$\,mbar is
accompanied by a total induced charge decrease of $11\%$ and an efficiency
loss below $1\%$. This trend is included in the digitizer algorithm.

\begin{figure}[h]
 \begin{center}
 \begin{subfigure}[b]{0.31\textwidth}
    \includegraphics[width=\textwidth]{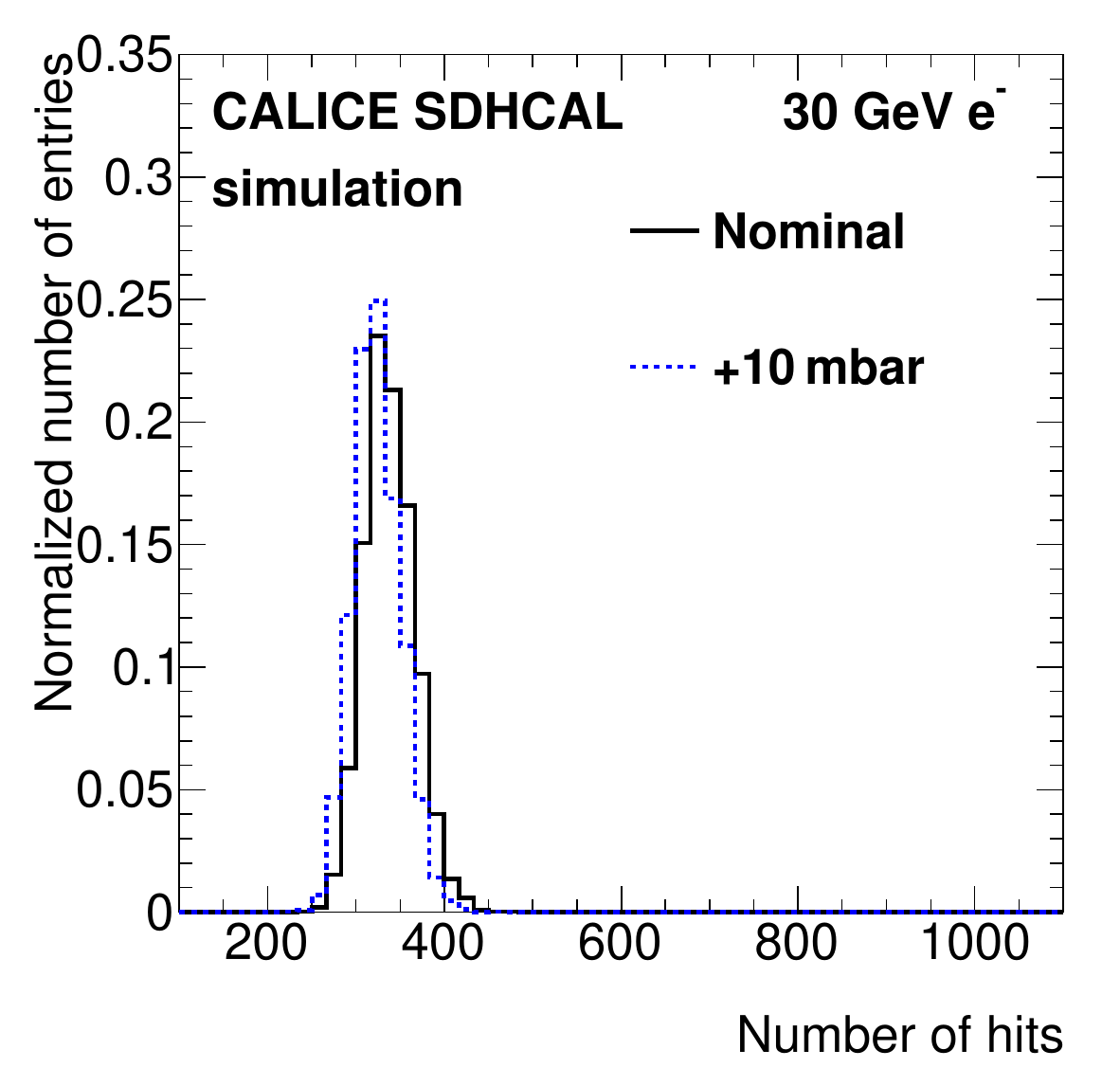}
    \caption{}
    \label{fig:nh.vs.p.ele.1}
  \end{subfigure}
  \begin{subfigure}[b]{0.31\textwidth}
    \includegraphics[width=\textwidth]{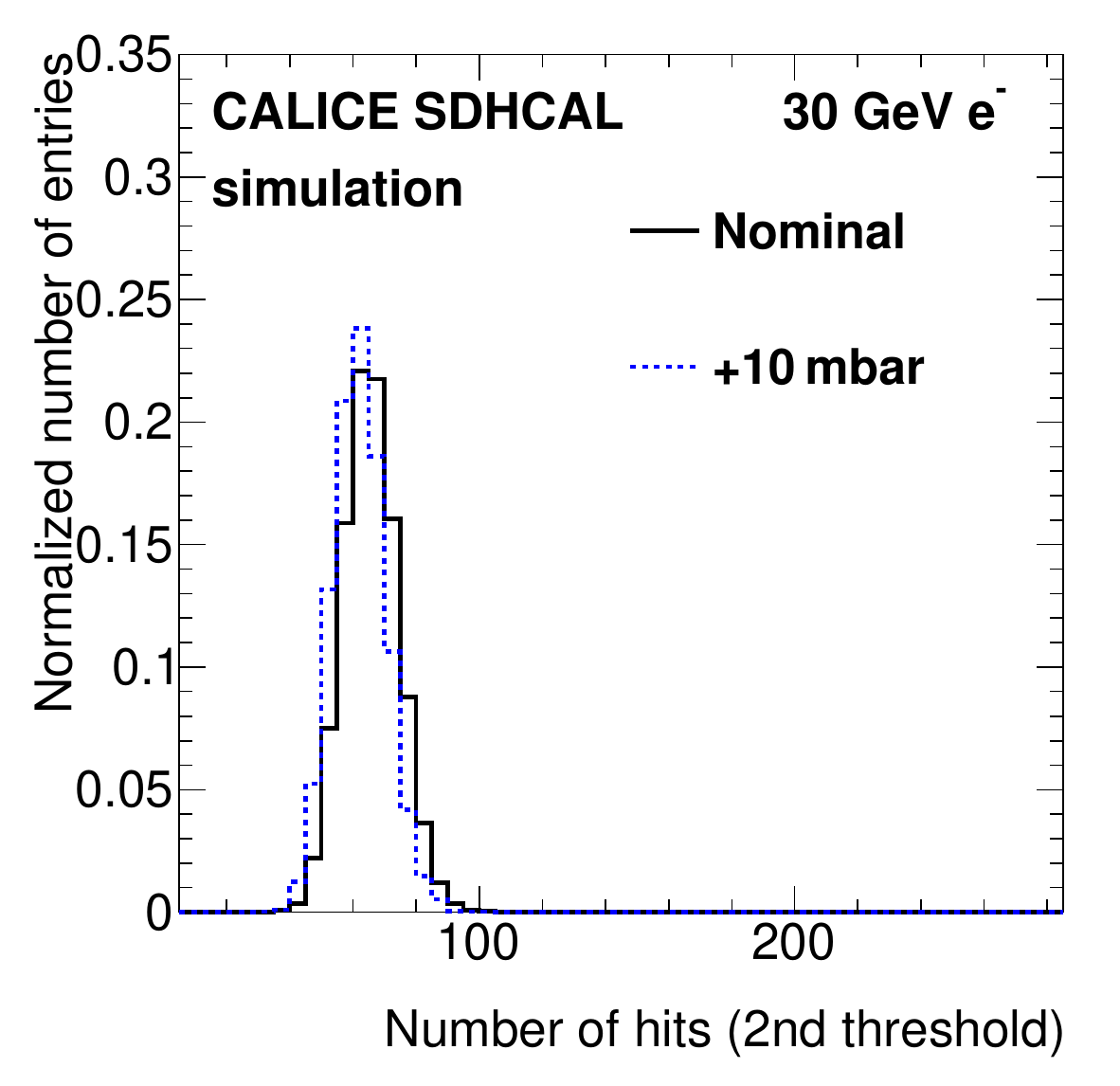}
    \caption{}
    \label{fig:nh.vs.p.ele.2}
  \end{subfigure}
  \begin{subfigure}[b]{0.31\textwidth}
    \includegraphics[width=\textwidth]{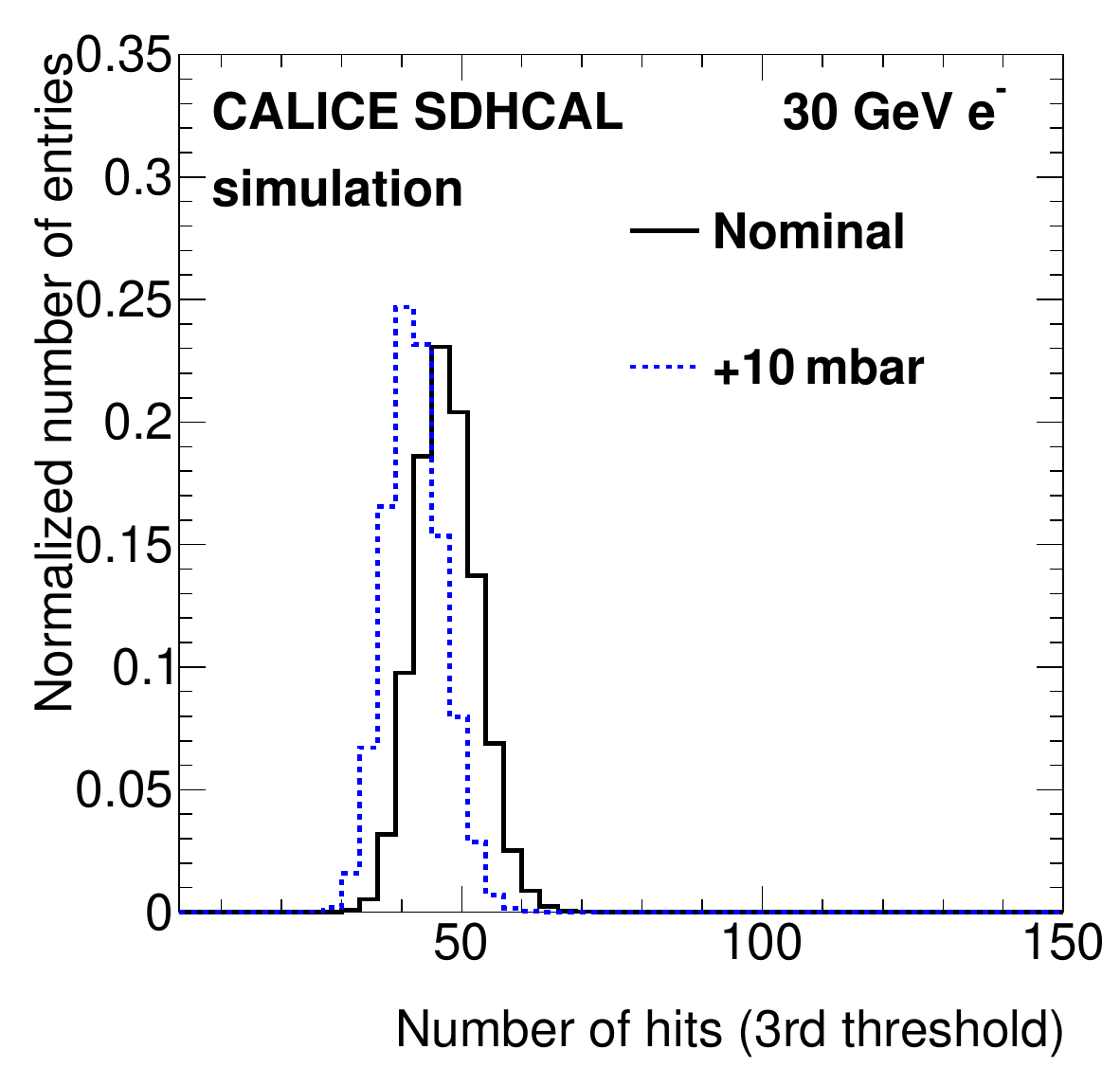}
    \caption{}
    \label{fig:nh.vs.p.ele.3}
  \end{subfigure}
 \begin{subfigure}[b]{0.31\textwidth}
    \includegraphics[width=\textwidth]{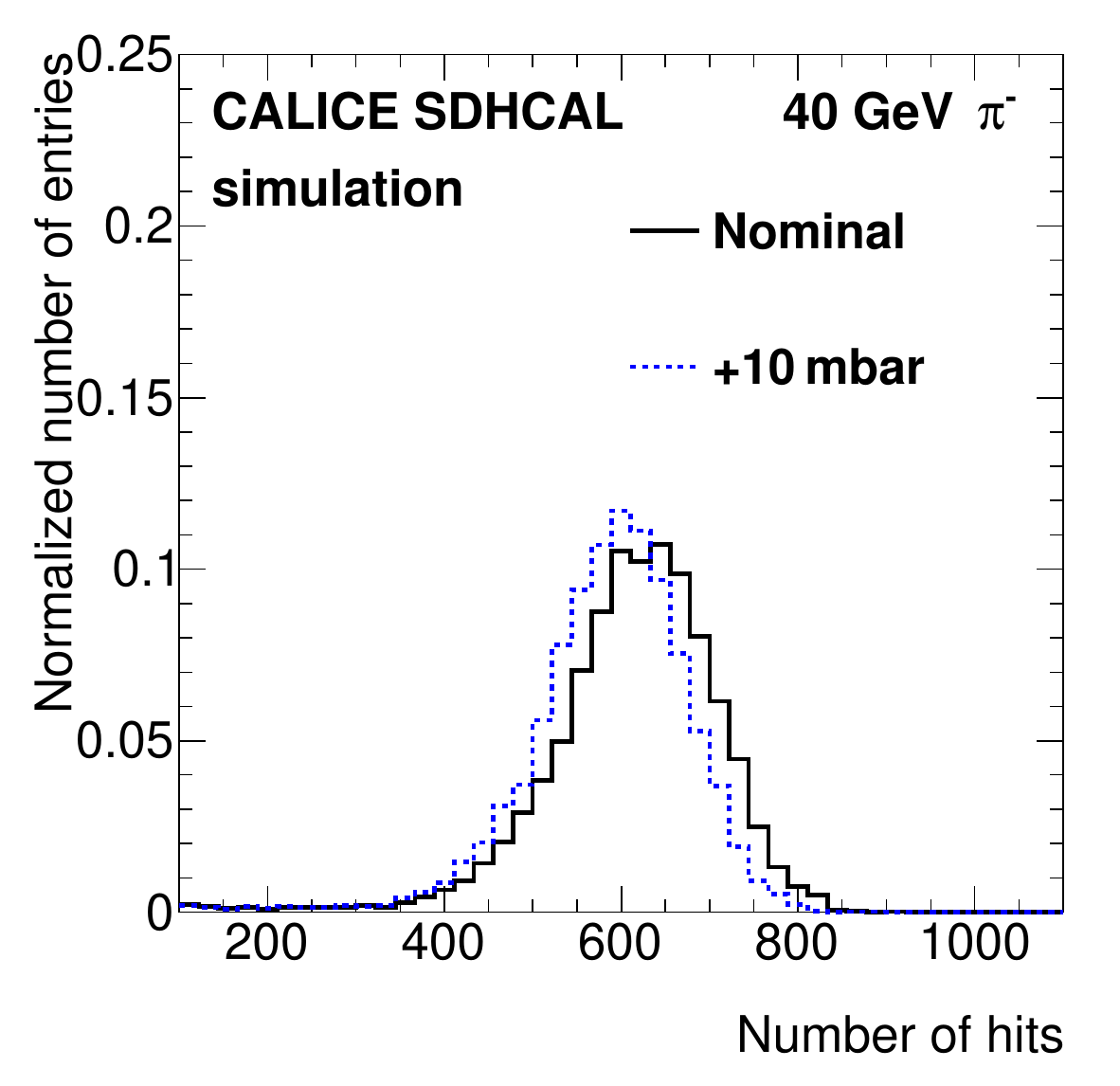}
    \caption{}
    \label{fig:nh.vs.p.1}
  \end{subfigure}
  \begin{subfigure}[b]{0.31\textwidth}
    \includegraphics[width=\textwidth]{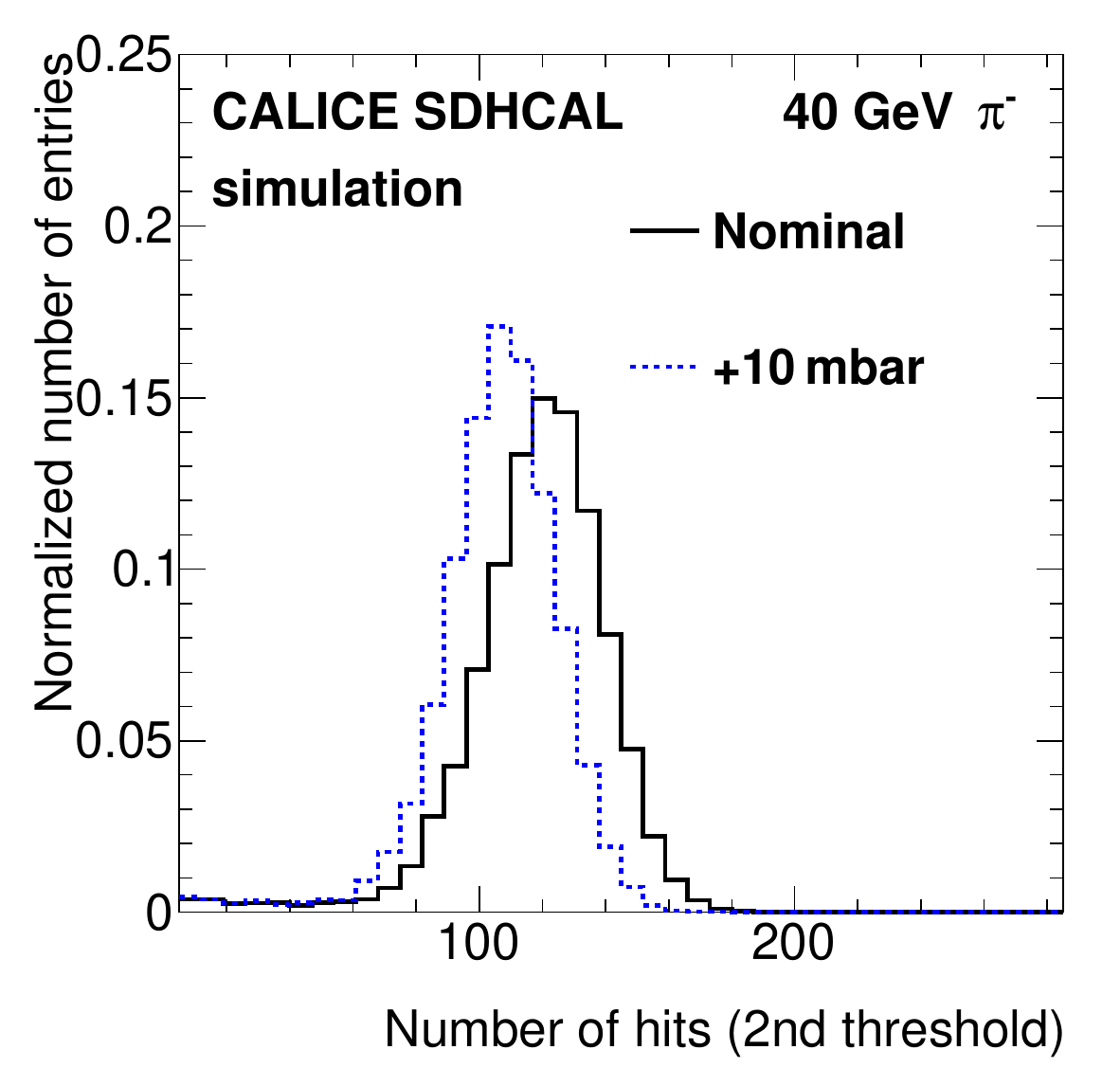}
    \caption{}
    \label{fig:nh.vs.p.2}
  \end{subfigure}
  \begin{subfigure}[b]{0.31\textwidth}
    \includegraphics[width=\textwidth]{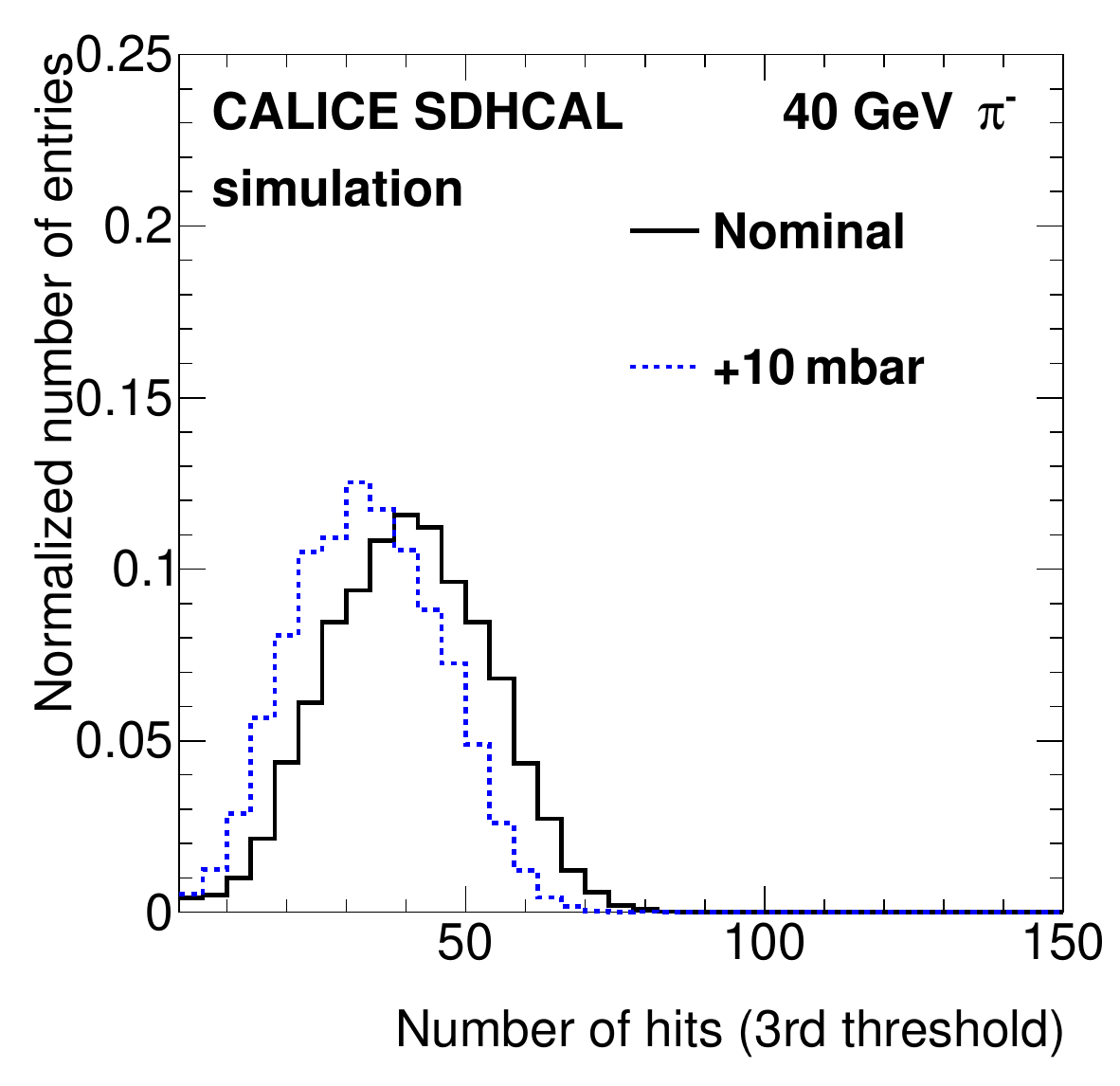}
    \caption{}
    \label{fig:nh.vs.p.3}
  \end{subfigure}
  \caption{Distribution of the number of hits passing the first
    threshold at $0.1$\,pC~(\subref{fig:nh.vs.p.ele.1},\subref{fig:nh.vs.p.1}), the second
    threshold at $5$\,pC~(\subref{fig:nh.vs.p.ele.2},\subref{fig:nh.vs.p.2}) and the third
    threshold at $15$\,pC~(\subref{fig:nh.vs.p.ele.3},\subref{fig:nh.vs.p.3}) for simulated  
    30\,GeV electrons (top) or 40\,GeV
    pions (bottom). The full GEANT4 simulation was performed with digitization modeling using the nominal pressure (solid black histograms) or
    a modeling where the pressure is increased by $+10$\,mbar
    (dashed blue histogram).\label{fig:nh.vs.p}} 
  \end{center}
\end{figure}

\begin{figure}[h]
 \begin{center}
 \begin{subfigure}[b]{0.49\textwidth}
    \includegraphics[width=\textwidth]{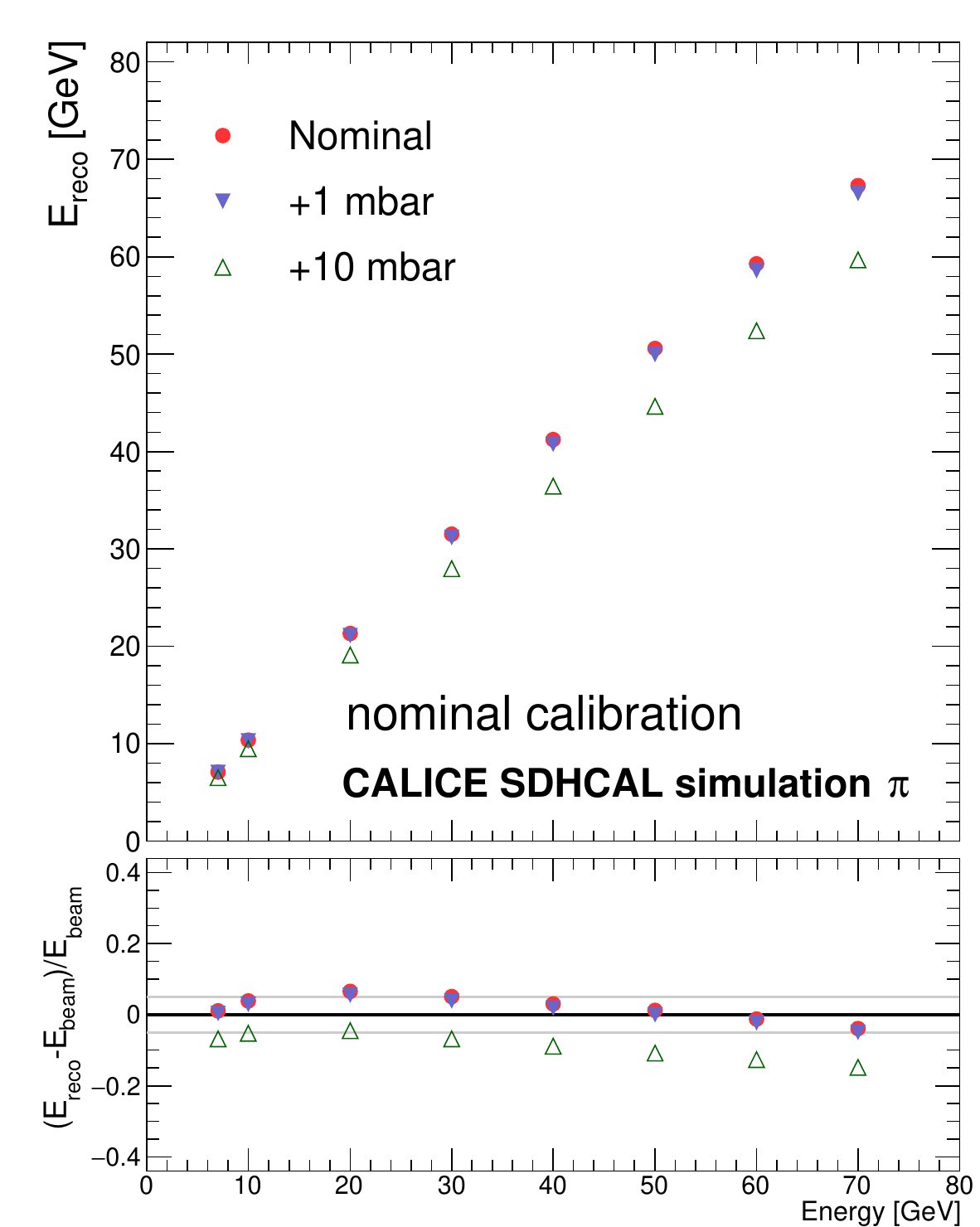}
    \caption{}
    \label{fig:ereco.pion.p.bias}
  \end{subfigure}
 \begin{subfigure}[b]{0.49\textwidth}
    \includegraphics[width=\textwidth]{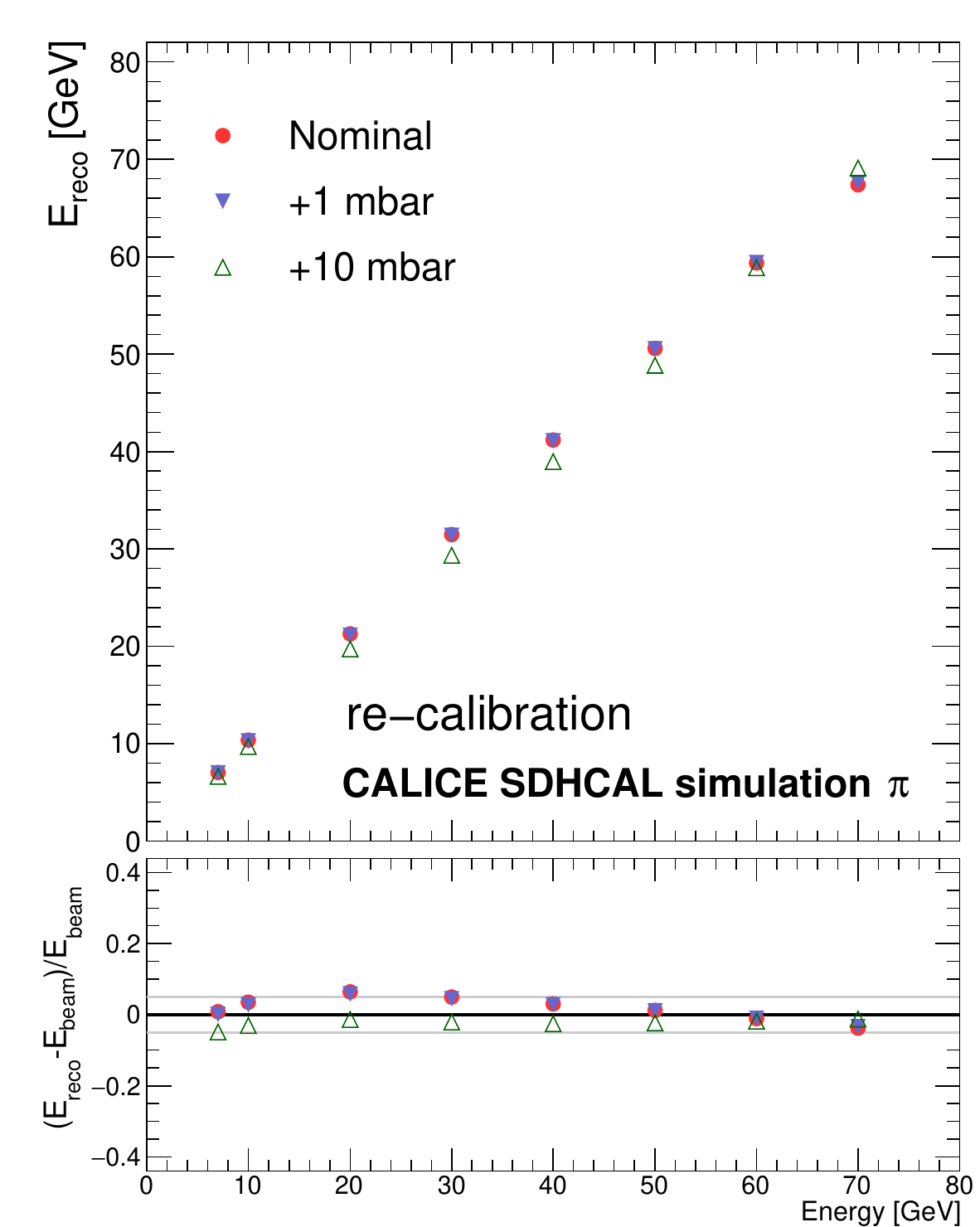}
    \caption{}
    \label{fig:ereco.pion.p.optimal}
  \end{subfigure}
  \caption{Pion reconstructed energy as a function of the generated energy for
    simulations with the GRPCs at the nominal pressure (red circles) and a
    pressure variation by $+10$\,mbar (blue filled triangles). The energy
    reconstruction factors are calibrated using the nominal simulation for all
    cases~(\subref{fig:ereco.pion.p.bias}) or re-calibrated for each of the
    three simulations~(\subref{fig:ereco.pion.p.optimal}).\label{fig:ereco.lin.p}}
\end{center}
\end{figure}

 The dependence of the
number of hits to a pressure variation is illustrated in
Figure~\ref{fig:nh.vs.p}. A pressure increase of $10$\,mbar leads to a variation
of the average number of hits of $-4.5\%$. The energy is reconstructed
using the nominal factors and applied to simulated samples assuming different
gas pressures. In this case the reconstructed energy of $40$\,GeV pions is
biased by 12\%. When the energy calibration reconstruction factors are optimised on each
simulation, the energy linearity is improved as shown in
Figure~\ref{fig:ereco.lin.p}. 

It was previously shown that adjusting the applied high voltage, $V$, by a value $\Delta{V}$ such as 
\begin{equation}
\Delta{V}/V=\Delta{T}/T
\label{eq:hvscale}
\end{equation}
where $\Delta{T}/T$ is the relative temperature variation,
allows maintaining an approximately stable
efficiency~\cite{rpcvst}. The simulation is used to test
  the efficiency of this method.

A dedicated simulation was performed where both temperature and the applied electric field
amplitude are changed. It was assumed that the high voltage is scaled
following eq.~\eqref{eq:hvscale}. As shown on Figure~\ref{fig:qeff.vs.t}, the
stability of the GRPC response is then improved by the high voltage tuning. The trends in
the total induced charge and efficiency are reduced. However, a residual dependence to
the temperature is observed in the simulation. A temperature increase of
$5\degreecelsius$ leads to a decrease of the total induced charge by $5.3\%$
in a detector where the high voltage is scaled, whereas it
would increase by $22\%$ in a detector where the high voltage is constant.

The effect of the temperature on the GRPC signal is linear while a non-linear
dependence to the high voltage is expected as reported in ref.~\cite{cms}. An online
automatized high voltage adjustment would require a tuning specific to this detector. 

%% file: latex/section-bfield.tex
\subsection{Impact of magnetic field}

The SDHCAL technology proposal intends to equip a full-size hadronic
calorimeter in a large scale detector like ILD. In this context, it has to
be operated under a strong magnetic field of a few Tesla. 
The magnetic field affects the electron attachment and multiplication coefficients in the GRPC as well as
their velocity and diffusion length. The Townsend coefficients, the diffusion length and the
electron velocity are stable within 1 to 3\% when the magnetic field is
increased from 0 to 4~Tesla.
Two magnetic field configurations are considered: perpendicular and
longitudinal to the GRPCs as expected in the forward and barrel regions.

\begin{figure}[t]
 \begin{center}
 \begin{subfigure}[b]{0.495\textwidth}
    \includegraphics[width=\textwidth]{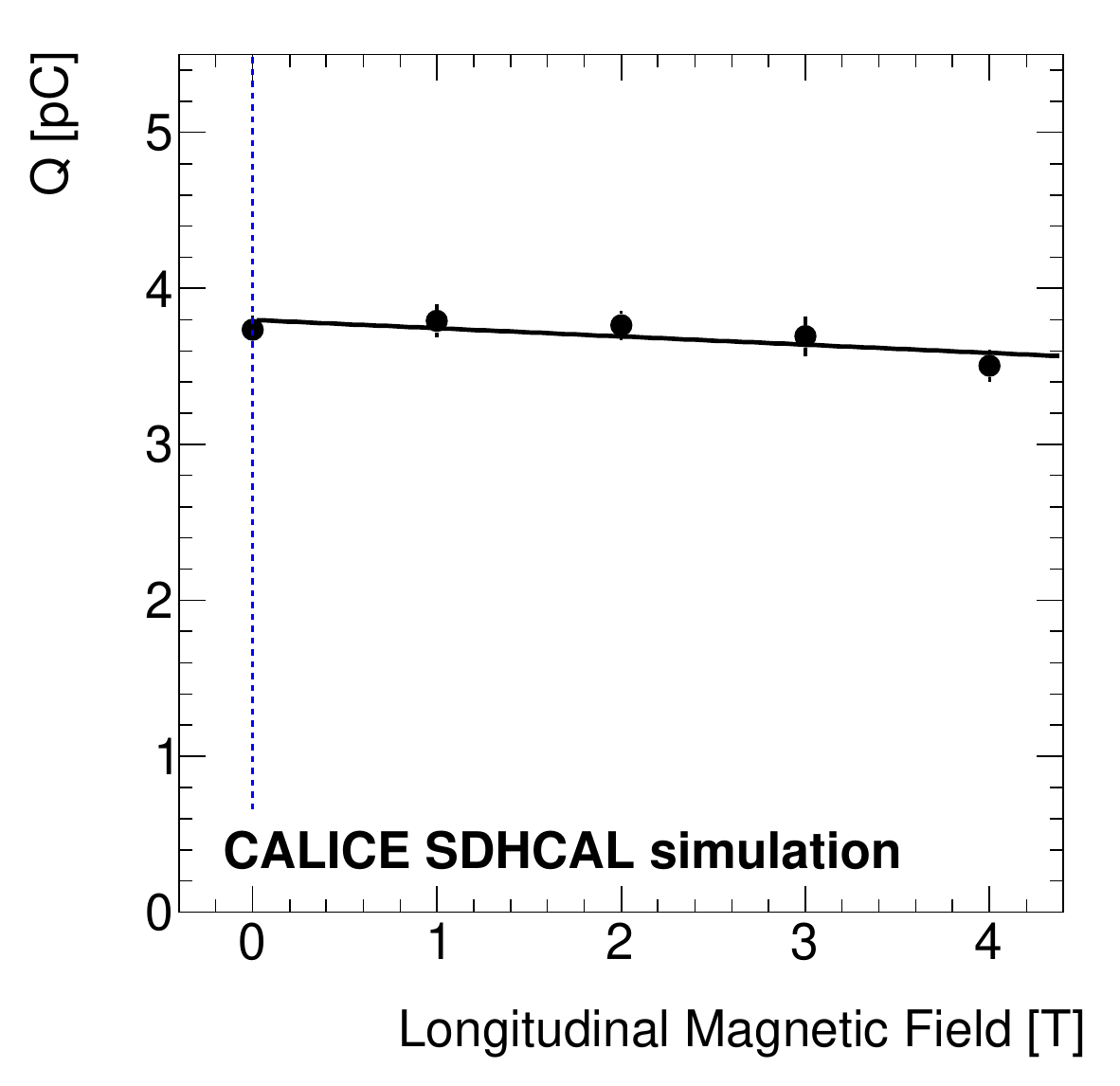}
    \caption{}
    \label{fig:q.vs.tb}
  \end{subfigure}
  \begin{subfigure}[b]{0.495\textwidth}
    \includegraphics[width=\textwidth]{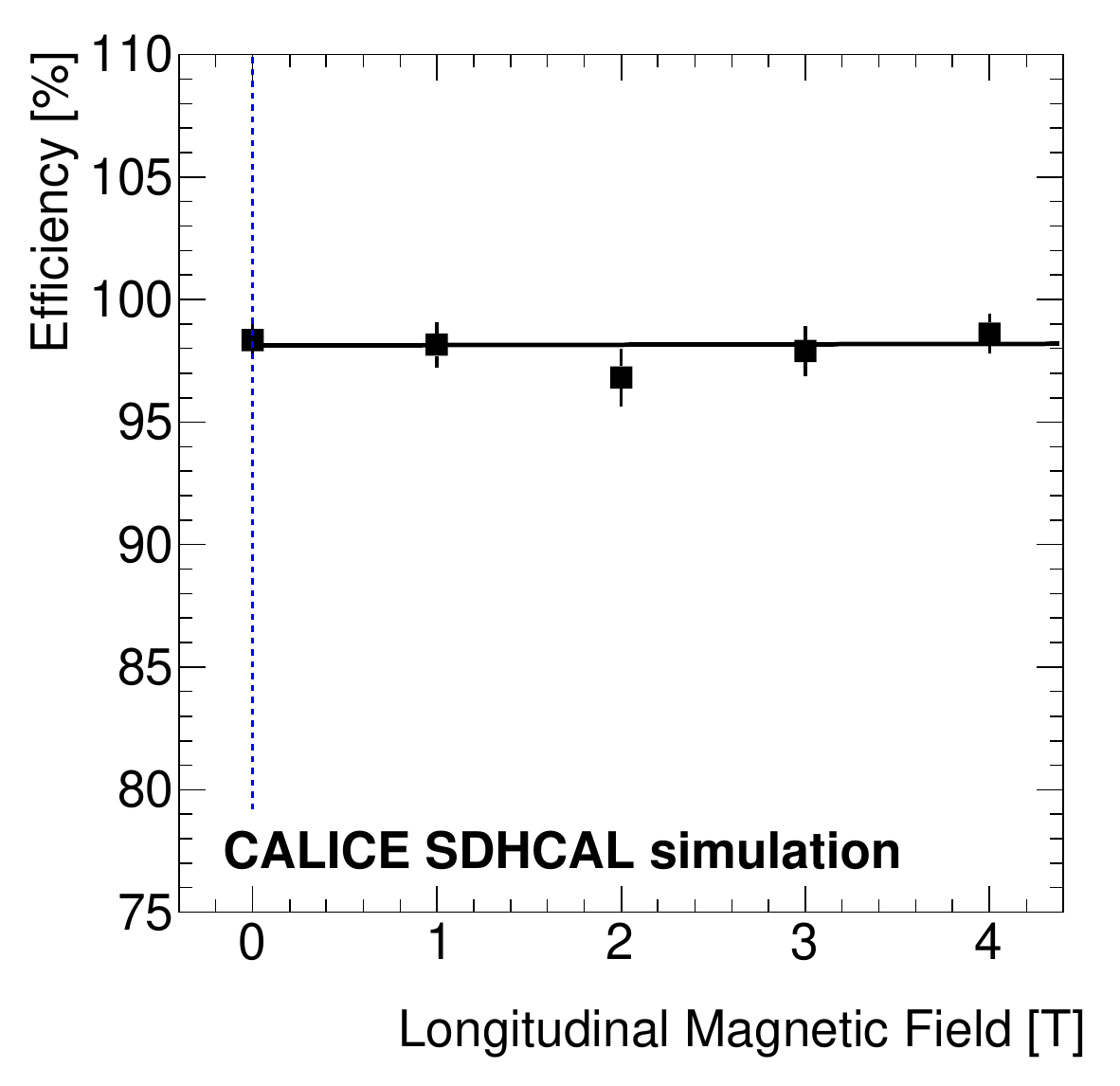}
    \caption{}
    \label{fig:eff.vs.tb}
  \end{subfigure}
  \caption{ Mean total induced charge (\subref{fig:q.vs.tb}) and efficiency
    (\subref{fig:eff.vs.tb}) as a function of the magnetic field
    amplitude, longitunidal to the GRPC, for hits 
    initiated by simulated 100\,GeV muons. The nominal magnetic field amplitude in the simulation is
  indicated by a vertical dashed blue line. The linear function that fits
  the variation is shown with a solid line.}
  \end{center}
\end{figure}

Events induced by 100\,GeV muons are
produced under different magnetic field configurations. The total induced charge
and efficiency are given as a function of the amplitude of a longitudinal field on
Figures~\ref{fig:q.vs.tb} and~\ref{fig:eff.vs.tb}. No significant effect on
the GRPC performances is observed at the hit level. This agrees with the test reported in ref.~\cite{powerpulsing}.
Thus no dependence on the magnetic field is implemented in the digitizer algorithm.

%% file: latex/section-gas.tex
\subsection{Impact of gas mixture}

A composition of the gas mixture combined to the high voltage is at the core of
the amplification process. Variations in the gas mixture lead to changes on
the gain. 
The fractions of \ch{SF6} and \ch{CO2} components were varied and the expected
signal and efficiencies are estimated for different fractions. In the
following, a variation of $5\%$ in \ch{SF6} was considered for the
 stability study. This variation is considered as pessimistic compared to the
 precision of typical flowmeters used to control the gaz mixture which is
 $0.6\%$~\cite{flowmeter}. 

\begin{figure}[t!]
 \begin{center}
  \begin{subfigure}[b]{0.327\textwidth}
    \includegraphics[width=\textwidth]{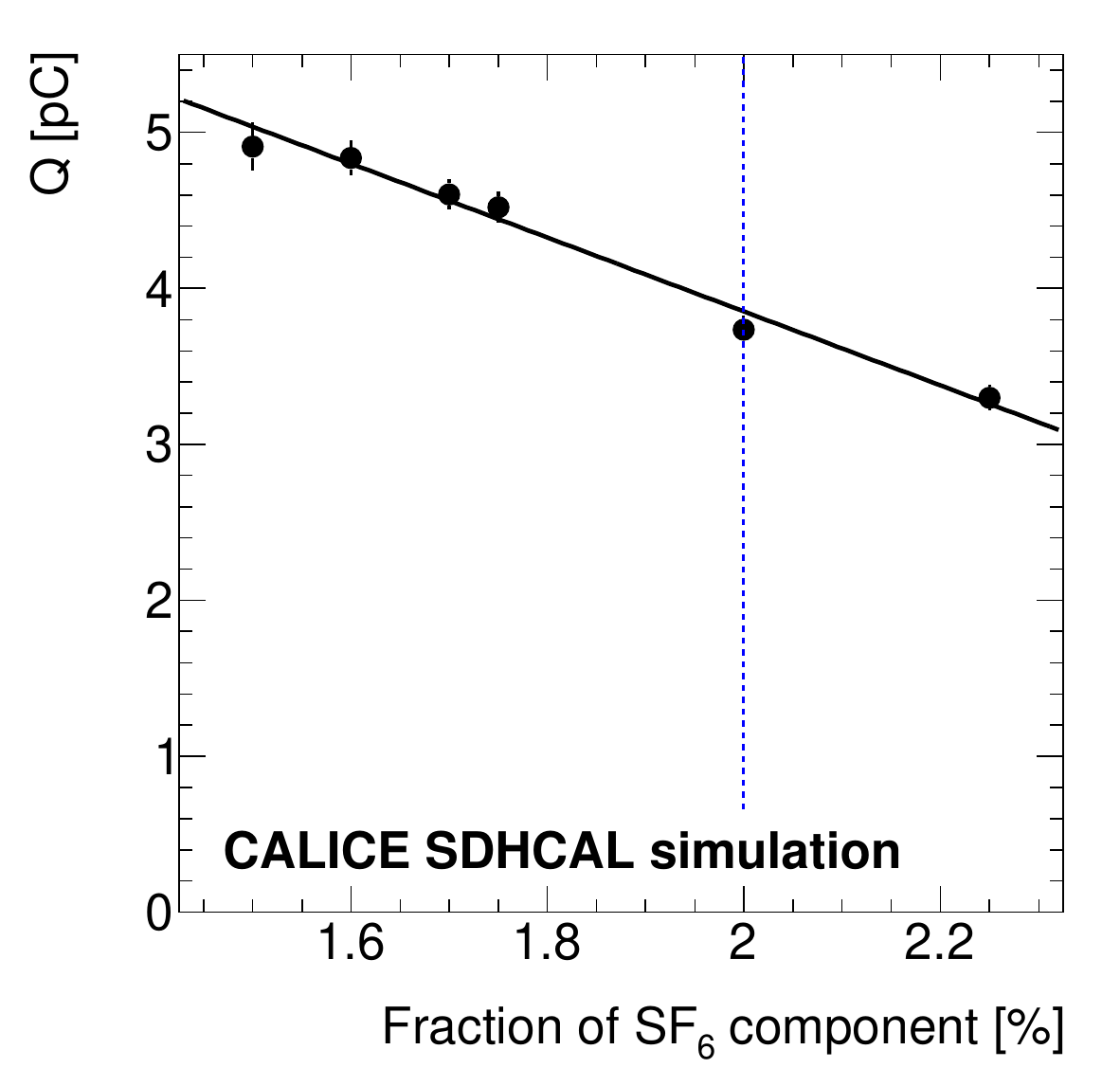}
    \caption{}
    \label{fig:q.vs.sf6}
  \end{subfigure}
   \begin{subfigure}[b]{0.327\textwidth}
    \includegraphics[width=\textwidth]{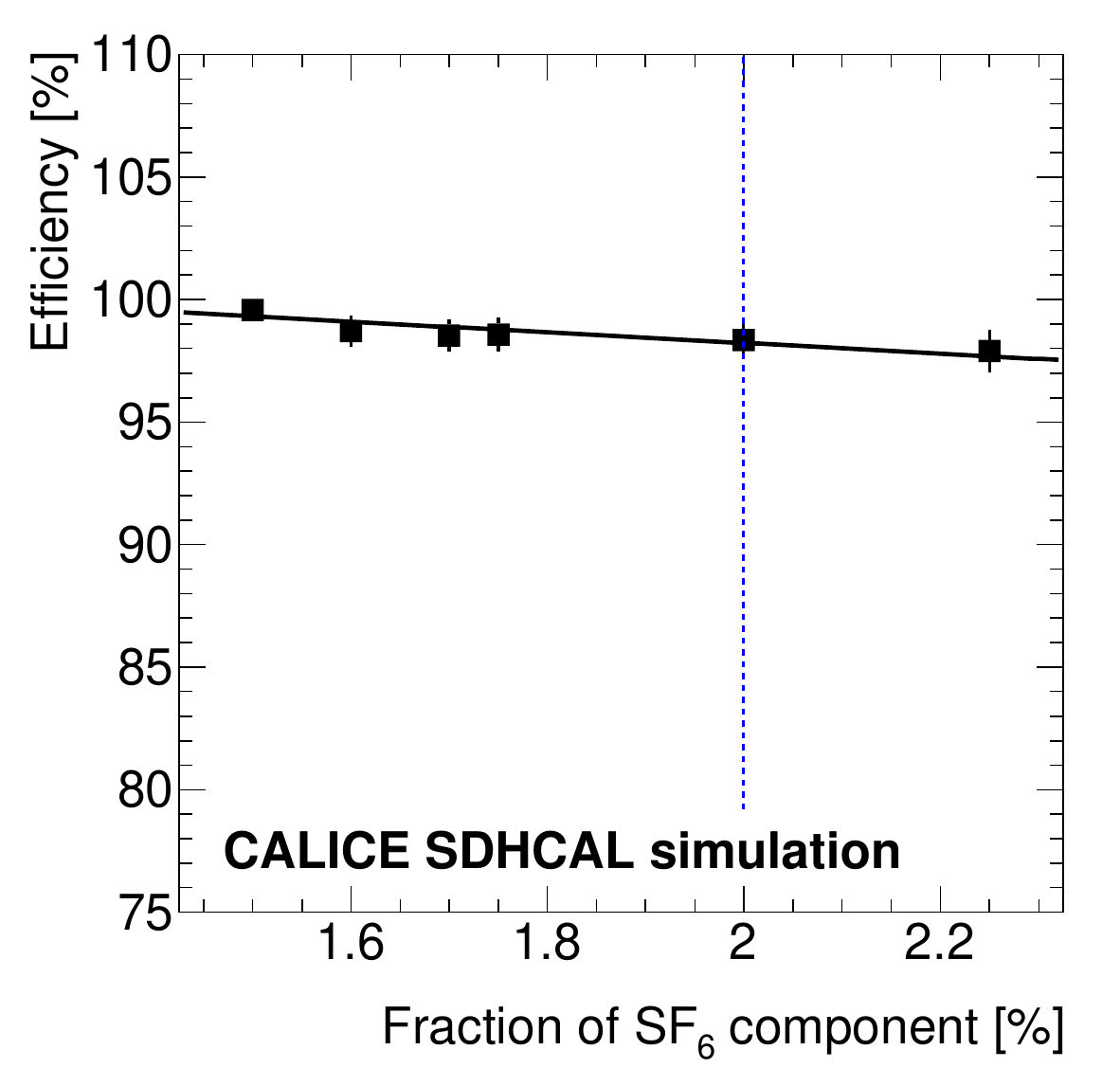}
    \caption{}
    \label{fig:eff.vs.sf6}
  \end{subfigure}
  \begin{subfigure}[b]{0.327\textwidth}
    \includegraphics[width=\textwidth]{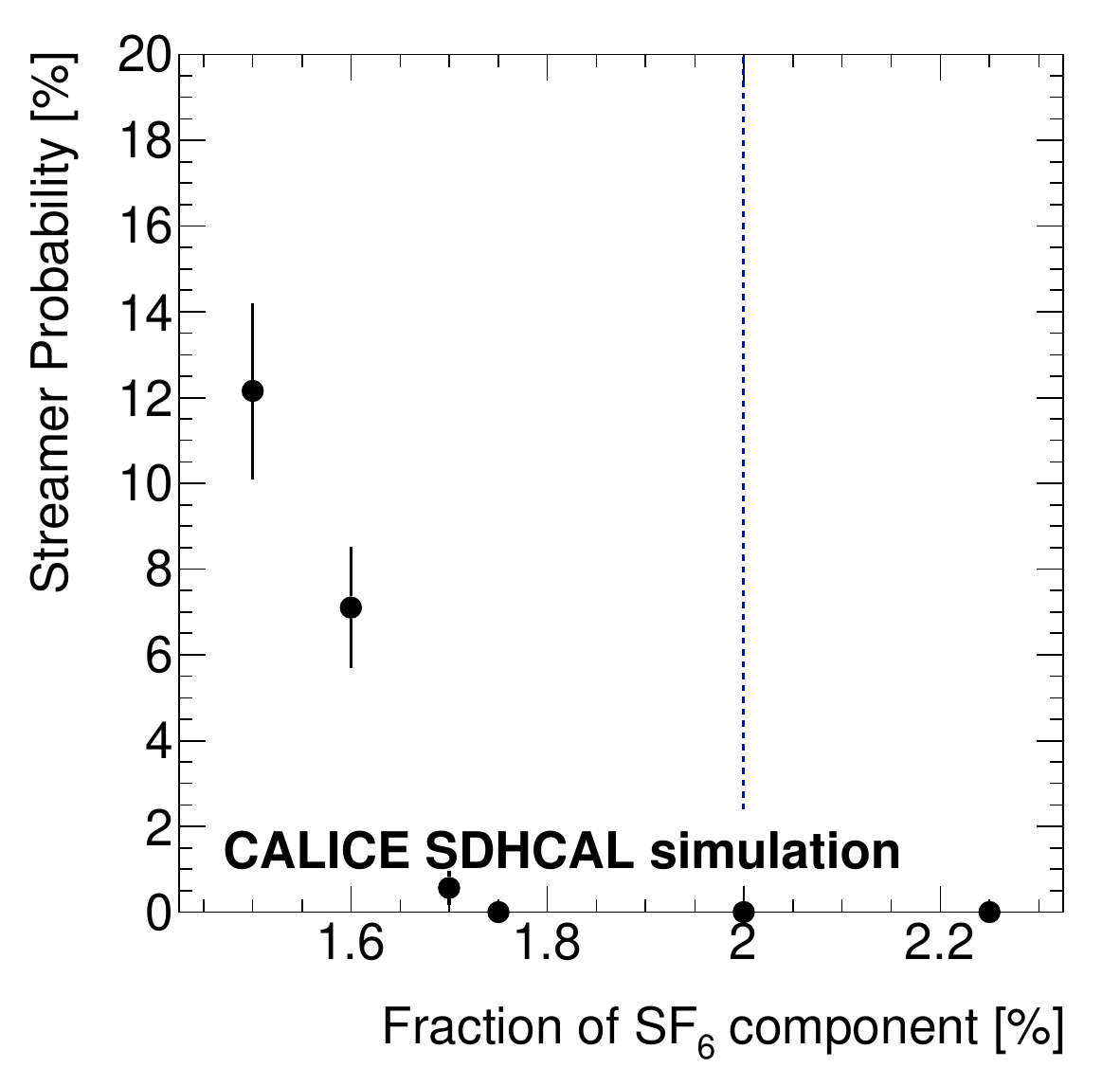}
    \caption{}
    \label{fig:str.vs.sf6}
  \end{subfigure}
  \caption{    Mean total induced charge (\subref{fig:q.vs.sf6}), efficiency
    (\subref{fig:eff.vs.sf6}) and streamer probability (\subref{fig:str.vs.sf6}) as a function of the fraction of \ch{SF6} for hits
    initiated by simulated 100\,GeV muons. The nominal \ch{SF6} fraction is
  indicated by a vertical dashed blue line. The linear function that fits to
  the variation is represented by a black solid line.\label{fig:sf6}}
  \end{center}
\end{figure}

It is found that the total induced charge drops by 6\% if the
amount of \ch{SF6} is varied by $5\%$ (which corresponds to the variation in
the total \ch{SF6} fraction by $0.1\%$),
 while the efficiency is also reduced (cf. Figure~\ref{fig:sf6}).
A reduction of \ch{SF6} is accompanied by an increase of the high charge
probability that is interpreted as an increased risk of streamers as seen in
Figure~\ref{fig:str.vs.sf6}.
This effect was implemented in the digitizer algorithm and used in full GEANT4
simulations of electron and pion showers.

\begin{figure}[t!]
 \begin{center}
 \begin{subfigure}[b]{0.31\textwidth}
    \includegraphics[width=\textwidth]{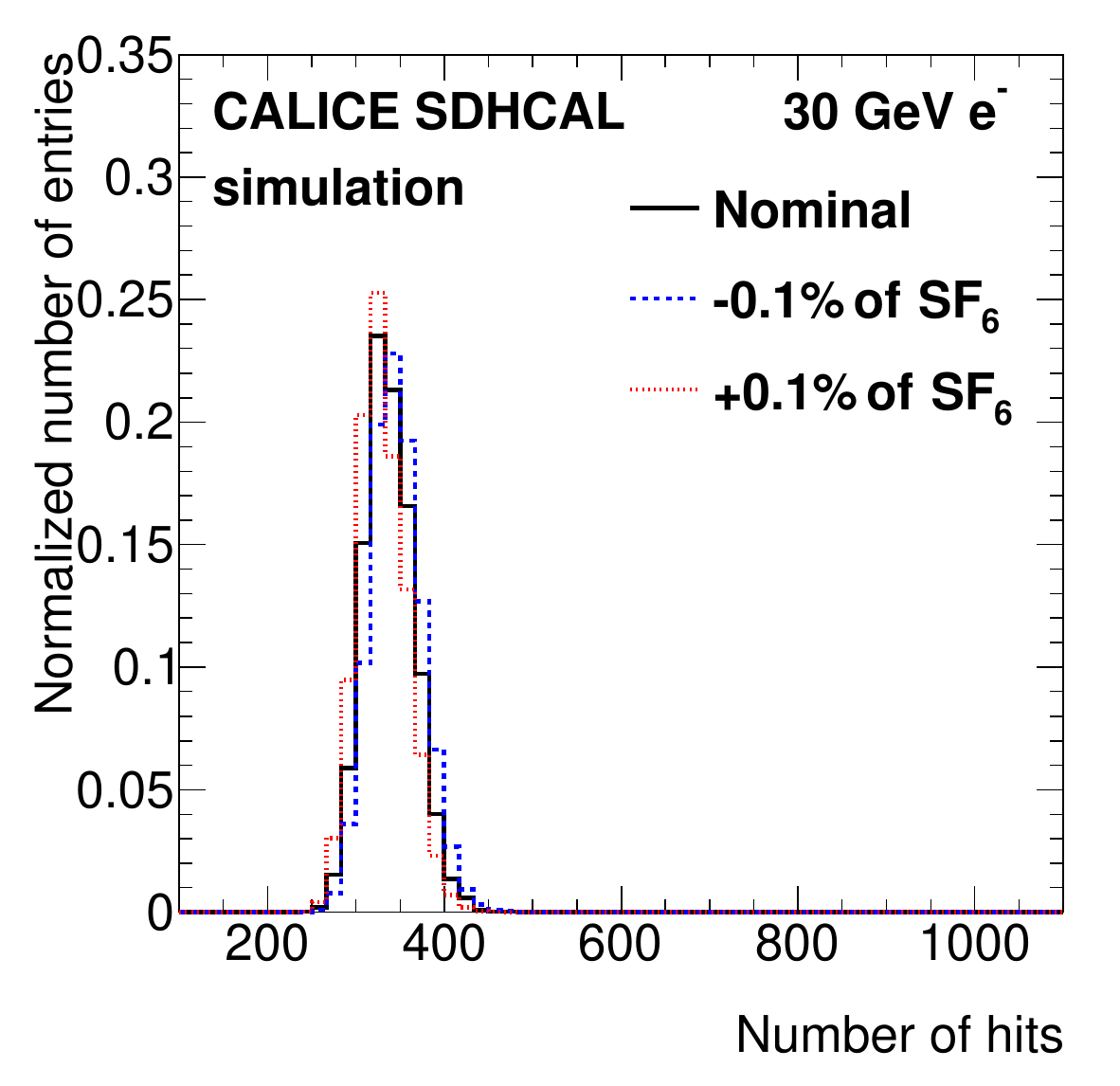}
    \caption{}
    \label{fig:nh.vs.sf.1}
  \end{subfigure}
  \begin{subfigure}[b]{0.31\textwidth}
    \includegraphics[width=\textwidth]{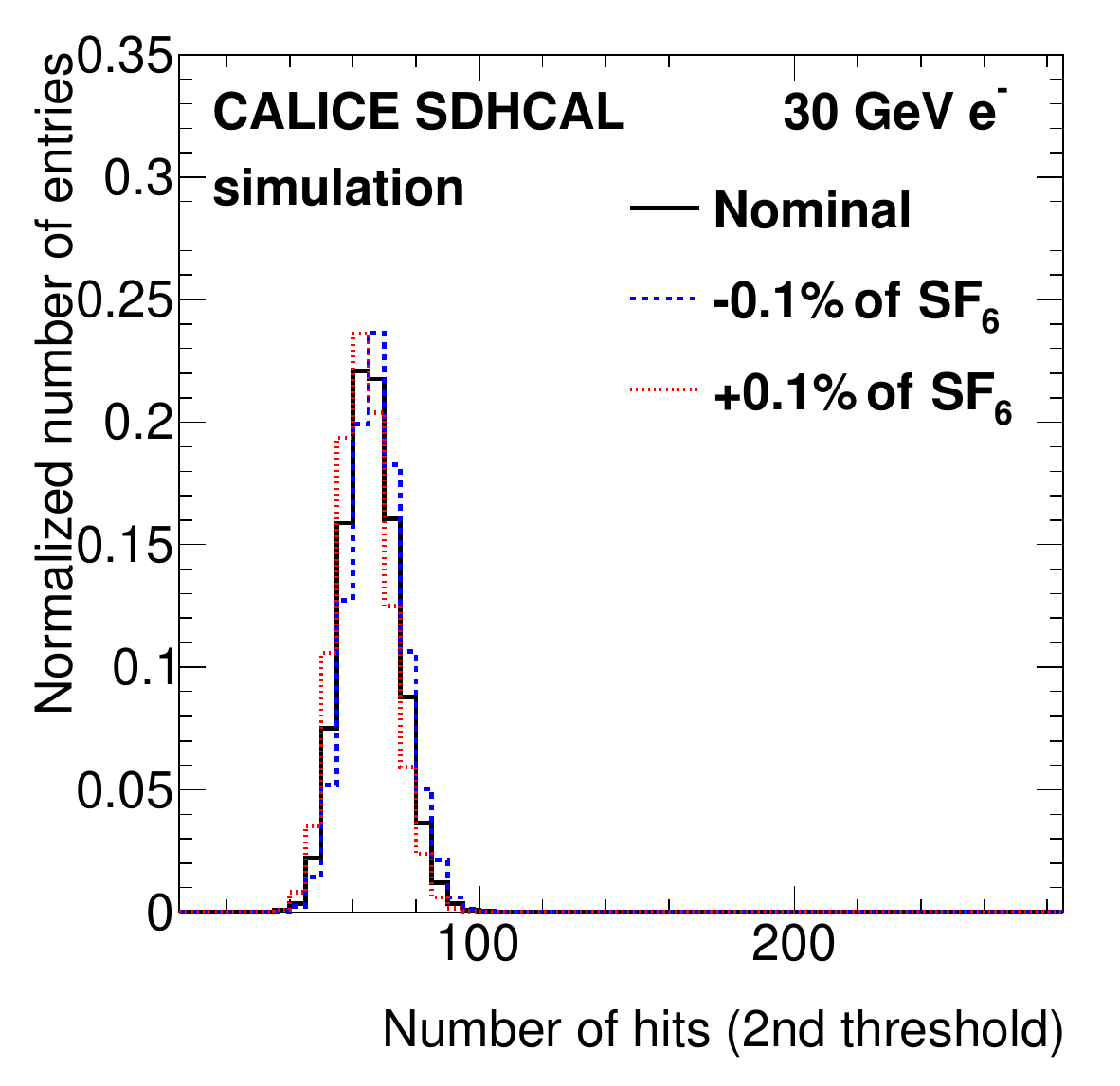}
    \caption{}
    \label{fig:nh.vs.sf.2}
  \end{subfigure}
  \begin{subfigure}[b]{0.31\textwidth}
    \includegraphics[width=\textwidth]{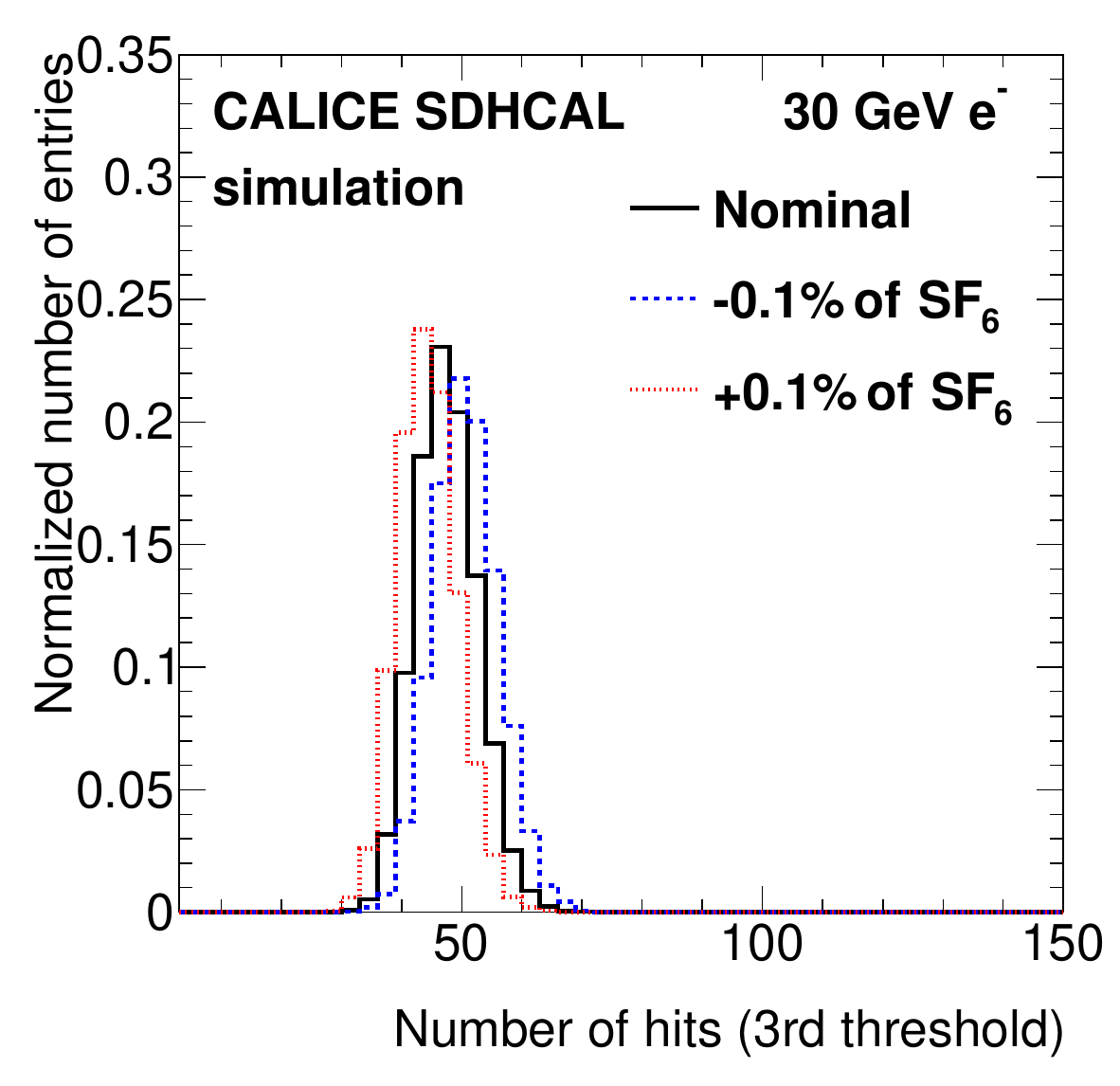}
    \caption{}
    \label{fig:nh.vs.sf.3}
  \end{subfigure}
  \begin{subfigure}[b]{0.31\textwidth}
    \includegraphics[width=\textwidth]{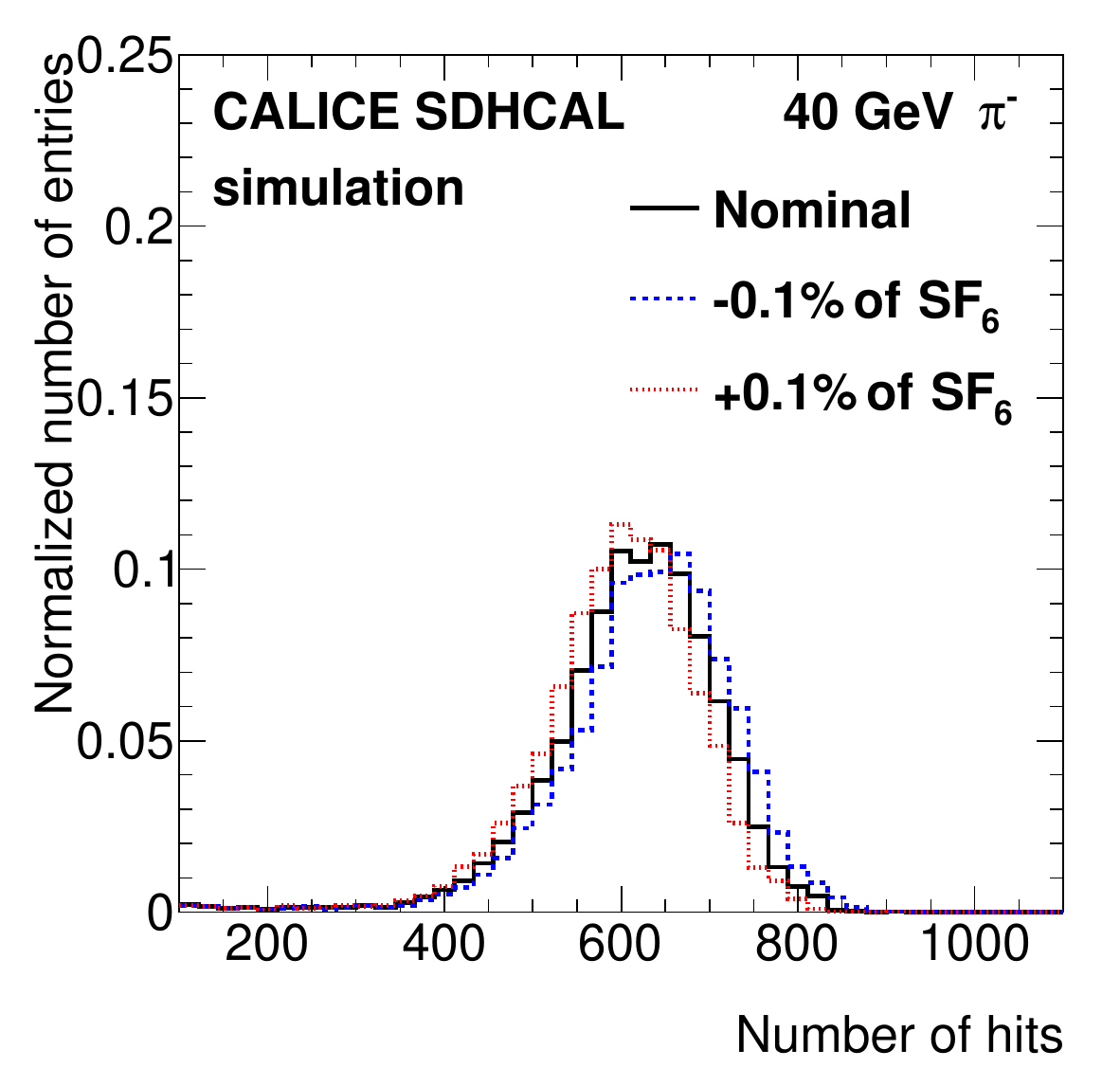}
    \caption{}
    \label{fig:nh.vs.sf.pi1}
  \end{subfigure}
  \begin{subfigure}[b]{0.31\textwidth}
    \includegraphics[width=\textwidth]{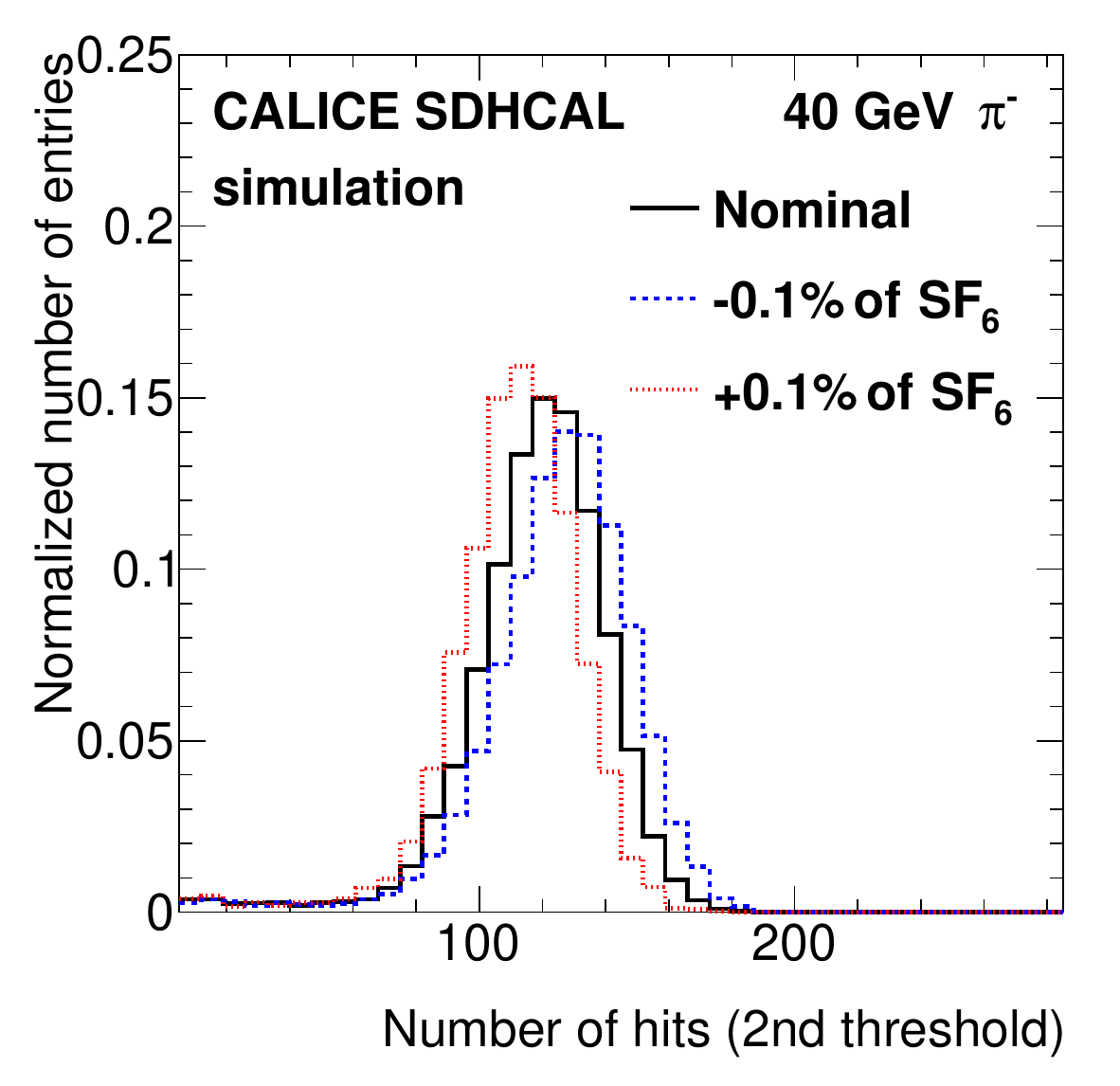}
    \caption{}
    \label{fig:nh.vs.sf.pi2}
  \end{subfigure}
  \begin{subfigure}[b]{0.31\textwidth}
    \includegraphics[width=\textwidth]{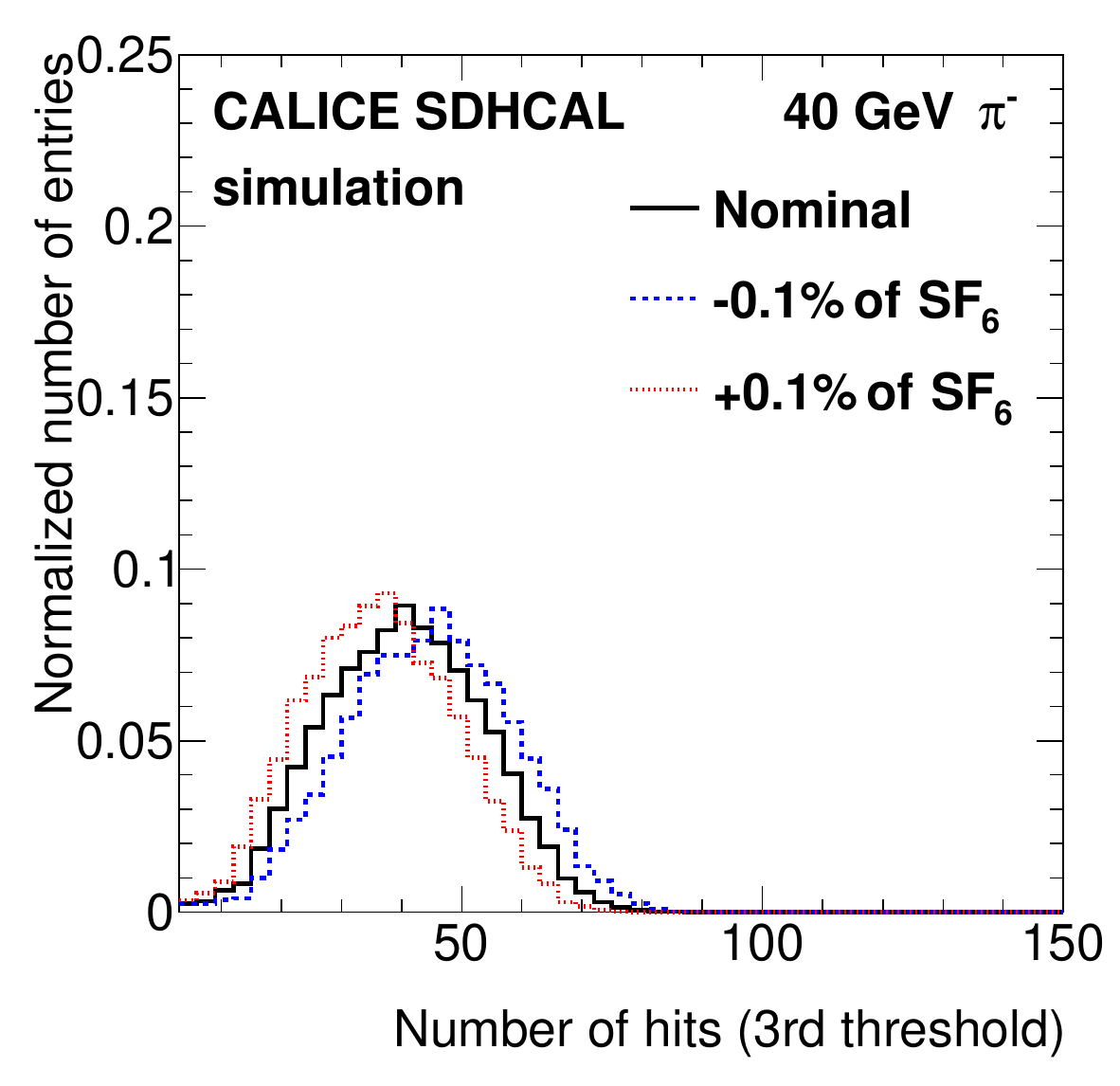}
    \caption{}
    \label{fig:nh.vs.sf.pi3}
  \end{subfigure}
  \caption{Distribution of the number of hits passing the first
    threshold at $0.1$\,pC~(\subref{fig:nh.vs.sf.1},\subref{fig:nh.vs.sf.pi1}), the second
    threshold at $5$\,pC~(\subref{fig:nh.vs.sf.2},\subref{fig:nh.vs.sf.pi2}) and the third
    threshold at $15$\,pC~(\subref{fig:nh.vs.sf.3},\subref{fig:nh.vs.sf.pi3}) for simulated  
    30\,GeV electrons (top) or 40\,GeV
    pions (bottom). The full GEANT4 simulation was performed with digitization
    modeling using the nominal \ch{SF6} component fraction (solid black histograms) and
    the \ch{SF6} fraction decreased by $0.1\%$ (dashed blue histogram) or increased by
    $0.1\%$ (dotted red histogram).\label{fig:nh.vs.sf}} 
  \end{center}
\end{figure}

\begin{figure}[t]
 \begin{center}
 \begin{subfigure}[b]{0.49\textwidth}
    \includegraphics[width=\textwidth]{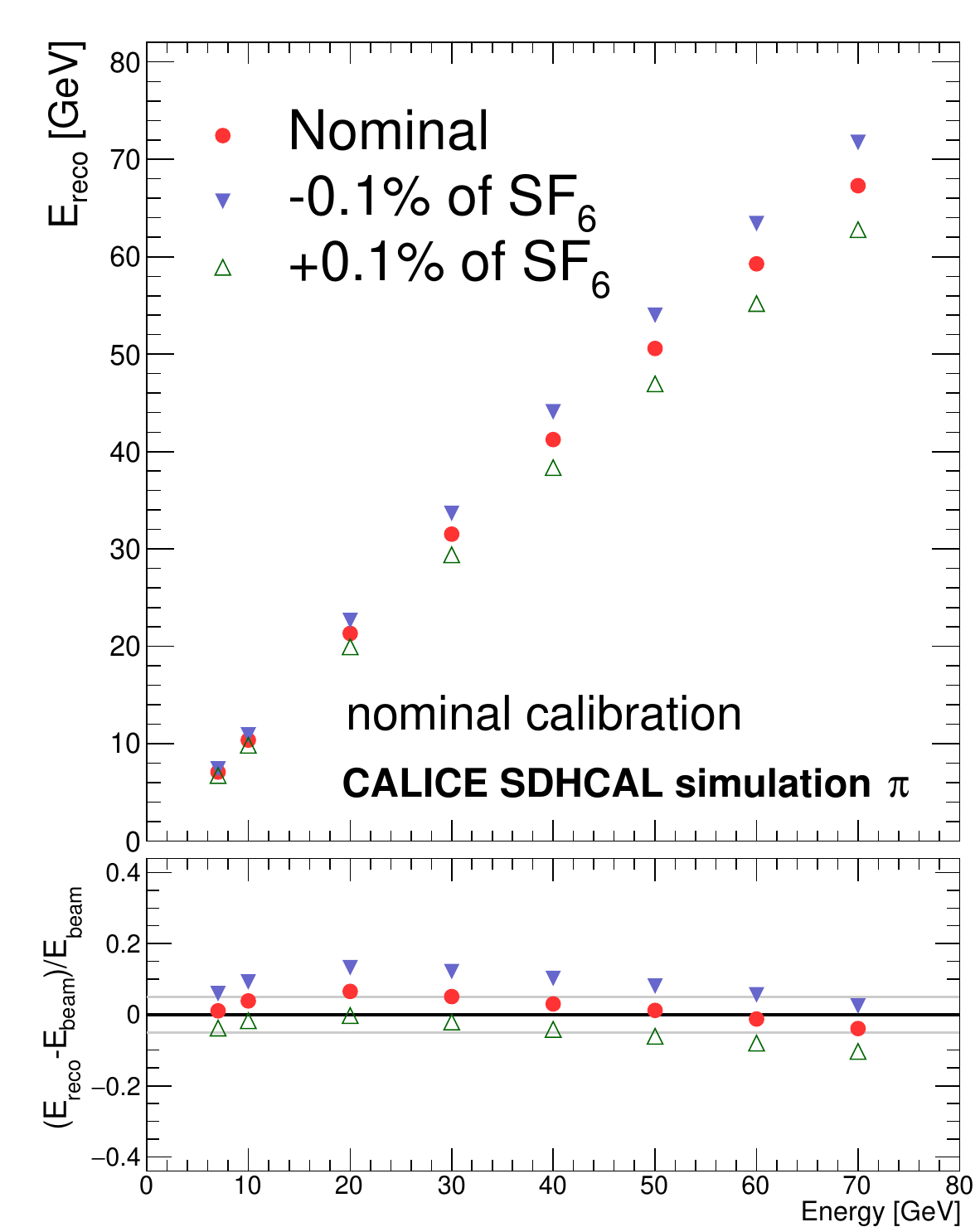}
    \caption{}
    \label{fig:ereco.pion.sf6.bias}
  \end{subfigure}
 \begin{subfigure}[b]{0.49\textwidth}
    \includegraphics[width=\textwidth]{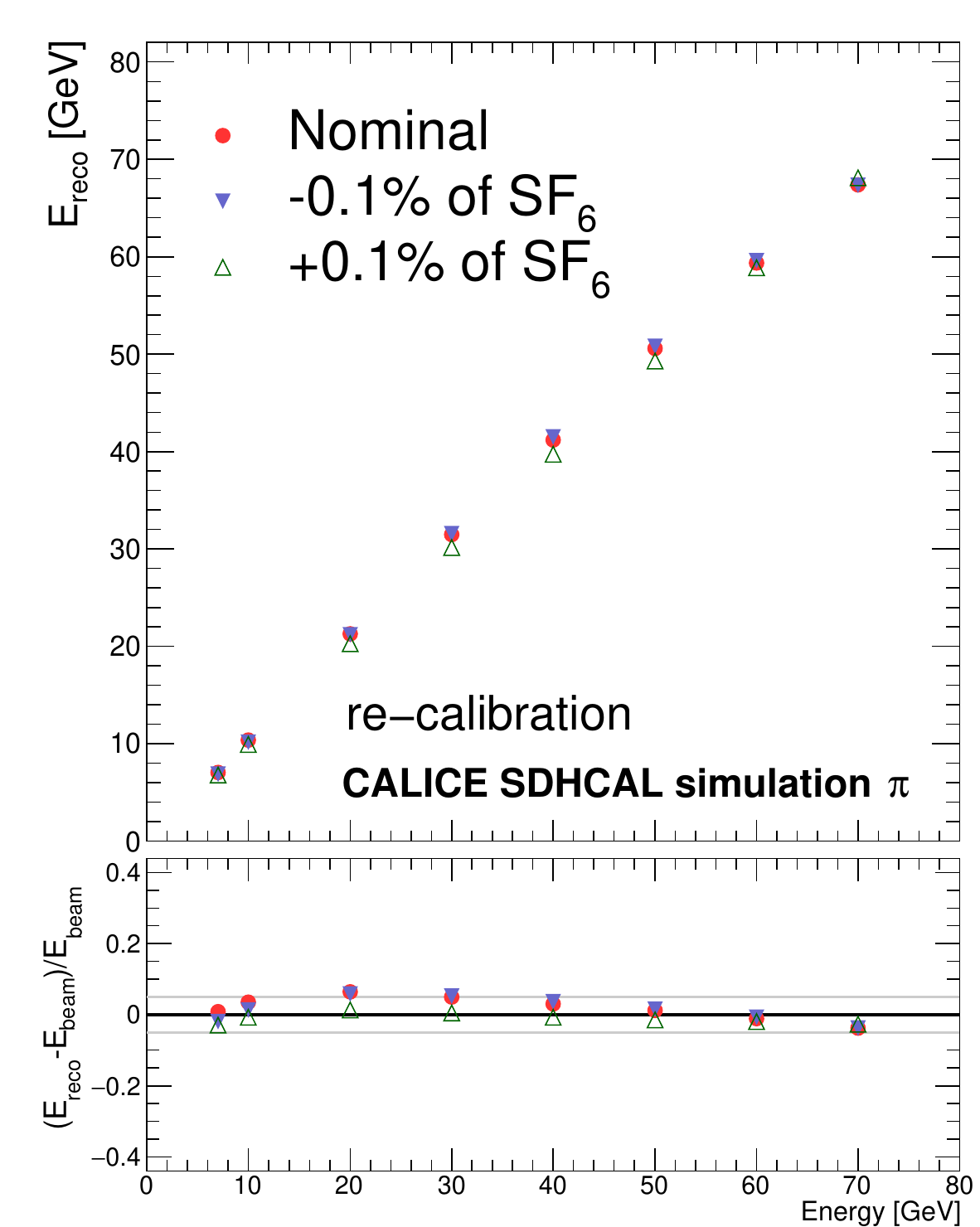}
    \caption{}
    \label{fig:ereco.pion.sf6.optimal}
  \end{subfigure}
  \caption{Pion reconstructed energy as a function of the generated energy for
    simulations with the GRPCs using the nominal \ch{SF6} component fraction (red circles),
    the \ch{SF6} component fraction decreased by $0.1\%$ (blue filled
    triangles) and increased by $+0.1\%$ (green open triangles). The energy
    calibration reconstruction factors are optimized using the nominal simulation for all cases~(\subref{fig:ereco.pion.t.bias}) or re-calibrated for each of the three simulations~(\subref{fig:ereco.pion.t.optimal}).\label{fig:ereco.lin.sf6}}
  \end{center}
\end{figure}

 This effect leads to a variation in the number of hits as
shown in Figure~\ref{fig:nh.vs.sf}. The impact on the energy reconstruction was
studied using the same approach as 
described in the previous sections. The energy is reconstructed, using the
nominal factors and from simulated events with different \ch{SF6} fractions.
 It is shown in Figure~\ref{fig:ereco.lin.sf6}
that in the absence of re-calibration, the reconstructed energy
of a 40\,GeV pion varies by $\pm20\%$ if the relative amount \ch{SF6} is varied by 
$\mp5\%$. 

For a relative variation of $\pm20\%$ of the \ch{CO2} component, i.e.\ between 4 and
6\% of the gas mixture, 
no significant effect is observed within the precision of the simulation.
However, the full impact of \ch{CO2}, especially on secondary avalanches, is
not modeled in this simulation. 

%% file: latex/data.tex
\section{Comparison with Beam Test Data}
\label{sec:data}
Among all the effects that can induce a sizable variation in the detector
response, temperature, pressure and gap width variation were classified as the most significant ones. 
The test beam data collected in 2015 allow correlating the detector response with the
temperature or pressure. However, the experimental setup cannot provide monitoring of
the mechanical structure at a scale of $10\,\micro\meter$. Furthermore, the high voltage
was adjusted in some data-taking periods.

In this section, a period where the high voltage was stable is considered. The
prototype was exposed to a beam of pions at six different energies. 
Trends are observed within the runs and are associated with changes in the
temperature or pressure and unstable
beam intensities that lead to evolving saturations during the run. 
The temperature was measured at the outer side of three
chambers which can lead to a bias with respect to the actual GRPC gas
temperature. The largest temperature difference between two runs is
$2.1\degreecelsius$. The atmospheric pressure was measured and the largest
difference between two runs is $0.6\,\milli\mathrm{bar}$.

\begin{figure}[t]
 \begin{center}
  \begin{subfigure}[b]{0.495\textwidth}
    \includegraphics[width=\textwidth]{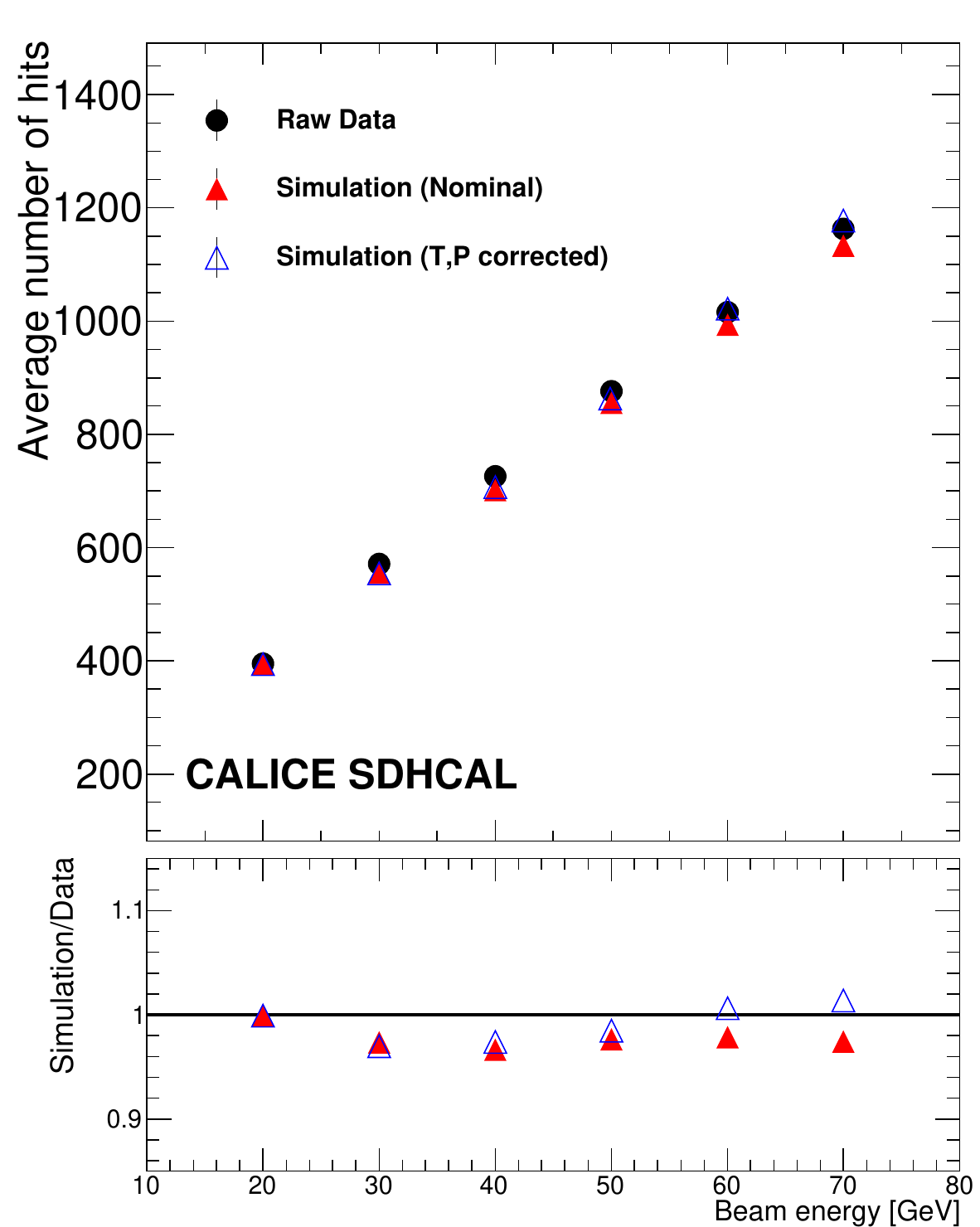}
    \caption{}
    \label{fig:nhit.linearity.2015}
  \end{subfigure}
  \begin{subfigure}[b]{0.495\textwidth}
    \includegraphics[width=\textwidth]{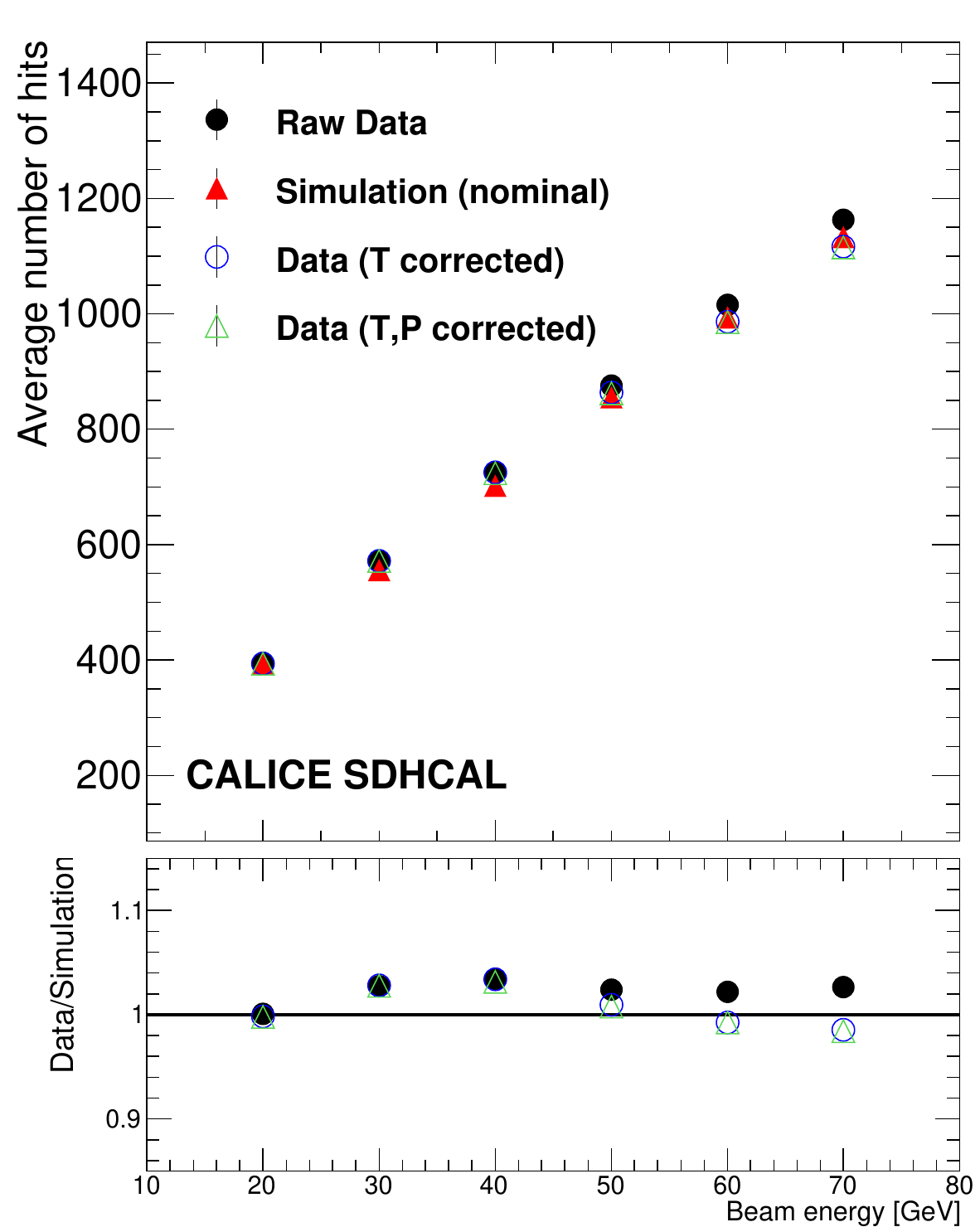}
    \caption{}
    \label{fig:nhit.linearity.2015.cor}
  \end{subfigure}
   \caption{ (\subref{fig:nhit.linearity.2015}) Evolution of the number of hits as a function of the
    pion beam energy. Uncorrected data (black dots) are compared to the nominal simulation that
    assumes stable data taking conditions
    (full red triangles) and to a simulation including the observed temperatures and pressures (empty
    triangles). (\subref{fig:nhit.linearity.2015.cor}) An event by event
    correction is applied to the data (black circles) in order to compensate
    for temperature variations (empty circles),
    temperature and pressure variations (empty triangles), compared to the
    nominal simulation (full red triangles).
    \label{fig:nhit.linearity}} 
  \end{center}
\end{figure}

The linearity of the detector response is compared to the simulated one, as seen in Figure~\ref{fig:nhit.linearity}.
Two simulations were used in Figure~\ref{fig:nhit.linearity.2015}. In the first one, stable data-taking conditions were
considered. The second simulated sample was produced using the average
observed temperature and pressure, run by run, as input leading to a better
agreement.  

Although the purpose of the simulation is to predict the stability of the detector and not to correct the data, the overall impact of the temperature on the detector response was also
checked by including an event by event correction to the number of hits. This
correction is deduced from the simulation and is a function of the temperature
and of the pressure associated to the event. 

The correction improves the agreement between data and simulation as seen on Figure~\ref{fig:nhit.linearity.2015.cor}.

\begin{figure}[t]
 \begin{center}
   \includegraphics[width=0.6\textwidth]{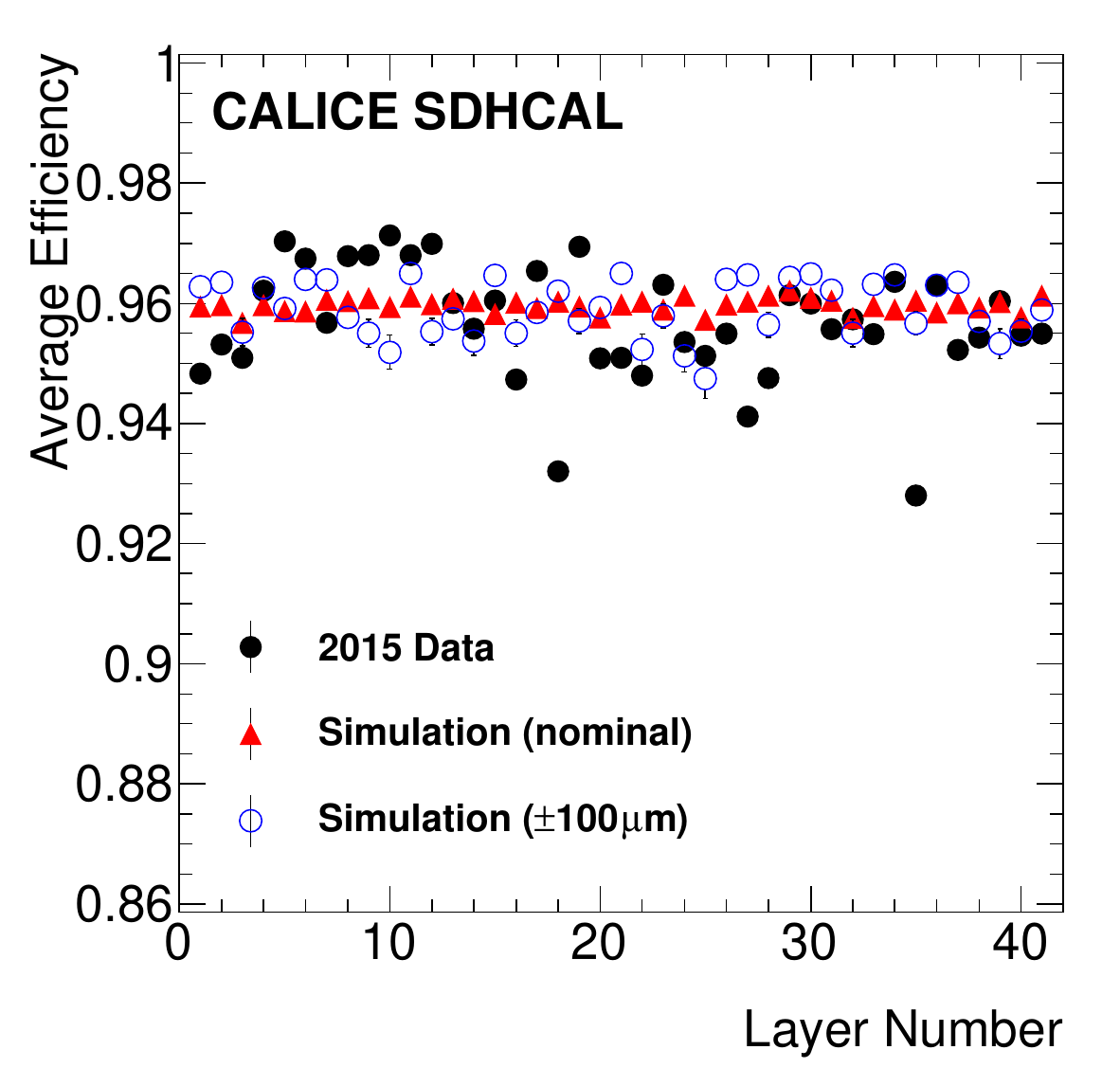}
   \caption{ Average efficiency estimated for each layer of the SDHCAL
     prototype using beam test data (black filled circles), simulation with
     digitization modeling using the nominal geometry (red triangles) and a
     geometry where the gaps are randomly varied within $\pm100\,\micro\meter$
     (blue open circles). 
    \label{fig:eff.2015}} 
  \end{center}
\end{figure}

Among all the simulated effects presented in this document, only the gap
inflation can produce random layer-to-layer variations in the detector response.
The efficiency was estimated from simulated and observed pion showers using
reconstructed tracks. The layer efficiencies are reported in
Figure~\ref{fig:eff.2015}. The digitizer algorithm accounts for the overall observed
efficiency but does not include any data-driven mapping of it. 
It was assumed
that the charge collection and readout is homogeneous. The layer-to-layer
variations observed in the nominal simulation are due to statistical
fluctuations. Simulating a $\pm100\,\micro\meter$ gap width 
tolerance introduces a sizable spread in the efficiency. However, this spread is
still smaller than what is observed in the data. The
layer-to-layer variations observed in the data reflects the inhomogeneity of the
gap widths superimposed with potential other effects not modeled in this study, like dead channels, inhomogeneous
readout or inhomogeneous layer paintings. 

%% file: latex/conclusion.tex
\section{Conclusion}

The SDHCAL digitizer algorithm was extended to include dependencies on the particle types, temperature, pressure, gap width deformations, magnetic field and gas mixture changes. It was used to model the detector response, including different scenarios. 
It was shown that when the SDHCAL technology is used with a purely digital
approach, its key performances are quite stable  regarding data-taking
conditions and potential detector inhomogeneities. The detector efficiency
variation is at the percent level while the typical variation in the total
number of hits is smaller than 5\%.
When exploiting the semi-digital information, its performances are
affected, especially when the detector is confronted with mechanical or
temperature variations. However, it was also shown that the energy linearity
could be restored via frequent calibrations with data. Online detector-based corrections
were also implemented by changing the high voltage to mitigate some
environmental effects like temperature and pressure variations. The
performances of these corrections need to be studied. 